%% file: ms.tex
\documentclass[letterpaper,11pt]{book}
\usepackage[top=3cm,bottom=3.22cm,left=4.29cm,right=4.29cm,asymmetric]{geometry}
\usepackage[T1]{fontenc}
\usepackage[utf8]{inputenc}
\usepackage{lmodern}
\usepackage{courier}
\usepackage{csquotes}
\usepackage{tabto}

\input{math.tex}
\input{proof.tex}
\input{colours.tex}
\input{orcid.tex}

\usepackage{hyperref}
\usepackage{graphicx}
\usepackage[british]{babel}
\RequirePackage[style=ieee,backend=bibtex]{biblatex}

\makeatletter
\def\blx@maxline{77}
\makeatother
\makeatletter
\newenvironment{chapquote}[2][2em]
  {\setlength{\@tempdima}{#1}
   \def\chapquote@author{#2}
   \parshape 1 \@tempdima \dimexpr\textwidth-2\@tempdima\relax%
   \itshape}
  {\par\normalfont\hfill--\ \chapquote@author\hspace*{\@tempdima}\par\bigskip}
\makeatother

\makeatletter
\newenvironment{abstract}{
  \null\vfil
  \thispagestyle{empty}
  \@beginparpenalty\@lowpenalty
  \begin{center}%
    \bfseries Abstract \@endparpenalty\@M
  \end{center}%
  }{
  \par\vfil\null
}
\makeatother

\title{\Huge \textbf{Efficient Recognition} \\ \textbf{of Graph Languages} \\[0.5cm] \huge Project Report \\[0.5cm]}
\author{\large{\textsc{Graham Campbell}\thanks{Supported by a Vacation Internship from the Engineering and Physical Sciences Research Council (EPSRC) in the UK.} \orcidicon{0000-0002-6767-2747} and \textsc{Detlef Plump} \orcidicon{0000-0002-1148-822X}} \\[0.1cm]
\large{\textsc{Department of Computer Science}} \\[0.1cm]
\large{\textsc{University of York, United Kingdom}} \\[0.1cm]
}
\date{September 2019\\~\\Revised February 2020}

\begin{document}

\frontmatter
\maketitle

\clearpage
\thispagestyle{empty}
\cleardoublepage

\begin{abstract}
Graph transformation is the rule-based modification of graphs, and is a discipline dating back to the 1970s. In general, to match the left-hand graph of a fixed rule within a host graph requires polynomial time, but to improve matching performance, D\"orr proposed to equip rules and host graphs with distinguished root nodes. This model was implemented by Plump and Bak, but unfortunately, such rules are not invertible. We address this problem by defining rootedness using a partial function into a two-point set rather than pointing graphs with root nodes, meaning derivations are natural double pushouts. Moreover, we give a sufficient condition on rules to give constant time rule application on graphs of bounded degree, and that, the graph class of trees can be recognised in linear time, given an input graph of bounded degree. Finally, we define a new notion of confluence up to garbage and non-garbage critical pairs, showing it is sufficient to require strong joinability of only the non-garbage critical pairs to establish confluence up to garbage. Finally, this new result, presented for conventional graph transformation systems, can be lifted to our rooted setting by encoding node labels and rootedness as looped edges.
\end{abstract}

\clearpage
\thispagestyle{empty}
\cleardoublepage
\phantomsection

\tableofcontents

\cleardoublepage
\phantomsection
\addcontentsline{toc}{chapter}{\listfigurename}
\listoffigures

\renewcommand{\listtheoremname}{List of Theorems}
\cleardoublepage
\phantomsection
\addcontentsline{toc}{chapter}{\listtheoremname}
\listoftheorems[ignoreall,show=theorem]

\input{content/0}

\mainmatter

\input{content/1}
\input{content/2}
\addtocontents{toc}{\protect\newpage}
\input{content/3}
\input{content/4}
\input{content/5}
\input{content/6}

\appendix
\input{content/a}
\input{content/b}
\addtocontents{toc}{\protect\newpage}
\input{content/c}
\input{content/d}
\input{content/e}

\begingroup
\sloppy
\cleardoublepage
\phantomsection
\addcontentsline{toc}{chapter}{Bibliography}
\printbibliography
\endgroup

\end{document}

%% file: math.tex
\usepackage[table,xcdraw]{xcolor}

\usepackage{amssymb,amsmath,amsthm,enumitem,tikz-cd,adjustbox,proof,float,stmaryrd,wasysym,wrapfig,multicol,thmtools,subcaption,pgfplots}
\usetikzlibrary{shapes,snakes,patterns}

\theoremstyle{definition}
\newtheorem{definition}{Definition}[chapter]
\newtheorem{proposition}{Proposition}[chapter]
\newtheorem{lemma}{Lemma}[chapter]
\newtheorem{theorem}{Theorem}[chapter]
\newtheorem{corollary}{Corollary}[chapter]
\newtheorem{remark}{Remark}[chapter]

\newtheorem{examplex}{Example}[chapter]
\newenvironment{example}
  {\pushQED{\qed}\examplex}
  {\popQED\endexamplex}

\usepackage{scalerel}
\usepackage{mathtools}
\newcommand{\defeq}{\,\,\,\vcentcolon=\,\,}
\newcommand\abs[1]{\ensuremath{\lvert#1\rvert}}

\newcommand{\overbar}[1]{\mkern 1.5mu\overline{\mkern-1.5mu#1\mkern-1.5mu}\mkern 1.5mu}

\newcommand\restr[2]{{
  #1\!\!\mid_{\scaleto{#2}{6pt}}
}}

\newenvironment{allintypewriter}{\ttfamily}{\par}

%% file: proof.tex
\makeatletter
\renewenvironment{proof}[1][\proofname]{\par
  \vspace{-\topsep}
  \pushQED{\qed}%
  \normalfont
  \topsep0pt \partopsep0pt 
  \trivlist
  \item[\hskip\labelsep
        \itshape
    #1\@addpunct{.}]\ignorespaces
}{%
  \popQED\endtrivlist\@endpefalse
  \addvspace{6pt plus 6pt} 
}
\makeatother

%% file: colours.tex
\definecolor{gp2green}{RGB}{69, 191, 156}
\definecolor{gp2blue}{RGB}{153, 187, 255}
\definecolor{gp2red}{RGB}{236, 107, 116}
\definecolor{gp2pink}{RGB}{239, 161, 193}
\definecolor{gp2grey}{RGB}{196, 192, 200}
\definecolor{performanceBlue}{RGB}{0, 136, 255}
\definecolor{performanceYellow}{RGB}{252, 199, 17}

%% file: orcid.tex
\usetikzlibrary{svg.path}

\definecolor{orcidlogocol}{HTML}{A6CE39}
\tikzset{
  orcidlogo/.pic={
    \fill[orcidlogocol] svg{M256,128c0,70.7-57.3,128-128,128C57.3,256,0,198.7,0,128C0,57.3,57.3,0,128,0C198.7,0,256,57.3,256,128z};
    \fill[white] svg{M86.3,186.2H70.9V79.1h15.4v48.4V186.2z}
                 svg{M108.9,79.1h41.6c39.6,0,57,28.3,57,53.6c0,27.5-21.5,53.6-56.8,53.6h-41.8V79.1z M124.3,172.4h24.5c34.9,0,42.9-26.5,42.9-39.7c0-21.5-13.7-39.7-43.7-39.7h-23.7V172.4z}
                 svg{M88.7,56.8c0,5.5-4.5,10.1-10.1,10.1c-5.6,0-10.1-4.6-10.1-10.1c0-5.6,4.5-10.1,10.1-10.1C84.2,46.7,88.7,51.3,88.7,56.8z};
  }
}

\newcommand\orcidicon[1]{\href{https://orcid.org/#1}{\!\textsuperscript{\mbox{\scalerel*{
\begin{tikzpicture}[yscale=-1,transform shape]
\pic{orcidlogo};
\end{tikzpicture}
}{|}}}}}

%% file: content/0.tex
\chapter{Executive Summary}

\section*{Motivation and Goals}

Graph transformation is the rule-based modification of graphs, and is a discipline dating back to the 1970s, with the \enquote{algebraic approach} invented at the Technical University of Berlin by Ehrig, Pfender, and Schneider \cite{Ehrig-Pfender-Schneider73a,Ehrig79a}. It is a comprehensive framework in which the local transformation of structures can be modelled and studied in a uniform manner \cite{Corradini-Montanari-Rossi-Ehrig-Heckel-Lowe97a,Ehrig-Heckel-Korff-Lowe-Ribeiro-Wagner-Corradini97a,Ehrig-Ehrig-Prange-Taentzer06a}. Applications in Computer Science are wide-reaching including compiler construction, software engineering and concurrent systems \cite{Corradini-Rossi-ParisiPresicce91a,Corradini-Ehrig-Kreowski-Rozenberg02a,Ehrig-Engels-ParisiPresicce-Rozenberg04a,Corradini-Ehrig-Montanari-Ribeiro-Rozenberg06a}.

The declarative nature of graph rewriting rules comes at a cost: In general, to match the left-hand graph \(L\) of a rule within a host graph \(G\) requires \(|G|^{|L|}\) time. To improve matching performance, D\"orr \cite{Dorr95a} proposed to equip rules and host graphs with distinguished (root) nodes, and to match roots in rules with roots in host graphs. This concept has been implemented in GP 2, allowing GP 2 specifications to rival the performance of traditional implementations in languages such as C \cite{Bak15a}. Automated refinement of graph rewriting specifications to rooted versions remains an open research problem.

Graph transformation with root nodes and relabelling is not yet well understood. With only relabelling, Habel and Plump have been able to recover many, but not all, of the standard results \cite{Habel-Plump02a,Habel-Plump12a}. Moreover, Bak and Plump's model suffers from the problem that derivations are not necessarily invertible. This motivates us to develop a new model of rooted graph transformation with relabelling which does not suffer this problem. If we have termination and invertibility, then we have an algorithm for testing graph language membership, and if we have confluence (and constant time matching), then we have an efficient algorithm too \cite{Dodds-Plump06a,Plump10a}.

Testing for \enquote{confluence} is not possible in general \cite{Plump93b}, however we can sometimes use \enquote{critical pair} analysis to show confluence. Confluence remains poorly understood, and while there are techniques for classifying \enquote{conflicts} \cite{Ehrig-Lambers-Orejas08a,Hristakiev18a,Lambers-Kosiol-Struber-Taentzer19a}, it is rarely possible to actually show confluence. Moreover, in general, confluence is stronger than required for language efficient membership testing, motivating a weaker definition of confluence.

Our method will be to use mathematical definitions and proofs, as is usual in theoretical computer science. We aim to:

\begin{enumerate}[itemsep=-0.4ex,topsep=-0.4ex]
\item Outline rooted DPO graph transformation with relabelling;
\item Repair the problem of lack of invertibility in rooted GT systems;
\item Develop a new example of linear time graph algorithm;
\item Develop new results for confluence analysis of GT systems.
\end{enumerate}

\section*{Results and Evaluation}

We regard this project as a success, both the original and extended versions, having achieved our four goals. Our first, second and third goals have been addressed by the first, second and third chapters, respectively, and the final goal by the next two chapters. We started by briefly reviewing the current state of graph transformation, with a particular focus on the \enquote{injective DPO} approach with relabelling and graph programming languages, establishing issues with the current approach to rooted graph transformation due to its \enquote{pointed} implementation. We also briefly reviewed DPO-based graph programming languages.

We address the lack of invertibility of rooted derivations by defining rootedness using a partial function onto a two-point set rather than pointing graphs with root nodes. We have shown rule application corresponds to \enquote{NDPOs}, how Dodds' complexity theory \cite{Dodds08a} applies in our system, and briefly discussed the equivalence of and refinement of GT systems. Developing a fully-fledged theory of correctness and refinement for (rooted) GT systems remains future work, as does establishing if the Parallelism and Concurrency theorems hold \cite{Ehrig-Golas-Habel-Lambers-Orejas14a,Habel-Plump12a}. Applications of our model to efficient graph class recognition are exciting due to the invertibility of derivations.

We have shown a new result that the graph class of trees can be recognised by a rooted GT system in linear time, given an input graph of bounded degree. Moreover, we have given empirical evidence by implementing the algorithm in GP\,2 and collecting timing results. The program and results were presented at CALCO 2019 \cite{Campbell-Courtehoute-Plump19b}. Overcoming the restriction of host graphs to be of bounded degree remains open research, as well as showing further case studies and applications.

We have defined a new notion of \enquote{confluence up to garbage} and \enquote{non-garbage critical pairs}, and shown that it is sufficient to require strong joinability of only the non-garbage critical pairs to establish confluence up to garbage. Moreover, we have lifted our results to our new framework for rooted graph transformation with relabelling via a faithful functor. We have applied this theory to Extended Flow Diagrams \cite{Farrow-Kennedy-Zucconi76a} and the encoding of partially labelled (rooted) GT systems as standard GT systems, performing non-garbage critical pair analysis on the encoded system. Further exploring the relationship between confluence up to garbage and closedness remains open work, as does improving the analysis of (non-garbage) critical pairs to allow us to decide confluence in more cases than currently possible via pair analysis.

\section*{Ethical Considerations}

This project is of a theoretical nature. As such, no human participants were required, and no confidential data has been collected. Moreover, there are no anticipated ethical implications of this work or its applications.

\chapter{Preface}

This report is an extended version of the BSc Thesis of Graham Campbell \cite{Campbell19a} which incorporates additional results developed over 10 weeks, funded by a Vacation Internship of the Engineering and Physical Sciences Research Council (EPSRC) and supervised by Dr. Detlef Plump. Part of Chapter \ref{chapter:treerecognision} was presented at CALCO 2019 as part of a co-authored paper looking at linear time algorithms in GP\,2 \cite{Campbell-Courtehoute-Plump19b}.

Compared to the original thesis, the focus has been even more on fast language recognition. We have provided a more detailed introduction and theoretical background, and more care has been taken when defining the new theory, due to the page limit being removed. We have separated the semantic definitions of bounded degree and bounded roots preservation from the syntactic condition, which gives clarity that the syntactic condition is both sufficient and necessary. We have also provided a linear time full binary tree recognition system on graphs of bounded degree.

Additionally, in the second half of the report, we have reworked the previous definitions of confluence up to garbage and closure, changing, give an result that closedness is undecidable in general, and an example showing that a system can be locally confluent up to garbage and terminating, but not confluent up to garbage. Finally, we give a formal encoding of our new rooted GT systems as standard GT systems, which allows us to recover the Local Church-Rosser Theorem and the Non-Garbage Critical Pair Lemma in our new setting.

The structure of the report is as follows. We:

\begin{enumerate}[itemsep=-0.4ex,topsep=-0.4ex]
\item Provide a quick introduction and outline the theoretical background;
\item Present our new framework for rooted DPO graph transformation with relabelling and invertible rules;
\item Look at complexity of derivations in our new framework, and show that the framework is sufficient to recognise trees in linear time;
\item Develop new results for confluence analysis of conventional GT systems;
\item Lift our new results to the rooted framework via a faithful functor;
\item Summarise and evaluate our results, and list future work.
\end{enumerate}

In the appendices, we have provided an overview of the notation used for sets, functions, relations, categories, and abstract reduction systems, and also some key results that we will be needing, from the literature. We also provide supplementary technical background on \enquote{conventional} graph transformation, and graph theory, including the definitions of various graph classes.

During the period covered by the grant, Campbell also worked on improving the GP\,2 Compiler. That work has not been summarised in this report, however is available in a separate report \cite{Campbell-Romo-Plump19a}.

%% file: content/1.tex
\chapter{Theoretical Background} \label{chapter:theoryintro}

\begin{chapquote}{David Hilbert, \textit{N Rose Mathematical Maxims and Minims (1988)}}
``Mathematics is a game played according to certain simple rules with meaningless marks on paper.''
\end{chapquote}

Graph transformation is the rule-based modification of graphs, and is a discipline dating back to the 1970s, with the \enquote{algebraic approach} invented at the Technical University of Berlin by Ehrig, Pfender, and Schneider \cite{Ehrig-Pfender-Schneider73a,Ehrig79a}. It is a comprehensive framework in which the local transformation of structures can be modelled and studied in a uniform manner \cite{Corradini-Montanari-Rossi-Ehrig-Heckel-Lowe97a,Ehrig-Heckel-Korff-Lowe-Ribeiro-Wagner-Corradini97a,Ehrig-Ehrig-Prange-Taentzer06a}.

In this chapter, we will review the rewriting of totally labelled graphs with relabelling, and Bak and Plump's modifications adding \enquote{root} nodes \cite{Bak-Plump12a}. We will see how (rooted) graph transformation systems are instances of abstract reduction systems, and will look at graph programming languages.


\section{Graphs and Morphisms} \label{sec:maingraphs}

There are various definitions of a \enquote{graph}. In particular, we are interested in graphs where edges are directed and parallel edges are permitted. We will start by introducing \enquote{unlabelled} graphs. We will then build on these with (partially) labelled (rooted) graphs and morphisms.

\begin{definition} \label{def:graph}
We can formally define a \textbf{concrete graph} as:
\begin{align*}
G = (V, E, s: E \to V, t: E \to V)
\end{align*}
where \(V\) is a \textbf{finite} set of \textbf{vertices}, \(E\) is a \textbf{finite} set of \textbf{edges}. We call \(s: E \to V\) the \textbf{source} function, and \(t: E \to V\) the \textbf{target} function.
\end{definition}

\begin{definition} \label{def:graphsize}
If \(G\) is a \textbf{concrete graph}, then \(\abs{G} = \abs{V_G} + \abs{E_G}\).
\end{definition}

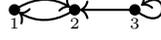
\begin{figure}[H]
\centering
\noindent
\input{fig/1/eg1}
\vspace{-0.5em}
\caption{Example Concrete Graph}
\label{fig:eg1}
\end{figure}

\begin{example}
Consider the concrete graph \(G = (\{1, 2, 3\}, \{a, b, c, d\}, s, t)\) where \(s = \{(a, 1), (b, 2), (c, 3), (d, 3)\}\), \(t = \{(a, 2), (b, 1), (c, 1), (d, 3)\}\) (treating functions as sets). Its graphical representation is given in Figure \ref{fig:eg1}. Note that the numbers are not \enquote{labels}, but \enquote{node ids}.
\end{example}

\vspace{-1.3em}

\begin{definition}\label{def:morph}
Given two concrete graphs \(G\) and \(H\), a \textbf{graph morphism} \(g: G \to H\) is a pair of maps \(g = (g_V: V_G \to V_H, g_E: E_G \to E_H)\) such that sources and targets are preserved. That is, \(\forall e \in E_G,\) \(g_V(s_G(e))\) \(= s_H(g_E(e))\) and \(g_V(t_G(e)) = t_H(g_E(e))\). Equivalently, both of the squares in Figure \ref{fig:comsqrsmorph} commute.
\end{definition}

\vspace{-0.8em}
\begin{figure}[H]
\centering
\noindent
\input{fig/1/morphism}
\vspace{-1.0em}
\caption{Graph Morphism Commuting Diagrams}
\label{fig:comsqrsmorph}
\end{figure}

\begin{definition}
A graph morphism \(g: G \to H\) is \textbf{injective}/\textbf{surjective} iff both \(g_V\) and \(g_E\) are injective/surjective as functions. We say \(g\) is an \textbf{isomorphism} iff it is both injective and surjective.
\end{definition}

\begin{example}
The identity morphism \((id_V, id_E)\) is an isomorphism between any graph and itself.
\end{example}

\begin{figure}[H]
\centering
\noindent
\input{fig/1/eg2}
\vspace{-0.6em}
\caption{Example Concrete Graphs}
\label{fig:eg2}
\end{figure}
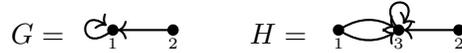

\begin{example} \label{eg:graphmorpheg1}
Consider the graphs in Figure \ref{fig:eg2}. There are four morphisms \(G \to H\), three of which are injective, none of which are surjective. There are actually also four morphisms \(H \to G\), three of which are surjective.
\end{example}

\vspace{-1.5em}

\begin{definition}
We say that graphs \(G, H\) are \textbf{isomorphic} iff there exists a \textbf{graph isomorphism} \(g: G \to H\), and we write \(G \cong H\). This naturally gives rise to \textbf{equivalence classes} \([G]\), called \textbf{abstract graphs}.
\end{definition}

\begin{proposition}
The \textbf{quotient} (Definition \ref{dfn:quotient}) of the \textbf{collection} of all \textbf{concrete graphs} with \(\cong\) is the \textbf{countable set} of all \textbf{abstract graphs}.
\end{proposition}


\section{Graph Transformation} \label{sec:gtintro}

There are various approaches to graph transformation, most notably the \enquote{edge replacement} \cite{Drewes-Kreowski-Habel97a}, \enquote{node replacement} \cite{Engelfriet-Rozenberg97a}, and \enquote{algebraic} approaches \cite{Corradini-Montanari-Rossi-Ehrig-Heckel-Lowe97a,Ehrig-Heckel-Korff-Lowe-Ribeiro-Wagner-Corradini97a}. The two major approaches to algebraic graph transformation are the so called \enquote{double pushout} (DPO) approach, and the \enquote{single pushout} (SPO) approach. Because the DPO approach operates in a structure-preserving manner (rule application in SPO is without an interface graph, so there are no dangling condition checks), this approach is more widely used than the SPO \cite[p.9-14]{Ehrig-Ehrig-Prange-Taentzer06a,Ehrig-Heckel-Korff-Lowe-Ribeiro-Wagner-Corradini97a}. For this reason, we will focus only on the DPO approach with injective matching.

Given an unlabelled graph (Definition \ref{def:graph}), there are two common approaches to augmenting it with data: \textbf{typed graphs} and \textbf{totally labelled graphs}. We choose to work with the \textbf{labelled approach} (Section \ref{sec:graphpl}) because it is easy to understand and reason about, has a \textbf{relabelling} theory (Section \ref{section:relabelling}), and a \enquote{rooted} modification (Section \ref{section:rooted}). Details of the \textbf{typed approach} can be found in Section \ref{section:typed}. Note that typed (attributed) (hyper)graphs have a rich theory \cite{Lowe-Korff-Wagner93a,Corradini-Montanari-Rossi96a,Berthold-Fischer-Koch00a,Heckel-Kuster-Taentzer02a,Ehrig-Prange-Taentzer04a,Plump10a,Hristakiev-Plump16a,Peuser-Habel16a}.

A review of graph transformation of labelled graphs using the DPO approach with injective matching can be found in Appendix \ref{appendix:transformation}. We will also cover the definitions and results for our new type of system in Chapter \ref{chapter:newtheory}, so we will not repeat ourselves in this chapter by giving all of the detail again. Additionally, an example system and grammar can be found in Chapter \ref{chapter:treerecognision}.


\section{Adding Relabelling} \label{section:relabelling}

The origin of partially labelled graphs is from the desire to have \enquote{relabelling}. If the interface \(K\) is totally labelled, then any node which has context (incident edges) cannot be deleted, and so we must preserve its label to avoid breaking uniqueness of rule application. We can get around this problem with partial labelling of interface graphs, and thus with modest modifications to the theory for totally labelled graphs we allow rules to \enquote{relabel} nodes. We shall be using this foundation going forward. All the relevant definitions and theorems are in Appendix \ref{appendix:transformation}.

We are in fact using a restricted version of the theory presented by Habel and Plump \cite{Habel-Plump02a}, the restriction being that we allow the interface \(K\) to be partially labelled, but require \(L\), \(R\) and \(G\) to be totally labelled, ensuring that given a totally labelled input graph \(G\), the result graph \(H\) is also totally labelled. Thus, derivations are defined only on totally labelled graphs, but allow us to relabel nodes.

\begin{example}
Consider the following totally labelled \enquote{rule}, over the label alphabet \((\{1, 2\}, \{\Square\})\) where \(x, y\) are to be determined:

\begin{figure}[H]
\centering
\noindent
\input{fig/1/relabelling}
\vspace{-0.3em}
\caption{Relabelling Non-Example}
\end{figure}
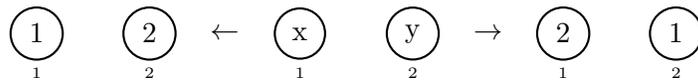

We want to swap the labels without deleting the nodes, because they may have context. There is no value we can choose for \(x\) or \(y\) such that the conditions to be a totally labelled graph morphism are satisfied. Now consider the setting where we allow the interface graph to have a partial node label map. We could simply not label the interface nodes, and then we have exactly what we want.
\end{example}


\section{Rooted Graph Transformation} \label{section:rooted}

Rooted graph transformation first appeared when D\"orr \cite{Dorr95a} proposed to equip rules and host graphs with distinguished (root) nodes, and to match roots in rules with roots in host graphs. Root nodes were first considered in the DPO setting by Dodds and Plump \cite{Dodds-Plump06a,Dodds08a}, and most recently, Bak and Plump \cite{Bak-Plump12a,Bak15a} have used rooted graph transformation in conjunction with the theory of partially labelled graph transformation in GP\,2.

The motivation for root nodes is to improve the complexity of finding a match of the left-hand graph \(L\) of a rule within a host graph \(G\). In general, linear time graph algorithms may, instead, take polynomial time when expressed as graph transformation systems \cite{Geiss-Batz-Grund-Hack-Szalkowski06a,Dodds-Plump06a,Bak-Plump12a,Campbell-Romo-Plump18a}. An excellent account of this is available in Part II of Dodds' Thesis \cite{Dodds08a}.

We can define rooted graphs in a pointed style, just as for typed graphs. An account of the theoretical modifications is provided in Section \ref{section:rootedgt}, using Bak's approach \cite{Bak-Plump12a}. This is different to Dodds' approach \cite{Dodds-Plump06a,Dodds08a} which implemented roots as a special node label. Bak's approach allows nodes to have genuine labels and also be rooted, and has been implemented in GP\,2.

We can formalise the problem of applying a rule:

\begin{definition}[Graph Matching Problem (GMP)] \label{dfn:gmp}
Given a graph \(G\) and a rule \(r = \langle L \leftarrow K \rightarrow R \rangle\), find the set of injective graph morphisms \(L \to G\).
\end{definition}

\begin{definition}[Rule Application Problem (RAP)] \label{dfn:rap}
Given a graph \(G\), a rule \(r = \langle L \leftarrow K \rightarrow R \rangle\), and an injective match \(g: L \to G\), find the result graph \(H\). That is, does it satisfy the \enquote{dangling condition}, and if so, construct \(H\).
\end{definition}

\begin{proposition} \label{prop:gmptime}
The GMP requires time \(O(\abs{G}^{\abs{L}})\) time given the assumptions in Figure \ref{fig:perfass}, derived from Dodds' PhD Thesis \cite{Dodds08a}. Moreover, given a match, one can decide if it is applicable in \(O(|r|)\) time. That is, the RAP requires \(O(|r|)\) time. \cite{Dodds08a}
\end{proposition}

To improve matching performance, one can add root nodes to rules and match roots in rules with roots in host graphs, meaning we need only consider subgraphs of bounded size for matching, vastly improving the time complexity. That is, given a graph \(G\) of bounded degree (Definition \ref{dfn:boundeddegree}) containing a bounded number of root nodes, and a rule of bounded size with \(L\) containing a single root node, then the time complexity of GMP is only constant \cite{Dodds08a}.

\vspace{-0.8em}
\begin{figure}[H]
\noindent
\input{fig/1/complexity}
\vspace{-1.0em}
\caption{Complexity Assumptions Table}
\label{fig:perfass}
\end{figure}
\vspace{-0.2em}

\begin{figure}[H]
\centering
\noindent
\input{fig/1/rooted}
\vspace{-0.2em}
\caption{Example Rooted Rule}
\label{fig:egrootedrule}
\end{figure}
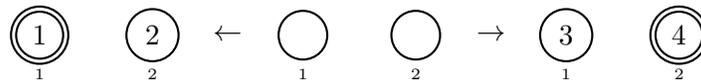

\begin{example}
Figure \ref{fig:egrootedrule} \enquote{moves the root node} and also \enquote{relabels} the nodes in the host graph. A \enquote{fast} rooted implementation of the 2-colouring problem is available at \cite{Bak-Plump12a}, showcasing root nodes in GP\,2.
\end{example}

\vspace{-0.8em}
\vspace{-0.4em}

We will revisit time complexity in Section \ref{section:complexitythms}, showing that if rules are of a certain type, then derivations take only constant time, allowing us to use only derivation length as a measure of time complexity, as in standard complexity analysis theory for (non-deterministic) Turing Machines, first considered by Hartmanis and Stearns in 1965 \cite{Hartmanis-Stearns65a}.


\section{Graph Languages}

Intuitively, a graph language is simply a set of graphs, just like a string language is a set of strings. Just like we can define string languages using string grammars, we can define graph languages using graph grammars, where we rewrite some start graph using a set of graph transformation rules. Derived graphs are then defined to be in the language if they are terminally labelled. We formally define graph languages in Section \ref{section:gt2}.

\begin{definition} \label{dfn:gl}
Let \(\mathcal{L}\) be some fixed label alphabet (Definition \ref{dfn:labelalphabet}). We let \(\mathcal{G}(\mathcal{L})\) be the \textbf{collection} of all totally labelled \textbf{abstract graphs}, and \(\widehat{\mathcal{G}}(\mathcal{L})\) be the \textbf{collection} of all totally labelled, totally rooted \textbf{abstract graphs}.
\end{definition}

\begin{proposition} \label{prop:graphuniverse}
Given some \(\mathcal{L}\), \(\mathcal{G}(\mathcal{L})\) and \(\widehat{\mathcal{G}}(\mathcal{L})\) are \textbf{countable sets}.
\end{proposition}

There are only countably many abstract graphs; an essential property when setting up graph languages. We would be in trouble if the universal language was not countable, or not even a set! Naming of elements from uncountable sets requires non-standard theory (such as Type-2 Computability Theory \cite{Weihrauch13a}), and we get into even more trouble when we want to name elements from sets larger than the continuum cardinality. When we make computability statements about languages, we are tacitly assuming a G{\"o}del numbering \cite{Godel31a}.

We choose to define languages as abstract graphs, so that they are countable sets, rather than proper classes. Moreover, when one talks of a finite graph language, we can genuinely be referring to its cardinality, and not the cardinality of its quotient under graph isomorphism. This approach is not standardised, and many authors prefer to define graph languages in terms of concrete graphs. The other obvious approach (used by Courcelle and Engelfriet \cite{Courcelle-Engelfriet12a}) is to restrict the ambient from which nodes and edges are drawn from to some countable set which is closed under pairing (to allow for the disjoint union construction when computing pushouts).

Uesu showed in 1978 that the DPO graph grammars can generate every \textbf{recursively enumerable} (Definition \ref{dfn:semidecidable}) set of graphs \cite{Uesu78a}. A large body of literature exists when it comes to context-free graph grammars, however the focus is on (hyper)edge replacement grammars \cite{Drewes-Kreowski-Habel97a} and node replacement grammars \cite{Engelfriet-Rozenberg97a}, which are less powerful than DPO-grammars. We have provided a formal definition of edge replacement grammars in Section \ref{section:edgereplgram}.


\section{Abstract Reduction Systems} \label{section:arsgt}

\textbf{Abstract reduction systems} (or simply \textbf{reduction systems} or \textbf{ARS}) are a much more general setting than \textbf{graph transformation systems} (\textbf{GT systems} or \textbf{GTS}), and model the step-wise transformation of objects (see Appendix \ref{appendix:ars}). These systems were studied for the first time by Newman in the early 40s \cite{Newman42a}. Turing Machines and GT systems clearly fit into this model of reduction. Moreover, the formal semantics of programming languages is often defined in terms of a step-wise computation relation.

\begin{example}
\((\mathbb{N}, >)\) is a \textbf{terminating} (Definition \ref{dfn:arsterm}), \textbf{finitely branching} (Definition \ref{dfn:branching}), \textbf{confluent} (Definition \ref{dfn:arsconf}) ARS (Definition \ref{dfn:ars0}).
\end{example}

\begin{example}
\((\mathbb{Z}, >)\) by comparison is not \textbf{terminating} or \textbf{finitely branching}, but it is \textbf{confluent}!
\end{example}

\begin{definition}
Let \(T = (\mathcal{L}, \mathcal{R})\) be a (rooted) \textbf{GTS}. Then \((\mathcal{G}(\mathcal{L}), \rightarrow_{\mathcal{R}})\) is the \textbf{induced ARS} defined by \(\forall [G], [H] \in \mathcal{G}(\mathcal{L}), [G] \rightarrow_{\mathcal{R}} [H]\) iff \(G \Rightarrow_{\mathcal{R}} H\).
\end{definition}

\begin{lemma} \label{lem:gtprops}
Consider the \textbf{ARS} \((\mathcal{G}(\mathcal{L}), \rightarrow)\) induced by a (rooted) \textbf{GTS}. Then \(\rightarrow\) is a \textbf{binary relation} (Definition \ref{def:binrel}) on \(\mathcal{G}(\mathcal{L})\). Moreover, it is \textbf{finitely branching} (Definition \ref{dfn:branching}) and \textbf{decidable} (Definition \ref{dfn:decidable}).
\end{lemma}

\begin{proof}
By Proposition \ref{prop:graphuniverse}, \(\mathcal{G}(\mathcal{L})\) is a countable set, and so \(\rightarrow\) is a countable set (by Theorem \ref{thm:setprod}), and is well-defined since derivations are unique up to isomorphism (Theorem \ref{thm:uniquederivations}). Finally, we have only finitely many rules, and for each rule, there can only exist finitely many matches \(L \to G\), so there can only ever be finitely many result graphs \(H\) (up to isomorphism) \(G \Rightarrow_{\mathcal{R}} H\) for any given \(G\).
\end{proof}

\begin{theorem}[Property Undecidability] \label{thm:undecidablegtsrealdeal}
Consider the \textbf{ARS} \((\mathcal{G}(\mathcal{L}), \rightarrow)\) induced by a (rooted) \textbf{GTS}. Then testing if \(\rightarrow\) is \textbf{terminating}, \textbf{acyclic}, or (\textbf{locally}) \textbf{confluent} is \textbf{undecidable} in general.
\end{theorem}

\begin{proof}
Testing for acyclicity or termination was shown to be undecidable in general by Plump in 1998 \cite{Plump98a}. Undecidability of (local) confluence checking was shown by Plump in 1993 \cite{Plump93b}, even for terminating GT systems \cite{Plump05a}.
\end{proof}

While confluence testing is undecidable, there are some instances in which one can decide confluence or non-confluence of GT systems. In 1970, Knuth and Bendix showed that confluence checking of terminating term rewriting systems is decidable \cite{Knuth-Bendix70a}. Moreover, it suffices to compute all \enquote{critical pairs} and check their joinability \cite{Huet80a,Baader-Nipkow98a,Terese03a}. Unfortunately, joinability of critical pairs does not imply local confluence of GT system, otherwise we could have contradicted our theorem! In 1993, Plump showed that \enquote{strong joinability} of all critical pairs is sufficient but not necessary to show local confluence \cite{Plump93b,Plump05a}. We have summarised these results in Section \ref{section:critpairs}.


\section{Graph Programming Languages}

GT systems naturally lend themselves to expressing computation by considering the normal forms of the input graph.

\begin{example} \label{eg:gtsemfunc}
Given a GT system \(T = (\mathcal{L}, \mathcal{R})\), consider the \textbf{state space} \(\Sigma = \mathcal{G}(\mathcal{L}) \cup \{\bot\}\) and the induced ARS \((\mathcal{G}(\mathcal{L}), \rightarrow_{\mathcal{R}})\). We may define the \textbf{semantic function} \(f_T: \mathcal{G}(\mathcal{L}) \to \mathcal{P}(\Sigma)\) by \(f_T([G]) = \{[H] \mid [H]\) is a normal form of \([G]\) with respect to \(\rightarrow_{\mathcal{R}}\} \cup \{\bot \mid\) there is an infinite reduction sequence starting from \([G]\}\) and \(f_T(\bot) = \{\bot\}\).
\end{example}

There are a number of GT languages and tools, such as
AGG \cite{Runge-Ermel-Taentzer11a},
GMTE \cite{Hannachi-Rodriguez-Drira-Pomares-Saul13a},
Dactl \cite{Glauert-Kennaway-Sleep91a},
GP\,2 \cite{Plump12a},
GReAT \cite{Agrawal-Karsai-Neema-Shi-Vizhanyo06a},
GROOVE \cite{Ghamarian-Mol-Rensink-Zambon-Zimakova12a},
GrGen.Net \cite{Jakumeit-Buchwald-Kroll10a},
Henshin \cite{Arendt-Biermann-Jurack-Krause-Taentzer10a},
PROGRES \cite{Schurr-Winter-Zundorf99a},
and PORGY \cite{Fernandez-Kirchner-Mackie-Pinaud14a}.
Habel and Plump \cite{Habel-Plump01a} show that such languages can be \enquote{computationally complete}:

\begin{proposition}
To be \textbf{computationally complete}, the three constructs:
\begin{enumerate}[itemsep=-0.4ex,topsep=-0.4ex]
\item Nondeterministic application of a rule from a set of rules (\(\mathcal{R}\));
\item Sequential composition (\(P1;\,P2\));
\item Iteration in the form that rules are applied as long as possible \(P\!\!\downarrow\).
\end{enumerate}
\noindent
are not only \textbf{sufficient}, but \textbf{necessary} (using DPO-based rule application).
\end{proposition}

\begin{example}
The semantics of some program \(P\) is a binary relation \(\rightarrow_P\) on some set of abstract (rooted) graphs \(\mathcal{G}\), inductively defined as follows:
\begin{enumerate}[itemsep=-0.4ex,topsep=-0.4ex]
\item \(\rightarrow_{\mathcal{R}} \defeq \rightarrow\) \: (where \(\rightarrow\) is the induced ARS relation on \(\mathcal{R})\).
\item \(\rightarrow_{P1;\,P2} \defeq \rightarrow_{P2} \circ \rightarrow_{P1}\).
\item \(\rightarrow_{P\!\downarrow} \defeq \{([G], [H]) \mid [G] \rightarrow_P^* [H] \text{ and } [H] \text{ is in normal form}\footnotemark\}\).\footnotetext{\([H]\) is in normal form iff it is not reducible using \(\rightarrow_P\)}
\end{enumerate}
\vspace{-1.6em}
\end{example}

\begin{remark}
While GT systems can \enquote{simulate} any Turing Machine, this does not make them \enquote{computationally complete} in the strong sense that any computable function on arbitrary graphs can be programmed.
\end{remark}

GP\,2 is an experimental rule-based language for problem solving in the domain of graphs, developed at York, the successor of GP \cite{Plump09a,Plump12a}. GP\,2 is of interest because it has been designed to support formal reasoning on programs \cite{Plump16a}, with a semantics defined in terms of partially labelled graphs, using the injective DPO approach with relabelling \cite{Habel-Muller-Plump01a,Habel-Plump02a}. Poskitt and Plump have set up the foundations for verification of GP\,2 programs \cite{Poskitt-Plump12a,Poskitt-Plump13a,Poskitt13a,Poskitt-Plump14a} using a Hoare-Style \cite{Hoare69a} system (actually for GP \cite{Manning-Plump08a,Plump09a}), Hristakiev and Plump have developed static analysis for confluence checking \cite{Hristakiev-Plump18a,Hristakiev18a}, and Bak and Plump have extended the language, adding root nodes \cite{Bak-Plump12a,Bak15a}. Plump has shown computational completeness \cite{Plump17a}.

GP\,2 uses a model of \enquote{rule schemata} with \enquote{application conditions}, rather than \enquote{rules} as we have seen up until now. The label alphabet used for both nodes and edges is \((\mathbb{Z} \cup {Char}^*)^* \times \mathcal{B}\). Roughly speaking, rule application works by finding an injective \enquote{premorphism} by ignoring labels, and then checking if there is an assignment of values such that after evaluating the label expressions, the morphism is label-preserving. The application condition is then checked, then rule application continues. \cite{Plump12a}

The formal semantics of GP\,2 is given in the style of Plotkin's structural operational semantics \cite{Plotkin04a}. Inference rules inductively define a small-step transition relation \(\rightarrow\) on configurations. The inference rules and definition of the semantic function \(\llbracket . \rrbracket: ComSeq \to \mathcal{G} \to \mathcal{P}(\mathcal{G} \cup \{fail, \bot\})\) were first defined in \cite{Plump12a}. Up-to-date versions can be found in \cite{Bak15a}.

%% file: fig/1/eg1.tex
\begin{tikzpicture}[every node/.style={align=center}]
    \node (a) at (0.8,0.0)    [draw, circle, thick, fill=black, scale=0.3] {\,};
    \node (b) at (0.0,-0.0)   [draw, circle, thick, fill=black, scale=0.3] {\,};
    \node (c) at (-0.8,-0.0)  [draw, circle, thick, fill=black, scale=0.3] {\,};

    \node (A) at (0.8,-0.18)  {\tiny{3}};
    \node (B) at (-0.8,-0.18) {\tiny{1}};
    \node (C) at (0.0,-0.18)  {\tiny{2}};

    \draw (b) edge[->,thick, bend left] (c)
          (c) edge[->,thick, bend left] (b)
          (a) edge[->,thick] (b)
          (a) edge[->,thick,in=35,out=-35,loop] (a);
\end{tikzpicture}

%% file: fig/1/morphism.tex
\begin{equation*}
\begin{tikzcd}
  E_G \arrow[r, "s_G"] \arrow[d, "g_E"]
    & V_G \arrow[d, "g_V"]
    & E_G \arrow[r, "t_G"] \arrow[d, "g_E"]
    & V_G \arrow[d, "g_V"] \\
  E_H \arrow[r, "s_H"]
    & V_H
    & E_H \arrow[r, "t_H"]
    & V_H
\end{tikzcd}
\end{equation*}

%% file: fig/1/eg2.tex
\begin{tikzpicture}[every node/.style={align=center}]
    \node (a) at (-0.2,-0.05)  {$G = $};
    \node (b) at (0.8,0.0)    [draw, circle, thick, fill=black, scale=0.3] {\,};
    \node (c) at (1.6,0.0)    [draw, circle, thick, fill=black, scale=0.3] {\,};

    \node (B) at (0.8,-0.18)  {\tiny{1}};
    \node (C) at (1.6,-0.18)  {\tiny{2}};

    \node (d) at (3.0,-0.05) {$H = $};
    \node (e) at (3.8,0.0)    [draw, circle, thick, fill=black, scale=0.3] {\,};
    \node (f) at (5.4,0.0)    [draw, circle, thick, fill=black, scale=0.3] {\,};
    \node (g) at (4.6,0.0)    [draw, circle, thick, fill=black, scale=0.3] {\,};

    \node (E) at (3.8,-0.18)  {\tiny{1}};
    \node (F) at (5.4,-0.18)  {\tiny{2}};
    \node (G) at (4.6,-0.18)  {\tiny{3}};

    \draw (b) edge[->,thick,in=145,out=215,loop] (b)
          (c) edge[->,thick] (b)
          (e) edge[->,thick, bend left] (g)
          (e) edge[->,thick, bend right] (g)
          (f) edge[->,thick] (g)
          (g) edge[->,thick,in=55,out=125,loop] (g);
\end{tikzpicture}

%% file: fig/1/relabelling.tex
\begin{tikzpicture}[every node/.style={inner sep=0pt, text width=6.5mm, align=center}]
    \node (a) at (0.0,0) [draw, circle, thick] {1};
    \node (b) at (1.5,0) [draw, circle, thick] {2};

    \node (c) at (2.5,0) {$\leftarrow$};

    \node (d) at (3.5,0) [draw, circle, thick] {x};
    \node (e) at (5.0,0) [draw, circle, thick] {y};

    \node (f) at (6.0,0) {$\rightarrow$};

    \node (g) at (7.0,0) [draw, circle, thick] {2};
    \node (h) at (8.5,0) [draw, circle, thick] {1};

    \node (A) at (0.0,-.52) {\tiny{1}};
    \node (B) at (1.5,-.52) {\tiny{2}};
    \node (D) at (3.5,-.52) {\tiny{1}};
    \node (E) at (5.0,-.52) {\tiny{2}};
    \node (G) at (7.0,-.52) {\tiny{1}};
    \node (H) at (8.5,-.52) {\tiny{2}};
\end{tikzpicture}

%% file: fig/1/complexity.tex
\begin{center}
\begin{tabular}{l|l|l}
Input                   & Output                                                                                       & Time           \\ \hline
label \(l\)             & The set \(Z\) of nodes with label \(l\).                                                     & \(O(\abs{Z})\) \\
node \(v\)              & Values \(\operatorname{deg}(v)\), \(\operatorname{indeg}(v)\), \(\operatorname{outdeg}(v)\). & \(O(1)\)       \\
node \(v\), label \(l\) & No. edges with source \(v\), label \(l\).                                                    & \(O(1)\)       \\
node \(v\), label \(l\) & No. edges with target \(v\), label \(l\).                                                    & \(O(1)\)       \\
node \(v\), label \(l\) & Set \(X\) of edges with source \(v\), label \(l\).                                           & \(O(\abs{X})\) \\
node \(v\), label \(l\) & Set \(X\) of edges with target \(v\), label \(l\).                                           & \(O(\abs{X})\) \\
graph \(G\)             & \(\abs{V_G}\) and \(\abs{E_G}\).                                                             & \(O(1)\)       \\
\end{tabular}
\end{center}

%% file: fig/1/rooted.tex
\begin{tikzpicture}[every node/.style={inner sep=0pt, text width=6.5mm, align=center}]
    \node (a) at (0.0,0) [draw, circle, thick, double, double distance=0.4mm] {1};
    \node (b) at (1.5,0) [draw, circle, thick] {2};

    \node (c) at (2.5,0) {$\leftarrow$};

    \node (d) at (3.5,0) [draw, circle, thick] {\,};
    \node (e) at (5.0,0) [draw, circle, thick] {\,};

    \node (f) at (6.0,0) {$\rightarrow$};

    \node (g) at (7.0,0) [draw, circle, thick] {3};
    \node (h) at (8.5,0) [draw, circle, thick, double, double distance=0.4mm] {4};

    \node (A) at (0.0,-.52) {\tiny{1}};
    \node (B) at (1.5,-.52) {\tiny{2}};
    \node (D) at (3.5,-.52) {\tiny{1}};
    \node (E) at (5.0,-.52) {\tiny{2}};
    \node (G) at (7.0,-.52) {\tiny{1}};
    \node (H) at (8.5,-.52) {\tiny{2}};
\end{tikzpicture}

%% file: content/2.tex
\chapter{A New Theory of Rooted Graph Transformation} \label{chapter:newtheory}

\begin{chapquote}{Georg Cantor, \textit{Doctoral thesis (1867)}}
``In mathematics the art of asking questions is more valuable than solving problems.''
\end{chapquote}

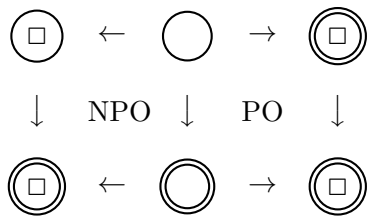
\begin{figure}[H]
\centering
\noindent
\input{fig/2/derivation}
\vspace{-0.4em}
\caption{Example Rooted Derivation}
\label{fig:egrd}
\end{figure}

Graph transformation with relabelling as described in Sections \ref{sec:gtintro}, \ref{section:relabelling}, \ref{section:gt1} and \ref{section:gt2} has desirable properties. It was shown by Habel and Plump in 2002 \cite{Habel-Plump02a} that derivations are natural double pushouts (Theorem \ref{theorem:uniquederivations}) and thus are invertible. Unfortunately, Bak and Plump's modifications to add root nodes (Sections \ref{section:rooted} and \ref{section:rootedgt}) mean that derivations no longer exhibit these properties. That is, only the right square of a derivation in a rooted GT system need be a natural pushout. This asymmetry is unfortunate, because derivations are no longer invertible. Figure \ref{fig:egrd} shows an example rooted derivation in which the right square is not a pullback.

We propose an alternative theory for rooted graph transformation with relabelling, with some more desirable properties. Critically, we \textbf{restore invertibility} of derivations (Corollary \ref{cor:invertiblederivations}), and remove some undesirable matching cases (Lemma \ref{lem:rootedmatching}), allowing us to prove a handy root node invariance result (Corollary \ref{cor:invariantroots}). Our restrictions are slightly more than Habel and Plump's original version of graph transformation with relabelling, however our interest is only in the cases that are practically useful. Similar restrictions are already in place within the graph programming language GP\,2 \cite{Plump12a}.


\section{Graphs and Morphisms}

Fix some common label alphabet (Definition \ref{dfn:labelalphabet}) \(\mathcal{L} = (\mathcal{L}_V, \mathcal{L}_E)\). In this section we define our new notions of graphs and morphisms.

\begin{definition}
A \textbf{graph} over \(\mathcal{L}\) is a tuple \(G = (V, E, s, t, l, m, p)\) where:
\begin{enumerate}[itemsep=-0.4ex,topsep=-0.4ex]
\item \(V\) is a \textbf{finite} set of \textbf{vertices};
\item \(E\) is a \textbf{finite} set of \textbf{edges};
\item \(s: E \to V\) is a \textbf{total} source function;
\item \(t: E \to V\) is a \textbf{total} target function;
\item \(l: V \to \mathcal{L}_V\) is a \textbf{partial} function, labelling the vertices;
\item \(m: E \to \mathcal{L}_E\) is a \textbf{total} function, labelling the edges;
\item \(p: V \to \mathbb{Z}_2\)\footnote{\(\mathbb{Z}_2\) is the quotient \(\mathbb{Z}/2\mathbb{Z} = \{0, 1\}\).} is a \textbf{partial} function, determining vertex rootedness.
\end{enumerate}
\end{definition}

\begin{definition}
If \(G\) is a \textbf{graph}, then we define its size \(\abs{G} = \abs{V_G} + \abs{E_G}\).
\end{definition}

\begin{definition}
A graph \(G\) is \textbf{totally labelled} iff \(l_G\) is total, and \textbf{totally rooted} if \(p_G\) is total. If \(G\) is both, then we call it a \textbf{TLRG}.
\end{definition}

\begin{remark}
A totally rooted graph need not have every node a root node, only \(p_G\) must be total. \(0\) denotes unrooted, and \(1\) rooted. When we draw graphs, we shall denote the absence of rootedness with \textbf{diagonal stripes}. If a node has a \textbf{double border}, it is rooted, otherwise, it is unrooted.
\end{remark}

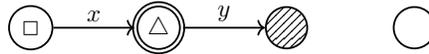
\begin{figure}[H]
\centering
\noindent
\scalebox{0.85}{\input{fig/2/eg}}
\vspace{-0.25em}
\caption{Example Graph}
\label{fig:eg-graph}
\end{figure}

\begin{example}
Let \(\mathcal{L} = (\{\Square, \triangle\}, \{x, y\})\). Then \(G = (V, E, s, t, l, m, p)\) is a graph over \(\mathcal{L}\) where \(V = \{1, 2, 3, 4\}\), \(E = \{1, 2\}\), \(s = \{(1, 1), (2, 2)\}\), \(t = \{(1, 2), (2, 3)\}\), \(l = \{(1, \Square), (2, \triangle)\}\), \(m = \{(1, x), (2, y)\}\), and \(p = \{(1, 0), (2, 1), (4, 0)\}\). Its graphical representation is shown in Figure \ref{fig:eg-graph}. \(G\) is neither totally rooted nor totally labelled, since node \(3\) has both undefined rootedness and no label, and node \(4\) also has no label.
\end{example}

\vspace{-1.2em}

\begin{definition}
A \textbf{graph morphism} between graphs \(G\) and \(H\) is a pair of functions \(g = (g_V: V_G \to V_H, g_E: E_G \to E_H)\) such that sources, targets, labels, and rootedness are preserved. That is:
\begin{enumerate}[itemsep=-0.4ex,topsep=-0.4ex]
\item \(\forall e \in E_G, \, g_V(s_G(e))= s_H(g_E(e))\);                 \tabto*{20em} [Sources]
\item \(\forall e \in E_G, \, g_V(t_G(e)) = t_H(g_E(e))\);                \tabto*{20em} [Targets]
\item \(\forall e \in E_G, \, m_G(e) = m_H(g_E(e))\);                     \tabto*{20em} [Edge Labels]
\item \(\forall v \in l_G^{-1}(\mathcal{L}_V), \, l_G(v) = l_H(g_V(v))\); \tabto*{20em} [Node Labels]
\item \(\forall v \in p_G^{-1}(\mathbb{Z}_2), \, p_G(v) = p_H(g_V(v))\).  \tabto*{20em} [Rootedness]
\end{enumerate}
\end{definition}

\begin{remark}
If \(G\) and \(H\) are \textbf{TLRG}s, then this is equivalent to the following diagram commuting (for \(s_G, s_H\) and \(t_G, t_H\) separately):

\vspace{-0.6em}
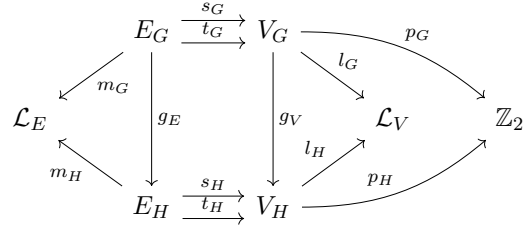
\begin{figure}[H]
\centering
\noindent
\input{fig/2/morphism}
\vspace{-0.8em}
\caption{Graph Morphism Commuting Diagrams}
\label{fig:graphmorphismcomdiag}
\vspace{0em}
\end{figure}
\end{remark}

An important property of morphisms is that if a node is labelled, then its image under the morphism must have the same label, and the same for the rootedness. However, if the label or rootedness of a node is not defined, then we do not specify anything about its image. Nodes with undefined labels may be mapped to nodes with undefined labels, but could also be mapped to labelled nodes, and similarly for rootedness. There are some circumstances where it is useful to additionally insist on undefinedness being preserved, which motivates our next definition.

\begin{definition}
A graph morphism \(g: G \to H\) is \textbf{undefinedness preserving} iff it is both:
\begin{enumerate}[itemsep=-0.4ex,topsep=-0.4ex]
\item \textbf{Label-undefinedness preserving}: \(l_G = l_H \circ g_V\);
\item \textbf{Root-undefinedness preserving}: \(p_G = p_H \circ g_V\).
\end{enumerate}
\end{definition}

\begin{remark}
The undefinedness preserving property is exactly requiring that the two right-most triangles in Figure \ref{fig:graphmorphismcomdiag} commute. It is thus a triviality that all morphisms between TLRG are undefinedness preserving.
\end{remark}

Just like unlabelled (totally labelled) graphs and morphisms, our new notion of graphs and morphisms forms a category where the graphs are the objects and the morphisms are the arrows.

\begin{proposition}
Our new notion of graphs and morphisms is a \textbf{locally small} (Definition \ref{def:catsmall}), \textbf{finitary} (Definition \ref{def:catfinitary}) \textbf{category} (Definition \ref{def:cat}).
\end{proposition}

\begin{definition}
A graph morphism \(g: G \to H\) is \textbf{injective} (\textbf{surjective}) iff the underlying functions \(g_V\), \(g_E\) are injective (surjective). We say that \(g\) is an \textbf{isomorphism} iff it is both \textbf{injective} and \textbf{surjective}, and \(g^{-1}: H \to G\) is also a graph morphism.
\end{definition}

\begin{proposition}
Morphisms are \textbf{injective} (\textbf{surjective}, \textbf{isomorphisms}) iff they are \textbf{monic} (\textbf{epic}, \textbf{isomorphisms}) in the category theoretic sense.
\end{proposition}

\begin{proposition}
A morphism is an \textbf{isomorphism} iff it is \textbf{injective}, \textbf{surjective}, and \textbf{undefinedness preserving}.
\end{proposition}

\begin{example}
There exist morphisms that are injective, surjective graph morphisms that are not isomorphisms. Just consider two discrete (no edges) graph with one node, the first with no node label, and the second with some label. There is an injective, surjective morphism from the first to second, but not in the other direction.
\end{example}

\begin{definition}
We say \(H\) is a \textbf{subgraph} of \(G\) iff there exists an \textbf{inclusion morphism} \(H \hookrightarrow G\). This happens iff \(V_H \subseteq V_G\), \(E_H \subseteq E_G\), \(s_H = \restr{s_G}{E_H}\), \(t_H = \restr{t_G}{E_H}\), \(m_H = \restr{m_G}{E_H}\), \(l_H \subseteq l_G\), \(p_H \subseteq p_G\) (treating functions as sets).
\end{definition}

\begin{remark} \label{remark:subgraphdfns}
There are two obvious (non-equivalent) definitions of subgraph. We have chosen here to define subgraphs to be exactly those that can can be included in the parent structure. This is particularly useful when it comes to defining rules later. The other definition (closer to that given by Habel and Plump for partially labelled graphs) would be to require the morphism to also be undefinedness preserving.
\end{remark}

\begin{definition}
We say that graphs \(G, H\) are \textbf{isomorphic} iff there exists a \textbf{graph isomorphism} \(g: G \to H\). This gives \textbf{equivalence classes} \([G]\) over \(\mathcal{L}\). We denote by \(\mathcal{G}^{\varoplus}(\mathcal{L})\) the collection of totally labelled, totally rooted \textbf{abstract graphs} over some fixed \(\mathcal{L}\).
\end{definition}

There are only countably many abstract graphs; an essential property when setting up graph languages. As we remarked in Chapter, \ref{chapter:theoryintro}, we would be in trouble if the universal language was not a countable set.

\begin{proposition}
\(\mathcal{G}^{\varoplus}(\mathcal{L})\) is a \textbf{countable set}.
\end{proposition}


\section{Rules and Derivations}

Fixing some common \(\mathcal{L} = (\mathcal{L}_V, \mathcal{L}_E)\), we define rules and derivations. We only concern ourselves, for now, with \enquote{linear rules} and injective matches \cite{Habel-Muller-Plump01a}. Moreover, some of the results we will be showing later don't hold in more general versions of this set up.

\begin{definition}
A \textbf{rule} \(r = \langle L \leftarrow K \rightarrow R \rangle\) consists of left/right \textbf{TLRGs} \(L\), \(R\), the interface \textbf{graph} \(K\), and \textbf{inclusions} \(K \hookrightarrow L\) and \(K \hookrightarrow R\).
\end{definition}

\begin{remark}
Insisting on inclusions is without loss of generality. We could have only required two injective graph morphisms, but assuming inclusions makes for neater proofs involving rules. Moreover, we can think of \(K\) as a subgraph of both \(L\) and \(R\).
\end{remark}

\begin{example}
See Figure \ref{fig:tree2}.
\end{example}

\begin{definition}
We define the \textbf{inverse rule} to be \(r^{-1} = \langle R \leftarrow K \rightarrow L \rangle\).
\end{definition}

\begin{definition}
If \(r = \langle L \leftarrow K \rightarrow R \rangle\) is a \textbf{rule}, then \(\abs{r} = max \{\abs{L},\abs{R}\}\).
\end{definition}

\begin{definition}
Given a \textbf{rule} \(r = \langle L \leftarrow K \rightarrow R \rangle\) and a \textbf{TLRG} \(G\), we say that an \textbf{injective} morphism \(g: L \hookrightarrow G\) satisfies the \textbf{dangling condition} iff no edge in \(G \setminus g(L)\) is incident to a node in \(g(L \setminus K)\).
\end{definition}

\begin{definition} \label{dfn:ruleapplication}
To \textbf{apply} a rule \(r = \langle L \leftarrow K \rightarrow R \rangle\) to some \textbf{TLRG} \(G\), find an \textbf{injective} graph morphism \(g: L \hookrightarrow G\) satisfying the \textbf{dangling condition}, then:

\begin{enumerate}[itemsep=-0.4ex,topsep=-0.4ex]
\item Delete \(g(L \setminus K)\) from \(G\). For each unlabelled node \(v\) in \(K\), make \(g_V(v)\) unlabelled, and for each node \(v\) in \(K\) with undefined rootedness, make \(g_V(v)\) have undefined rootedness, giving \textbf{intermediate graph} \(D\).
\item Add disjointly \(R \setminus K\) to D, keeping their labels and rootedness. For each unlabelled node \(v\) in \(K\), label \(g_V(v)\) with \(l_R(v)\), and for each node with undefined rootedness \(v\) in \(K\), make \(g_V(v)\) have rootedness \(p_R(v)\), giving the \textbf{result graph} \(H\).
\end{enumerate}

\noindent
If the \textbf{dangling condition} fails, then the rule is not applicable using the \textbf{match} \(g\). We can exhaustively check all matches to determine applicability.
\end{definition}

\begin{definition}
We write \(G \Rightarrow_{r,g} M\) for a successful application of \(r\) to \(G\) using match \(g\), obtaining result \(M \cong H\). We call this a \textbf{direct derivation}. We may omit \(g\) when it is not relevant, writing simply \(G \Rightarrow_{r} M\).
\end{definition}

\begin{remark}
Formally, a direct derivation is actually the diagram Figure \ref{fig:comsqrs}.
\end{remark}

\begin{definition}
For a given set of rules \(\mathcal{R}\), we write \(G \Rightarrow_{\mathcal{R}} H\) iff \(H\) is \textbf{directly derived} from \(G\) using any of the rules from \(\mathcal{R}\).
\end{definition}

\begin{definition}
We write \(G \Rightarrow_{\mathcal{R}}^{+} H\) iff \(H\) is \textbf{derived} from \(G\) in one or more \textbf{direct derivations}, and \(G \Rightarrow_{\mathcal{R}}^{*} H\) iff \(G \cong H\) or \(G \Rightarrow_{\mathcal{R}}^{+} H\).
\end{definition}


\section{Foundational Theorems}

We will show that gluing and deletions correspond to natural pushouts and natural pushout complements, respectively. Thus, derivations are invertible.

\vspace{-0.8em}
\begin{figure}[H]
\centering
\noindent
\input{fig/2/comsqrs}
\vspace{-0.8em}
\caption{Commuting Squares}
\label{fig:comsqrs}
\end{figure}

Our first technical lemma ensures that pullbacks exist in our new setting.

\begin{lemma} \label{lem:pullbacksexist}
Given \textbf{graph morphisms} \(g: L \to G\) and \(c: D \to G\), there exist a \textbf{graph} \(K\) and \textbf{graph morphisms} \(b: K \to L\), \(d: K \to D\) such that the resulting square is a \textbf{pullback} (Definition \ref{dfn:pullback}).
\end{lemma}

\begin{proof}
The constructions are exactly as in Lemma 1 of \cite{Habel-Plump02a}, with the rootedness function defined analogously to the node labelling function. Trivial modifications to the proof give the result.
\end{proof}

Our second technical lemma, roughly speaking, ensures that pullbacks exist in our new setting, in the case \(K \to D\) is injective and that we don't have any confusion created by conflicting labels or rootedness. In the totally labelled case (without root nodes), we don't need to worry about such confusion.

\begin{lemma} \label{lem:pushoutsexist}
Let \(b : K \to R\), \(d: K \to D\) be \textbf{graph morphisms} such that:
\begin{enumerate}[itemsep=-0.4ex,topsep=-0.4ex]
\item \(d\) is \textbf{injective},
\item \(\forall v \in V_R, \abs{l_R(\{v\}) \cup l_D(d_V(b_V^{-1}(\{v\})))} \leq 1\),
\item \(\forall v \in V_R, \abs{p_R(\{v\}) \cup p_D(d_V(b_V^{-1}(\{v\})))} \leq 1\).
\end{enumerate}
Then, there exist a \textbf{graph} \(H\) and \textbf{graph morphisms} \(h: R \to H\), \(c: D \to H\) such that the resulting square is a \textbf{pushout} (Definition \ref{dfn:pushout}).
\end{lemma}

\begin{proof}
The constructions are exactly as in Lemma 2 of \cite{Habel-Plump02a}, with the rootedness function defined analogously to the node labelling function.
\end{proof}

Our third technical lemma says that as long as \(K \to D\) is injective, and no additional labels or roots are introduced in \(D\), then the left square is a pullback if it is a pushout.

\begin{lemma}
Given two \textbf{graph morphisms} \(b: K \to L\) and \(d: K \to D\) such that \(b\) is \textbf{injective} and \(L\) is a \textbf{TLRG}, then the pushout (1) is \textbf{natural} (Definition \ref{dfn:npo}) iff \(l_D(d_V(V_K \setminus l_K^{-1}(\mathcal{L}_V))) = \emptyset = p_D(d_V(V_K \setminus p_K^{-1}(\mathbb{Z}_2)))\).
\end{lemma}

\begin{proof}
Let square (1) in Figure \ref{fig:comsqrs} be a natural pushout with graph morphisms \(g: L \to G\) and \(c: D \to G\). Once again, we can proceed as in Lemma 3 of \cite{Habel-Plump02a} with the obvious modifications. Similar for the other direction.
\end{proof}

Our final technical lemma shows a correspondence between the existence of natural pushout complements and matches satisfying the dangling condition. Moreover, that the intermediate graph \(D\) is essentially unique (unique up to isomorphism).

\begin{lemma}
Let \(g: L \to G\) be an \textbf{injective graph morphism} and \(K \to L\) an \textbf{inclusion morphism}. Then, there exist a \textbf{graph} \(D\) and \textbf{morphisms} \(K \to D\) and \(D \to G\) such that the square (1) is a \textbf{natural pushout} iff \(g\) satisfies the \textbf{dangling condition}. Moreover, in this case, \(D\) is unique up to isomorphism.
\end{lemma}

\begin{proof}
Proceed as in Lemma 4 of \cite{Habel-Plump02a} with the obvious modifications.
\end{proof}

We now present our first major theorem. The explicit construction of derivations given earlier corresponds exactly to a NDPO diagram. Moreover, the result graph is essentially unique and the input is a \textbf{TLRG} iff the result is.

\begin{theorem}[Derivation Theorem] \label{thm:derivationtheorem}
Given a rule \(\langle L \leftarrow K \rightarrow R \rangle\) and an \textbf{injective graph morphism} \(g: L \to G\), then there exists a \textbf{natural DPO} diagram as above iff \(g\) satisfies the \textbf{dangling condition}. In this case, \(D\) and \(H\) are unique up to isomorphism, and rule application exactly corresponds to that given in Definition \ref{dfn:ruleapplication}. Moreover, if \(G \Rightarrow_r H\), then \(G\) is a \textbf{TLRG} iff \(H\) is a \textbf{TLRG}.
\end{theorem}

\begin{proof}
Proceed as in Theorem 1 of \cite{Habel-Plump02a} with the obvious modifications. Totality of labelling is given by Theorem 2 of \cite{Habel-Plump02a}, and totality of rootedness is given by replacing all occurrences of the labelling function with the rootedness function in the proof.
\end{proof}

An important consequence of having both squares as natural pushouts is that we have a symmetry giving invertibility of derivations. This symmetry is unique to this new approach to rooted graph transformation. In Bak's approach (Appendix \ref{section:rootedgt}), derivations are not, in general, invertible (Figure \ref{fig:egrd}). In Bak's system, the intermediate graph \(D\) must not have a root if we want to invert the derivation.

\begin{corollary} \label{cor:invertiblederivations}
Derivations are \textbf{invertible}. That is \(G \Rightarrow_{r} H\) iff \(H \Rightarrow_{r^{-1}} G\).
\end{corollary}

\begin{proof}
By the last theorem, \(G \Rightarrow_{r} H\) means we have a match \(g: L \to G\), and a comatch \(h: R \to H\), and so by symmetry, we have the result.
\end{proof}

It absolutely cannot be overstated how important invertibility is! It will be the foundation of so many arguments within the rest of the theory and examples, in particular when it comes to language recognition. Moreover, if it weren't for invertibility, there is no way that we could perform an encoding in Chapter \ref{chapter:confluenceanalysis2} to recover the Critical Pair Lemma!

Finally, we can now show our root node invariance result. By comparison, in Bak's system, non-root nodes could be matched against root nodes.

\begin{lemma} \label{lem:rootedmatching}
Let \(G\) be a \textbf{TLRG}, and \(r = \langle L \leftarrow K \rightarrow R \rangle\) a rule. Then root nodes in \(L\) can only be matched against root nodes in \(G\), and similarly for non-root nodes.
\end{lemma}

\begin{proof}
Immediate from the definitions.
\end{proof}

\begin{corollary} \label{cor:invariantroots}
Let \(G\) be a \textbf{TLRG}, and \(r = \langle L \leftarrow K \rightarrow R \rangle\) a rule such that \(L\) and \(R\) both contain \(k\) root nodes, for some fixed \(k \in \mathbb{N}\). Then any \textbf{TLRG} \(H\) derived from \(G\) using \(r\) contains \(n\) root nodes iff \(G\) contains \(n\) root nodes.
\end{corollary}

\begin{proof}
By Lemma \ref{lem:rootedmatching} (non-)roots in \(L\) can only be identified with (non-)roots in \(G\), and by symmetry the same for \(R\) in \(H\). By Theorem \ref{thm:derivationtheorem}, NDPO existence corresponds to Definition \ref{dfn:ruleapplication}, so, \(\abs{p_G^{-1}(\{1\})} = \abs{p_H^{-1}(\{1\})}\).
\end{proof}
\vspace{-1.6em}


\section{Equivalence of Rules} \label{sec:ruleequiv}

We now consider equivalence of rules, starting by formalising what it means to say that two rules are isomorphic, and then we will show that we can find a normal form for rules, unique up to isomorphism. In the next sections, we will see these notions of equivalence can be extended to GT systems in a natural way. Derivation compatibility will prove to be an important notion.

\begin{definition} \label{dfn:ruleiso}
Given \textbf{rules} \(r_1 = \langle L_1 \leftarrow K_1 \rightarrow R_1 \rangle\), \(r_2 = \langle L_2 \leftarrow K_2 \rightarrow R_2 \rangle\). We call \(r_1\) and \(r_2\) \textbf{isomorphic} iff there exists isomorphisms \(f: L_1 \to L_2\), \(g: R_1 \to R_2\) such that \(\restr{f}{K_1} = \restr{g}{K_1}\) and \(f(K_1) = K_2\). Write \(r_1 \cong r_2\).
\end{definition}

The above notion of \textbf{rule isomorphism} is an \textbf{equivalence}, and gives rise to \textbf{abstract rules} \([r]\). Moreover, the collection of all abstract rules over some fixed alphabet forms a countable set. This is a nice little result, since the collection of all rules is actually not a set.

\begin{definition} \label{dfn:rulenormequiv}
Given a \textbf{rule} \(r = \langle L \leftarrow K \rightarrow R \rangle\), define its \textbf{normal form} \(r\!\!\downarrow  = \langle L \leftarrow K' \rightarrow R \rangle\) where \(K' = (V_K, \emptyset, \emptyset, \emptyset, \emptyset, \emptyset, \emptyset)\). We say two rules \(r_1\), \(r_2\) are \textbf{normalisation equivalent} iff \(r_1\!\!\downarrow \cong r_2\!\!\downarrow\). We write \(r_1 \simeq r_2\).
\end{definition}

Clearly, this gives us a \textbf{coarser} notion of \textbf{equivalence} for rules than the notion of \textbf{isomorphism}. This notation of equivalence is important because it has the property that all concrete representatives behave the same, in the sense of Theorem \ref{thm:compatderivations}.

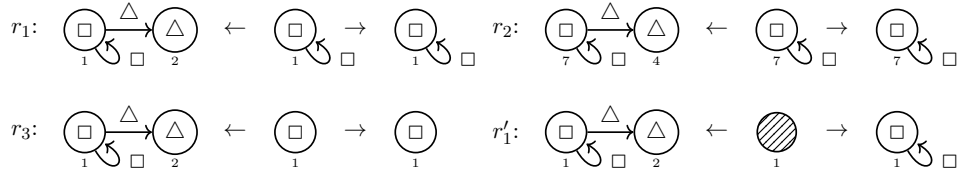
\begin{figure}[H]
\centering
\noindent
\scalebox{.8}{\input{fig/2/ruleiso}}
\vspace{-1.8em}
\caption{Example (Non-)Isomorphic Rules}
\label{fig:ruleiso}
\end{figure}

\begin{example}
Consider the rules over \((\{\Square, \triangle\}, \{\Square, \triangle\})\) as given in Figure \ref{fig:ruleiso}. Clearly \(r_1\) and \(r_2\) are isomorphic, but \(r_3\) is not isomorphic to either. Rule \(r_1\) has normal form \(r_1'\).
\end{example}

\vspace{-0.8em}
\vspace{0em}

\begin{theorem}[Compatible Derivations] \label{thm:compatderivations}
Given a \textbf{rule} \(r = \langle L \leftarrow K \rightarrow R \rangle\) and its normal form \(r\!\!\downarrow = \langle L \leftarrow K' \rightarrow R \rangle\), then for all TLRGs \(G\), \(H\), \(G \Rightarrow_{r} H\) iff \(G \Rightarrow_{r\!\downarrow} H\).
\end{theorem}

\vspace{-1.4em}
\begin{figure}[H]
\centering
\noindent
\input{fig/2/derivations}
\vspace{-0.7em}
\caption{Derivations Diagram}
\end{figure}
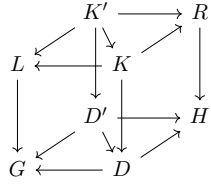

\begin{proof}
Consider some fixed graph \(G\). The set of injective morphisms \(g: L \to G\) satisfying the dangling condition must be identical for both rules since \(L\) is the same and so is \(V_K\). Then, by the explicit construction of \(H\) given by Definition \ref{dfn:ruleapplication}, \(G \Rightarrow_{r,g} H\) iff \(G \Rightarrow_{r\!\downarrow,g} H\).
\end{proof}

\begin{remark}
Normal forms for rules is not actually a new observation, and is the foundation of rule schemata in GP\,2 \cite{Plump12a}. Moreover, maximising the number of edges in the interface of rules leads to a reduction of the number of critical pairs (Section \ref{section:critpairs}) of a GT system \cite{Hristakiev-Plump18a}.
\end{remark}


\section{Graph Transformation Systems}

We can now define graph transformation systems using our new definitions of graphs and rules. Next, we will look at different notions of equivalence for such system. This will become important in Chapter \ref{chapter:confluenceanalysis2}, when we rely on derivation compatibility.

\begin{definition}
A \textbf{graph transformation system} \(T = (\mathcal{L}, \mathcal{R})\), consists of a label alphabet \(\mathcal{L} = (\mathcal{L}_V, \mathcal{L}_E)\), and a \textbf{finite} set \(\mathcal{R}\) of rules over \(\mathcal{L}\).
\end{definition}

\begin{definition}
Given a \textbf{graph transformation system} \(T = (\mathcal{L}, \mathcal{R})\), we define the inverse system \(T^{-1} = (\mathcal{L}, \mathcal{R}^{-1})\) where \(\mathcal{R}^{-1} = \{r^{-1} \mid r \in \mathcal{R}\}\).
\end{definition}

\begin{definition}
Given a \textbf{graph transformation system} \(T = (\mathcal{L}, \mathcal{R})\), a subalphabet of \textbf{non-terminals} \(\mathcal{N}\), and a \textbf{start graph} \(S\) over \(\mathcal{L}\), then a \textbf{graph grammar} is the system \(\pmb{G} = (\mathcal{L}, \mathcal{N}, \mathcal{R}, S)\).
\end{definition}

A graph is terminally labelled whenever none of the nodes or edges are labelled by non-terminals. Clearly the property of being terminally labelled is preserved under isomorphism, so we can lift it to a property of abstract graphs. We can then define the language generated by a grammar as follows:

\begin{definition}
Given a \textbf{graph grammar} \(\pmb{G}\) as defined above, we say that a graph \(G\) is \textbf{terminally labelled} iff \(l(V) \cap \mathcal{N}_V = \emptyset\) and \(m(E) \cap \mathcal{N}_E = \emptyset\). Thus, we can define the \textbf{graph language} generated by \(\pmb{G}\):
\begin{align*}
\pmb{L}(\pmb{G}) = \{[G] \mid S \Rightarrow_{\mathcal{R}}^{*} G, G \text{ terminally labelled}\}
\end{align*}
\end{definition}

One can think of graph transformation systems in terms of grammars that define languages. If they are terminating, then membership testing is decidable, but in general, non-deterministic in the sense that a deterministic algorithm must backtrack if it produces a normal form not equal to the start graph, to determine if another derivation sequence could have reached it. If the system is confluent too, then the algorithm becomes deterministic. 

\begin{definition}[Language Recognition] \label{dfn:langrec}
\(T = (\mathcal{L}, \mathcal{R})\) recognises a language \(\pmb{L}\) over \(\mathcal{P} \subseteq \mathcal{L}\) iff \(\exists [S] \in \pmb{L}\) such that \(\forall [G] \in \mathcal{G}^{\varoplus}(\mathcal{P})\), \([G] \in \pmb{L}\) iff \(G \Rightarrow_{\mathcal{R}}^{*} S\).
\end{definition}

\begin{theorem}[Membership Test] \label{thm:membershiptest}
Given a \textbf{grammar} \(\pmb{G} = (\mathcal{L}, \mathcal{N}, \mathcal{R}, S)\), \([G] \in \pmb{L}(\pmb{G})\) iff \(G \Rightarrow_{\mathcal{R}^{-1}}^* S\) and \(G\) is terminally labelled. That is, \((\mathcal{L}, \mathcal{R}^{-1})\) recognises \(\pmb{L}(\pmb{G})\) over \(\mathcal{L} \setminus \mathcal{N}\).
\end{theorem}

\begin{proof}
The key point is that rules and derivations are invertible, which means that if \(S\) can be derived from \(G\) using the reverse rules, then \(G\) can be derived from \(S\) using the original rules so is in the language. If \(S\) cannot be derived from \(G\), then \(G\) cannot be in the language since that would imply there was a derivation sequence from \(S\) to \(G\) which we could invert to give a contradiction. By construction, we only consider terminally labelled \(G\).
\end{proof}

\begin{remark}
If \(T = (\mathcal{L}, \mathcal{R})\) recognising a language \(\pmb{L}\) is terminating, then membership testing for \(\pmb{L}\) is decidable. Moreover, if \(T\) is also confluent, then recognition is efficient in the sense that we need not backtrack. The algorithm is simply to compute a normal form, and test if it is isomorphic to \(S\).
\end{remark}

Just like in Section \ref{section:arsgt}, we can define the \textbf{induced ARS} of a GT system.

\begin{definition} \label{dfn:gtars}
Let \(T = (\mathcal{L}, \mathcal{R})\) be a \textbf{GT system}. Then \((\mathcal{G}^{\varoplus}(\mathcal{L}), \rightarrow_{\mathcal{R}})\) is the \textbf{induced ARS} defined by \(\forall [G], [H] \in \mathcal{G}^{\varoplus}(\mathcal{L}), [G] \rightarrow_{\mathcal{R}} [H]\) iff \(G \Rightarrow_{\mathcal{R}} H\). We may also use the notation \(\rightarrow_{T}\) in place of \(\rightarrow_{\mathcal{R}}\) whenever the ambient label alphabet is not clear.
\end{definition}

\begin{lemma} \label{lem:gtars}
Consider the \textbf{ARS} \((\mathcal{G}^{\varoplus}(\mathcal{L}), \rightarrow_{\mathcal{R}})\) induced by a \textbf{GT system}. Then \(\rightarrow_{\mathcal{R}}\) is a \textbf{binary relation} (Definition \ref{def:binrel}) on \(\mathcal{G}^{\varoplus}(\mathcal{L})\). Moreover, it is \textbf{finitely branching} (Definition \ref{dfn:branching}) and \textbf{decidable} (Definition \ref{dfn:decidable}).
\end{lemma}

\begin{remark}
This does not, in general, imply that \(\rightarrow_{\mathcal{R}}^*\) is decidable. We say that \(T\) is (\textbf{locally}) \textbf{confluent} (\textbf{terminating}) iff its \textbf{induced ARS} is.
\end{remark}

\begin{lemma} \label{lem:gtmonoembedding}
A \textbf{GT system} is \textbf{terminating} iff there is a \textbf{monotone} (Definition \ref{dfn:monotone}) embedding from the \textbf{induced ARS} into \((\mathbb{N}, >)\).
\end{lemma}

\begin{proof}
By Lemma \ref{lem:gtars}, the induced ARS of a GT system is finitely branching, so the result follows from Lemma \ref{lem:monoembedding}.
\end{proof}

When we were thinking about reversing the rules to produce a membership test, we were thinking about them as a computation device. There was an input graph, and output graph(s) (and possibly non-termination). We defined a semantic function in Example \ref{eg:gtsemfunc}. We can do this again in our new setting.

\begin{definition}
Given \(T = (\mathcal{L}, \mathcal{R})\), define \textbf{state space} \(\Sigma = \mathcal{G}^{\varoplus}(\mathcal{L}) \cup \{\bot\}\) and \textbf{induced ARS} \((\mathcal{G}^{\varoplus}(\mathcal{L}), \rightarrow_{\mathcal{R}})\). Define the \textbf{semantic function} \(f_T: \mathcal{G}^{\varoplus}(\mathcal{L}) \to \mathcal{P}(\Sigma)\) by \(f_T([G]) = \{[H] \mid [H]\) is a normal form of \([G]\) w.r.t. \!\(\rightarrow_{\mathcal{R}}\} \cup \{\bot \mid\) there is an infinite reduction sequence starting from \([G]\) w.r.t. \!\(\rightarrow_{\mathcal{R}}\}\) and \(f_T(\bot) = \{\bot\}\).
\end{definition}

\begin{proposition}
Given \(T = (\mathcal{L}, \mathcal{R})\), then \(\forall [G] \in \mathcal{G}^{\varoplus}(\mathcal{L})\), \(\abs{f_T([G])} \geq 1\). Moreover, if \(T\) is confluent and terminating, then \(\forall [G] \in \mathcal{G}^{\varoplus}(\mathcal{L})\), \(\abs{f_T(G)} = 1\).
\end{proposition}

\begin{proposition}
Given \(T = (\mathcal{L}, \mathcal{R})\), in general it is undecidable if \([H] \in f_T([G])\) for any TLRG \(G, H\), and also if \(\bot \in f_T([G])\).
\end{proposition}


\section{Equivalence of GT Systems} \label{sec:gtequiv}

Building on the work from Section \ref{sec:ruleequiv}, we can ask when two graph transformations are equivalent, or rather, when they are distinct. We will give various notions of equivalence, and show there is a hierarchy of inclusion, as each notion is more and more general than the last.

\begin{definition}
Two \textbf{GT systems} \(T_1 = (\mathcal{L}, \mathcal{R}_1)\), \(T_2 = (\mathcal{L}, \mathcal{R}_2)\) are:
\begin{enumerate}[itemsep=-0.4ex,topsep=-0.4ex]
\item \textbf{Isomorphic} (\(T_1 \cong T_2\)) iff their quotients (Definition \ref{dfn:quotient}) under rule isomorphism (Definition \ref{dfn:ruleiso}) are equal (\(\mathcal{R}_1 /\!\!\cong \,\,= \mathcal{R}_2 /\!\!\cong\));
\item \textbf{Normalisation equivalent} (\(T_1 \simeq_N T_2\)) iff their quotients under rule equivalence (Definition \ref{dfn:rulenormequiv}) are equal (\(\mathcal{R}_1 /\!\!\simeq \,\,= \mathcal{R}_2 /\!\!\simeq\));
\item \textbf{Step-wise equivalent} (\(T_1 \simeq_S T_2\)) iff their \textbf{induced ARSs} (Definition \ref{dfn:gtars}) are equal (\(\rightarrow_{T_1} \,=\,\, \rightarrow_{T_2}\));
\item \textbf{Semantically equivalent} (\(T_1 \simeq_F T_2\)) iff their \textbf{semantic functions} are equal (\(f_{T_1} = f_{T_2}\)).
\end{enumerate}
\end{definition}

\begin{proposition}
Each of the above notions are equivalences.
\end{proposition}

\begin{remark} \label{rem:compatderivations}
A way of looking at \textbf{step-wise equivalence} is to say that it is a formalisation of what it means to have \textbf{derivation compatibility} in the sense of Theorem \ref{thm:compatderivations}.
\end{remark}

\begin{proposition}
This notion of isomorphism gives rise to \textbf{abstract graph transformation systems} \([T]\) over some fixed label alphabet \(\mathcal{L}\). Let \(\mathcal{T}(\mathcal{L})\) denote the collection of all such classes. Then, \(\mathcal{T}(\mathcal{L})\) is a \textbf{countable set}.
\end{proposition}

\begin{remark}
Clearly \textbf{isomorphism} and \textbf{normalisation equivalence} are well behaved. That is, it is \textbf{decidable} to check if two GT systems share the same class. The same is not true of \textbf{semantic equivalence}.
\end{remark}

\begin{theorem}[GT System Equivalence] \label{thm:gtequiv}
GT system \textbf{isomorphism} is finer than \textbf{normalisation equivalence} is finer than \textbf{step-wise equivalence} is finer than \textbf{semantic equivalence}. That is, \(T_1 \cong T_2\) implies \(T_1 \simeq_N T_2\) implies \(T_1 \simeq_S T_2\) implies \(T_1 \simeq_F T_2\). Moreover, there exist GT systems in which the implication direction is strictly one-way.
\end{theorem}

\begin{figure}[H]
\centering
\input{fig/2/gtequiv}
\vspace{-0.2em}
\caption{Example Rules Demonstrating Non-Equivalence}
\label{fig:gtequiv}
\end{figure}
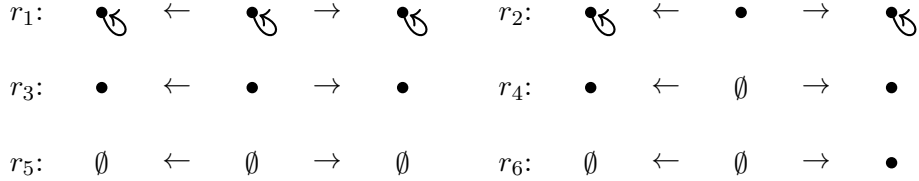

\begin{proof}
Let \(T_1\), \(T_2\) be GT systems over some \(\mathcal{L}\), with rule sets \(\mathcal{R}_1\), \(\mathcal{R}_2\). Within this proof, rules \(r_1, r_2, r_3, r_4, r_5, r_6\) can be found in Figure \ref{fig:gtequiv}.

Suppose \(T_1 \cong T_2\). Then the \(\cong\)-classes of \(\mathcal{R}_1\) correspond to those of \(\mathcal{R}_2\). Clearly, if we find the normal form of each class, then the correspondence between these classes of normal forms is preserved. So \(T_1 \simeq_N T_2\). To see the inclusion is strict, consider the two systems \((\mathcal{L}, \{r_1\})\), \((\mathcal{L}, \{r_2\})\). They are non-isomorphic, but are normalisation equivalent.

Next suppose \(T_1 \simeq_N T_2\). Then by Theorem \ref{thm:compatderivations}, the choice of representative element from each class is irrelevant, that is, the derivations possible are identical. Now, since the \(\simeq_N\)-classes of \(\mathcal{R}_1\) and \(\mathcal{R}_2\) are identical, combining all possible derivations from the classes leaves us with identical possible derivations for each. Thus, it is immediate that the induced ARS is identical. To see the inclusion is strict, consider the two systems \((\mathcal{L}, \{r_3\})\), \((\mathcal{L}, \{r_3, r_4\})\). They are not normalisation equivalent, but are step-wise.

Finally, suppose \(T_1 \simeq_S T_2\). Then the induced ARS relations \(\rightarrow_{\mathcal{R}_1}, \rightarrow_{\mathcal{R}_2}\) are equal, so clearly \(f_{T_1} = f_{T_2}\). To see the inclusion is strict, consider the two systems \((\mathcal{L}, \{r_5\})\), \((\mathcal{L}, \{r_6\})\) are not step-wise equivalent since \(r_5\) is always applicable with no effect, but \(r_6\) is also always applicable, adding a new node. They are, however, semantically equivalent since their semantic functions both evaluate to \(\{\bot\}\) on all inputs.

Thus, we have shown each forward implication, and that there exist cases where the reverse implication is false.
\end{proof}

In general, we might be interested in more than proving just equivalence. That is, when does one GT system \enquote{refine} the other. The (stepwise) refinement of programs was originally proposed by Dijkstra \cite{Dijkstra68a,Dijkstra72a} and Wirth \cite{Wirth71a}. Thinking in terms of GT systems, one may want to consider compatibility of the semantic function. Development of a refinement calculus that behaves properly with rooted GT systems remains open research.

%% file: fig/2/derivation.tex
\begin{tikzpicture}[every node/.style={inner sep=0pt, text width=6.5mm, align=center}]
    \node (a) at (0.0,0.0) [draw, circle, thick] {$\Square$};
    \node (b) at (1.0,0.0) {$\leftarrow$};
    \node (c) at (2.0,0.0) [draw, circle, thick] {\,};
    \node (d) at (3.0,0.0) {$\rightarrow$};
    \node (e) at (4.0,0.0) [draw, circle, thick, double, double distance=0.4mm] {$\Square$};

    \node (f) at (0.0,-1.0) {$\big\downarrow$};
    \node (g) at (1.0,-1.0) {NPO};
    \node (h) at (2.0,-1.0) {$\big\downarrow$};
    \node (i) at (3.0,-1.0) {PO};
    \node (j) at (4.0,-1.0) {$\big\downarrow$};

    \node (k) at (0.0,-2.0) [draw, circle, thick, double, double distance=0.4mm] {$\Square$};
    \node (l) at (1.0,-2.0) {$\leftarrow$};
    \node (m) at (2.0,-2.0) [draw, circle, thick, double, double distance=0.4mm] {\,};
    \node (n) at (3.0,-2.0) {$\rightarrow$};
    \node (o) at (4.0,-2.0) [draw, circle, thick, double, double distance=0.4mm] {$\Square$};
\end{tikzpicture}

%% file: fig/2/eg.tex
\begin{tikzpicture}[every node/.style={inner sep=0pt, text width=6.5mm, align=center}]
    \node (a) at (0,0) [draw, circle, thick] {\(\Square\)};
    \node (b) at (2,0) [draw, circle, thick, double, double distance=0.4mm] {\(\triangle\)};
    \node (c) at (4,0) [draw, circle, thick, pattern=north east lines] {\,};
    \node (d) at (6,0) [draw, circle, thick] {\,};

    \draw (a) edge[->,thick] node[above, yshift=2.5pt] {\(x\)} (b)
          (b) edge[->,thick] node[above, yshift=2.5pt] {\(y\)} (c);
\end{tikzpicture}

%% file: fig/2/morphism.tex
\begin{tikzpicture}
\node[scale=0.9] (a) at (0,0){
\begin{tikzcd}
  \text{}
    & E_G \arrow[r, shift left=1ex, "s_G"] \arrow[r, shift right=1ex, "t_G"] \arrow[dd, "g_E"] \arrow[ld, "m_G"]
    & V_G \arrow[dd, "g_V"] \arrow[rd, "{l_G}"] \arrow[rrd, bend left=20, "{p_G}"]
    & \text{}
    & \text{} \\
  \mathcal{L}_E 
    & \text{}
    & \text{}
    & \mathcal{L}_V
    & \mathbb{Z}_2 \\
  \text{}
    & E_H \arrow[r, shift left=1ex, "s_H"] \arrow[r, shift right=1ex, "t_H"] \arrow[lu, "m_H"]
    & V_H \arrow[ru, "{l_H}"] \arrow[rru, bend right=20, "{p_H}"]
    & \text{}
    & \text{}
\end{tikzcd}
};
\end{tikzpicture}

%% file: fig/2/comsqrs.tex
\begin{tikzpicture}
\node[scale=1.2] (a) at (0,0){
\begin{tikzcd}
  L \arrow[d, ""] \arrow[dr, phantom, "(1)"]
  & K \arrow[l, ""] \arrow[d, ""] \arrow[r, ""] \arrow[dr, phantom, "(2)"]
  & R \arrow[d, ""] \\
  G 
    & D \arrow[l, ""] \arrow[r, ""]
    & H
\end{tikzcd}
};
\end{tikzpicture}

%% file: fig/2/ruleiso.tex
\begin{tikzpicture}[every node/.style={inner sep=0pt, text width=6.5mm, align=center}]
    \node (a) at (0.0,0) {$r_1$:};
    \node (b) at (1.0,0) [draw, circle, thick] {$\Square$};
    \node (c) at (2.5,0) [draw, circle, thick] {$\triangle$};
    \node (d) at (3.5,0) {$\leftarrow$};
    \node (e) at (4.5,0) [draw, circle, thick] {$\Square$};
    \node (f) at (5.5,0) {$\rightarrow$};
    \node (g) at (6.5,0) [draw, circle, thick] {$\Square$};

    \node (B) at (1.0,-.52) {\tiny{1}};
    \node (C) at (2.5,-.52) {\tiny{2}};
    \node (E) at (4.5,-.52) {\tiny{1}};
    \node (G) at (6.5,-.52) {\tiny{1}};

    \draw (b) edge[->,thick] node[above, yshift=2.5pt] {$\triangle$} (c)
          (b) edge[->,in=-30,out=-60,loop,thick] node[right, yshift=1.5pt] {$\Square$} (b)
          (e) edge[->,in=-30,out=-60,loop,thick] node[right, yshift=1.5pt] {$\Square$} (e)
          (g) edge[->,in=-30,out=-60,loop,thick] node[right, yshift=1.5pt] {$\Square$} (g);

    \node (h) at (8.0,0)  {$r_2$:};
    \node (i) at (9.0,0)  [draw, circle, thick] {$\Square$};
    \node (j) at (10.5,0) [draw, circle, thick] {$\triangle$};
    \node (k) at (11.5,0) {$\leftarrow$};
    \node (l) at (12.5,0) [draw, circle, thick] {$\Square$};
    \node (m) at (13.5,0) {$\rightarrow$};
    \node (n) at (14.5,0) [draw, circle, thick] {$\Square$};

    \node (I) at (9.0,-.52)  {\tiny{7}};
    \node (J) at (10.5,-.52) {\tiny{4}};
    \node (L) at (12.5,-.52) {\tiny{7}};
    \node (N) at (14.5,-.52) {\tiny{7}};

    \draw (i) edge[->,thick] node[above, yshift=2.5pt] {$\triangle$} (j)
          (i) edge[->,in=-30,out=-60,loop,thick] node[right, yshift=1.5pt] {$\Square$} (i)
          (l) edge[->,in=-30,out=-60,loop,thick] node[right, yshift=1.5pt] {$\Square$} (l)
          (n) edge[->,in=-30,out=-60,loop,thick] node[right, yshift=1.5pt] {$\Square$} (n);

    \node (a) at (0.0,-1.7) {$r_3$:};
    \node (b) at (1.0,-1.7) [draw, circle, thick] {$\Square$};
    \node (c) at (2.5,-1.7) [draw, circle, thick] {$\triangle$};
    \node (d) at (3.5,-1.7) {$\leftarrow$};
    \node (e) at (4.5,-1.7) [draw, circle, thick] {$\Square$};
    \node (f) at (5.5,-1.7) {$\rightarrow$};
    \node (i) at (6.5,-1.7) [draw, circle, thick] {$\Square$};

    \node (B) at (1.0,-2.22) {\tiny{1}};
    \node (C) at (2.5,-2.22) {\tiny{2}};
    \node (E) at (4.5,-2.22) {\tiny{1}};
    \node (I) at (6.5,-2.22) {\tiny{1}};

    \draw (b) edge[->,thick] node[above, yshift=2.5pt] {$\triangle$} (c)
          (b) edge[->,in=-30,out=-60,loop,thick] node[right, yshift=1.5pt] {$\Square$} (b);

    \node (j) at (8.0,-1.7)  {$r_1'$:};
    \node (k) at (9.0,-1.7)  [draw, circle, thick] {$\Square$};
    \node (l) at (10.5,-1.7) [draw, circle, thick] {$\triangle$};
    \node (m) at (11.5,-1.7) {$\leftarrow$};
    \node (n) at (12.5,-1.7) [draw, circle, thick, pattern=north east lines] {\,};
    \node (o) at (13.5,-1.7) {$\rightarrow$};
    \node (p) at (14.5,-1.7) [draw, circle, thick] {$\Square$};

    \node (K) at (9.0,-2.22)  {\tiny{1}};
    \node (L) at (10.5,-2.22) {\tiny{2}};
    \node (N) at (12.5,-2.22) {\tiny{1}};
    \node (P) at (14.5,-2.22) {\tiny{1}};

    \draw (k) edge[->,thick] node[above, yshift=2.5pt] {$\triangle$} (l)
          (k) edge[->,in=-30,out=-60,loop,thick] node[right, yshift=1.5pt] {$\Square$} (k)
          (p) edge[->,in=-30,out=-60,loop,thick] node[right, yshift=1.5pt] {$\Square$} (p);
\end{tikzpicture}

%% file: fig/2/derivations.tex
\begin{tikzpicture}[baseline= (a).base]
\node[scale=.75] (a) at (0,0){
\begin{tikzcd}[sep={1.2em,between origins}]
               &  &  & K' \arrow[rrrr] \arrow[llldd] \arrow[rdd] \arrow[dddd] &                                           &  &  & R \arrow[dddd] \\
               &  &  &                                                        &                                           &  &  &                \\
L \arrow[dddd] &  &  &                                                        & K \arrow[dddd] \arrow[llll] \arrow[rrruu] &  &  &                \\
               &  &  &                                                        &                                           &  &  &                \\
               &  &  & D' \arrow[rdd] \arrow[llldd] \arrow[rrrr]              &                                           &  &  & H              \\
               &  &  &                                                        &                                           &  &  &                \\
G              &  &  &                                                        & D \arrow[llll] \arrow[rrruu]              &  &  &               
\end{tikzcd}
};
\end{tikzpicture}

%% file: fig/2/gtequiv.tex
\begin{tikzpicture}[every node/.style={align=center}]
    \node (a) at (0.0,-0.05) {$r_1$:};
    \node (b) at (1.0,0.0)   [draw, circle, thick, fill=black, scale=0.3] {\,};
    \node (c) at (2.0,0.0)   {$\leftarrow$};
    \node (d) at (3.0,0.0)   [draw, circle, thick, fill=black, scale=0.3] {\,};
    \node (e) at (4.0,0.0)   {$\rightarrow$};
    \node (f) at (5.0,0.0)   [draw, circle, thick, fill=black, scale=0.3] {\,};

    \node (g) at (6.5,-0.05) {$r_2$:};
    \node (h) at (7.5,0.0)   [draw, circle, thick, fill=black, scale=0.3] {\,};
    \node (i) at (8.5,0.0)   {$\leftarrow$};
    \node (j) at (9.5,0.0)   [draw, circle, thick, fill=black, scale=0.3] {\,};
    \node (k) at (10.5,0.0)  {$\rightarrow$};
    \node (l) at (11.5,0.0)  [draw, circle, thick, fill=black, scale=0.3] {\,};

    \draw (b) edge[->,in=-20,out=-70,loop,thick] (b)
          (d) edge[->,in=-20,out=-70,loop,thick] (d)
          (f) edge[->,in=-20,out=-70,loop,thick] (f);

    \draw (h) edge[->,in=-20,out=-70,loop,thick] (h)
          (l) edge[->,in=-20,out=-70,loop,thick] (l);

    \node (a) at (0.0,-1.05) {$r_3$:};
    \node (b) at (1.0,-1.0)  [draw, circle, thick, fill=black, scale=0.3] {\,};
    \node (c) at (2.0,-1.0)  {$\leftarrow$};
    \node (d) at (3.0,-1.0)  [draw, circle, thick, fill=black, scale=0.3] {\,};
    \node (e) at (4.0,-1.0)  {$\rightarrow$};
    \node (f) at (5.0,-1.0)  [draw, circle, thick, fill=black, scale=0.3] {\,};

    \node (g) at (6.5,-1.05) {$r_4$:};
    \node (h) at (7.5,-1.0)  [draw, circle, thick, fill=black, scale=0.3] {\,};
    \node (i) at (8.5,-1.0)  {$\leftarrow$};
    \node (j) at (9.5,-1.0)  {$\emptyset$};
    \node (k) at (10.5,-1.0) {$\rightarrow$};
    \node (l) at (11.5,-1.0) [draw, circle, thick, fill=black, scale=0.3] {\,};

    \node (a) at (0.0,-2.05) {$r_5$:};
    \node (b) at (1.0,-2.0)  {$\emptyset$};
    \node (c) at (2.0,-2.0)  {$\leftarrow$};
    \node (d) at (3.0,-2.0)  {$\emptyset$};
    \node (e) at (4.0,-2.0)  {$\rightarrow$};
    \node (f) at (5.0,-2.0)  {$\emptyset$};

    \node (g) at (6.5,-2.05) {$r_6$:};
    \node (h) at (7.5,-2.0)  {$\emptyset$};
    \node (i) at (8.5,-2.0)  {$\leftarrow$};
    \node (j) at (9.5,-2.0)  {$\emptyset$};
    \node (k) at (10.5,-2.0) {$\rightarrow$};
    \node (l) at (11.5,-2.0) [draw, circle, thick, fill=black, scale=0.3] {\,};
\end{tikzpicture}

%% file: content/3.tex
\chapter{Fast Language Recognition} \label{chapter:treerecognision}

\begin{chapquote}{Edsger W. Dijkstra, \textit{Unknown source}}
``Computer Science is no more about computers than astronomy is about telescopes.''
\end{chapquote}

The Graph Matching Problem (Definition \ref{dfn:gmp}) and Rule Application Problem (Definition \ref{dfn:rap}) can be considered in our setting. We will see that if we have an input graph with bounded degree (Definition \ref{dfn:boundeddegree}) and a bounded number of root nodes, and a finite set of \enquote{fast} rules, then we can perform matching in constant time. Moreover, under mild constraints, for systems of fast rules, we need only analyse the derivation length to determine the time complexity of finding a normal form.

The language of all unlabelled trees is well-known to be expressible using classical graph transformation systems, using a single rule. The question of recognising trees \enquote{efficiently} is less understood. We present a GT system that can test if a graph is a tree in linear time, given the input is of \enquote{bounded degree}: a new result for graph transformation systems. We also show that this can easily be adapted to recognise only full binary trees, and can be translated easily into GP\,2.

Part of this chapter was presented at CALCO 2019 as part of a co-authored paper looking at linear time algorithms in GP\,2 \cite{Campbell-Courtehoute-Plump19b}.


\section{Complexity Theorems} \label{section:complexitythms}

We begin by defining the notion of a \enquote{fast} rule. Intuitively, requiring every connected component of a rule to contain a root node is going to allow an implementation to both match each root node in the host graph in constant time, and then match the rest of the subgraphs around the root nodes quickly.

\begin{definition}
We call a rule \(r = \langle L \leftarrow K \rightarrow R \rangle\) \textbf{fast} iff every connected component (Definition \ref{dfn:connectedcomponent}) of \(L\) contains a root node. Additionally, we call a GT system \textbf{fast} iff all its rules are \textbf{fast}.
\end{definition}

Just like in Lemma \ref{prop:gmptime}, we need to set up some assumptions about the complexity of various problems. We will assume that graphs are stored in a format such that the time complexities of various problems are as given in the table (Figure \ref{fig:rperfass}), based on Dodds' assumptions (Figure \ref{fig:perfass}) \cite{Dodds08a}. In our model, we have root nodes specified explicitly with a rootedness function, so we have updated the table to reflect this. Moreover, we have made our assumptions as weak as possible to see the results we need. An additional advantage of our assumptions is that they are realistic even when the alphabet is not finite, such as in GP\,2.

\begin{figure}[H]
\noindent
\vspace{-0.25em}
\input{fig/3/complexity}
\vspace{-1.0em}
\caption{Complexity Assumptions Table}
\label{fig:rperfass}
\vspace{1.0em}
\end{figure}

\begin{lemma}
Given a \textbf{TLRG} \(G\) of \textbf{bounded degree} containing a \textbf{bounded} number of root nodes, and a \textbf{fast} rule \(r\), then the GMP (Definition \ref{dfn:gmp}) requires \(O(\abs{r})\) time and produces \(O(\abs{r})\) matches.
\end{lemma}

\begin{proof}
Under the same assumption as in Dodds' Thesis \cite[p. 39]{Dodds08a}, this is easy to see, since there are only a constant number of subgraphs to consider. The full proof is a minor modification of Dodds' proof. The most notable difference is that in our model, root nodes are not encoded as a special label, rather, via the rootedness function. This is handled in our complexity assumptions table. Then, because we have a bounded number of root nodes in \(G\), we can then, ultimately, conclude \(O(\abs{r})\) time rather than \(O(\abs{V_G})\) time. 
\end{proof}

\begin{lemma}
Given a \textbf{TLRG} \(G\) of \textbf{bounded degree}, a rule \(r\), and an \textbf{injective match} \(g\), then RAP (Definition \ref{dfn:rap}) requires \(O(\abs{r})\) time.
\end{lemma}

\begin{proof}
Obvious modifications of the proofs in Dodds' Thesis.
\end{proof}

\begin{definition}
Define \(\operatorname{mx}: \mathcal{P}(\mathbb{N}) \to \mathbb{N}\) by \(\operatorname{mx}(\emptyset) = 0\) and \(\operatorname{mx}(S) = \operatorname{max}(S)\) where \(S \neq \emptyset\). We can then define \(\operatorname{mx\_deg}(G) = \operatorname{mx}(\{\operatorname{deg}_G(v) \mid v \in V_G\})\).
\end{definition}

\begin{definition}
Given an upper bound on the degree of nodes \(N\) and an upper bound on the number of root nodes \(M\), a GT system \(T = (\mathcal{L}, \mathcal{R})\) is:
\begin{enumerate}[itemsep=-0.4ex,topsep=-0.4ex]
\item \textbf{Bounded degree preserving} iff for all \textbf{TLRG} \(G, H\) such that \(\operatorname{mx\_deg}\) \((G) \leq N\), then \(G \Rightarrow_{\mathcal{R}} H\) implies that \(\operatorname{mx\_deg}(H) \leq N\);
\item \textbf{Bounded roots preserving} iff for all \textbf{TLRG} \(G, H\) such that \(\abs{p_G^{-1}(\{1\})}\) \(\leq M\), then \(G \Rightarrow_{\mathcal{R}} H\) implies that \(\abs{p_H^{-1}(\{1\})} \leq M\).
\end{enumerate}
\end{definition}

Luckily, these properties can be easily statically checked by looking at the shape of the rules (Lemma \ref{lem:staticcheck}).

\begin{definition}
Given an upper bound on the degree of nodes \(N \in \mathbb{N}\), a rule \(r = \langle L \leftarrow K \rightarrow R \rangle\) is:
\begin{enumerate}[itemsep=-0.4ex,topsep=-0.4ex]
\item \textbf{Bounded degree preserving} iff \(\forall v \in (V_R \setminus V_K), \operatorname{deg}_R(v) \leq N\) and \(\forall v \in V_K, \operatorname{deg}_R(v) \leq \operatorname{deg}_L(v)\);
\item \textbf{Bounded roots preserving} iff \(\abs{p_R^{-1}(\{1\})} \leq \abs{p_L^{-1}(\{1\})}\).
\end{enumerate}
\end{definition}

\begin{lemma} \label{lem:staticcheck}
A GT system is \textbf{bounded degree preserving} (\textbf{bounded roots preserving}) iff every rule is \textbf{bounded degree preserving} (\textbf{bounded roots preserving}).
\end{lemma}

\begin{proof}
The forwards direction can be seen by contradiction. Suppose that a rule were not preserving, then one could construct a graph were one could derive a graph of too large degree or too many root nodes. The reverse direction is obvious by Theorem \ref{thm:derivationtheorem} and Corollary \ref{cor:invariantroots}.
\end{proof}

\begin{remark}
Being \textbf{bounded degree preserving} does not mean that the maximum degree of a direct successor of a graph need be reduced, unlike for bounded root nodes. The rule that introduces a new node with a looped edge is \textbf{bounded degree preserving}, but one could imagine starting a derivation with the empty graph, then applying such a rule, taking the maximum degree of the graph from 0 to 1.
\end{remark}

\begin{theorem}[Fast Derivations] \label{thm:fastderivations}
Given a \textbf{TLRG} \(G\) of \textbf{bounded degree} containing a \textbf{bounded} number of root nodes, and a \textbf{fast} GT system \(T = (\mathcal{L}, \mathcal{R})\), then one can decide in \textbf{constant time} the \textbf{direct successors} (Definition \ref{dfn:ars2}) of \(G\), up to isomorphism.
\end{theorem}

\begin{proof}
Combine the above lemmas. There is a constant number of rules to apply. For each rule, a bounded number of matches are produced in constant time, and then the RAP takes constant time for each match.
\end{proof}

\begin{corollary}[Derivation-Complexity Lemma] \label{corollary:lineargt}
Given \(G\) as above, and \(T = (\mathcal{L}, \mathcal{R})\) a \textbf{fast}, \textbf{bounded degree preserving}, \textbf{bounded roots preserving}, \textbf{terminating} GT system with maximum derivation length \(N \in \mathbb{N}\), then one can find a \textbf{normal form} (Definition \ref{dfn:ars2}) of \(G\) in \(O(N)\) time.
\end{corollary}

\begin{proof}
By induction, the application of a rule satisfying the stated conditions will preserve the bound on the number of root nodes and the bound on the degree of the nodes. Thus, we have the result.
\end{proof}

Thus, we have shown that if we have a set of rules as per Corollary \ref{corollary:lineargt}, we need only consider the maximum length of derivations when reasoning about time complexity, as mentioned at the end of Section \ref{section:rooted}.


\section{Recognising Trees in Linear Time} \label{sec:linearrec}

A \textbf{tree} is a graph containing a node from which there is a unique \textbf{directed path} to each node in the graph (Proposition \ref{prop:treedfn}). It is easy to see that it is possible to generate the collection of all trees by inductively adding new leaf nodes to the discrete graph of size one. That is, the language of all unlabelled trees can be specified by DPO-based graph grammar. The question of \enquote{efficient} recognition of trees is less understood. In this section, we will show that it is possible to recognise the collection of all unlabelled trees in linear time. Moreover, at the end of the chapter, we will give an implementation of the algorithm in GP\,2, providing empirical evidence of the linear run time.

Writing a graph grammar that generates all unlabelled trees is straightforward. Simply start with the trivial tree (a single node), and arbitrarily add edges pointing to a new node, away from this start node.

\begin{example}[Tree Grammar]
\begin{figure}[H]
\centering
\noindent
\input{fig/3/tree1}
\vspace{-0.3em}
\caption{Tree Grammar Rules}
\label{fig:tree1}
\end{figure}
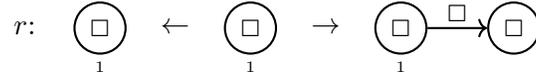

Let \(\pmb{TREE} = (\mathcal{L}, \mathcal{N}, S, \mathcal{R})\) where:
\begin{enumerate}[itemsep=-0.4ex,topsep=-0.4ex]
\item \(\mathcal{L} = (\{\Square\}, \{\Square\})\) where \(\Square\) denotes the empty label;
\item \(\mathcal{N} = (\emptyset, \emptyset)\);
\item \(S\) be the graph with a single node labelled with \(\Square\);
\item \(\mathcal{R} = \{r\}\).
\end{enumerate}

To see that this grammar generates the set of all trees, we must show that every graph in the language is a tree, and then that every tree is in the language. This is easy to see by induction.
\end{example}

Notice how the above construction has given us a decision procedure for testing if \([G] \in \pmb{L}(\pmb{TREE})\):

\begin{lemma}
The GT system \(((\{\Square\}, \{\Square\}), \{{r}^{-1}\})\) recognises \(\pmb{L}(\pmb{TREE})\), in the sense of Definition \ref{dfn:langrec}. That is, \([G] \in \pmb{L}(\pmb{TREE})\) iff \(G \Rightarrow_{{r}^{-1}} S\). Moreover, the system is confluent and terminating.
\end{lemma}

\begin{proof}
We first must comment that we are considering the GT system as a classical system as in Appendix \ref{appendix:transformation}, without root nodes. As such, we must adjust Definition \ref{dfn:langrec} accordingly. Now, by Theorem \ref{thm:membershiptest} (also adjusted accordingly), it is immediate that the GT system recognises the language of trees. To see that it is terminating is easy, since each rule is size reducing. Finally, local confluence can be seen via critical pair analysis (Section \ref{section:critpairs}) since there are no \enquote{critical pairs}. Thus, by Newman's Lemma (Theorem \ref{thm:newmanlem}), the system is confluent.
\end{proof}

While the algorithm may be fast in the sense that we need not backtrack, the process of finding a match at each derivation step is slow, since we must consider the entire host graph when finding a match. Next, we will see that rooted graph transformation rules can actually recognise trees in linear time.

\vspace{0.05em}
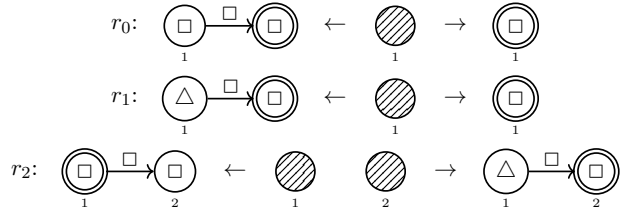
\begin{figure}[H]
\noindent
\centering
\input{fig/3/tree2}
\vspace{-0.3em}
\caption{Tree Recognition Rules}
\label{fig:tree2}
\vspace{-0.1em}
\end{figure}

Let \(\mathcal{L} = (\{\Square, \triangle\}, \{\Square\})\), and \(\mathcal{R} = \{r_0, r_1, r_2\}\). We are going to show that \(\mathcal{R}\) induces a linear time algorithm for testing if a graph is a tree. Intuitively, this works by pushing a special node (a \enquote{root} node) to the bottom of a branch, and then pruning. If we start with a tree and run this until we cannot do it anymore, we must be left with a single node. The triangle labels are necessary so that, in the case that the input graph is not a tree, we could \enquote{get stuck} in a directed cycle.

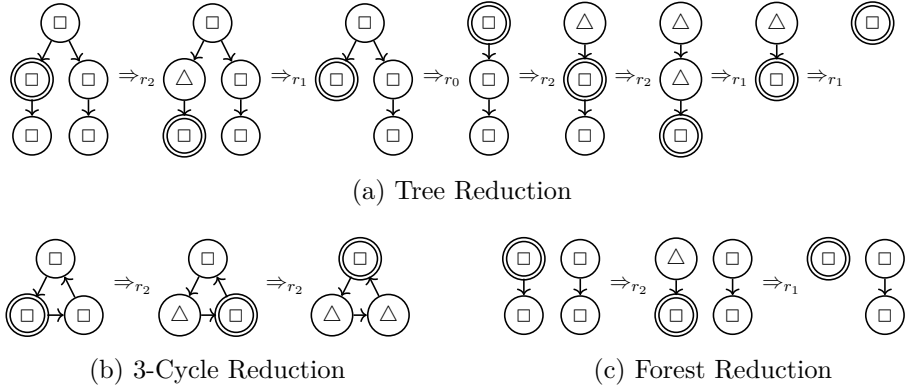
\begin{figure}[H]
\begin{subfigure}{1\textwidth}
    \centering
    \input{fig/3/reduction}
    \vspace{0.1em}
    \caption{Tree Reduction}
\end{subfigure}
\begin{subfigure}{.5\textwidth}
    \centering
    \vspace{2.0em}
    \input{fig/3/non-reduction1}
    \vspace{0.1em}
    \caption{3-Cycle Reduction}
\end{subfigure}
\begin{subfigure}{.5\textwidth}
    \centering
    \vspace{2.0em}
    \input{fig/3/non-reduction2}
    \vspace{0.1em}
    \caption{Forest Reduction}
\end{subfigure}
\vspace{0.6em}
\caption{Example Reductions}
\label{fig:tree-red}
\vspace{0.25em}
\end{figure}

\begin{example}
Figure \ref{fig:tree-red} shows a reduction of a tree and non-trees. Note, in particular, how the triangles prevent an infinite sequence of derivations moving the root node around the 3-cycle.
\end{example}

\vspace{-1.1em}
\noindent

\begin{definition}
Given a graph \(G\), we define \(G^{\ominus}\) to be exactly \(G\), but with every node unrooted, and everything labelled by \(\Square\). That is:
\vspace{-0.5em}
\begin{equation*}
G^{\ominus} = (V_G, E_G, s_G, t_G, V_G \times \{\Square\}, E_G \times \{\Square\}, V_G \times \{0\}).
\end{equation*}
\end{definition}

For the rest of this section, by \enquote{input graph}, we mean any TLRG containing exactly one \enquote{root} node, with edges and vertices all labelled \(\Square\). By \enquote{input tree}, we mean an \enquote{input graph} that is also a tree (Definition \ref{dfn:tree}). Next, we show some technical lemmata, giving names to the results that are directly used in our proof of main result (Theorem \ref{thm:treerecthm}).

\begin{lemma}[Tree Termination] \label{lem:treederlen}
The system \((\mathcal{L}, \mathcal{R})\) is terminating. Moreover, derivations have length at most \(2 \abs{V_G}\).
\end{lemma}

\begin{proof}
Let \(\# G = \abs{V_G}\), \(\Square G = \abs{\{v \in V_G \mid l_G(v) = \Square\}}\), for any TLRG \(G\). If \(G \Rightarrow_{r_0} H\) or \(G \Rightarrow_{r_1} H\), then \(\# G > \# H\) and \(\Square G > \Square H\). If \(G \Rightarrow_{r_2} H\) then \(\# G = \# H\) and \(\Square G > \Square H\). Thus, termination follows from Lemma \ref{lem:gtmonoembedding}, since \(\varphi([G]) = \# G\) is a monotone embedding into \((\mathbb{N}, >)\). To see the last part, notice that \(\Square G \leq \# G\) for all TLRGs \(G\), so the result is immediate since there are only \(2 \# G\) natural numbers less than \(2 \# G\).
\end{proof}

\begin{lemma} \label{lem:treegarbagesep}
If \(G\) is a tree and \(G \Rightarrow_{\mathcal{R}} H\), then \(H\) is a tree. If \(G\) is not a tree and \(G \Rightarrow_{\mathcal{R}} H\), then \(H\) is not a tree.
\end{lemma}

\begin{proof}
Clearly, the application of \(r_2\) preserves structure. Suppose \(G\) is a tree. \(r_0\) or \(r_1\) are applicable iff node 2 is matched against a leaf node due to the dangling condition. Upon application, the leaf node and its incoming edge is removed. Clearly the result graph is still a tree.

If \(G\) is not a tree and one of \(r_0\) or \(r_1\) is applicable, then we can see the properties of not being a tree are preserved. That is, if \(G\) is not connected, \(H\) is certainly not connected. If \(G\) had parallel edges, due to the dangling condition, they must exist in \(G \setminus g(L)\), so \(H\) has parallel edges. Similarly, cycles are preserved. Finally, if \(G\) had a node with incoming degree greater than one, then \(H\) must too, since the node in \(G\) that is deleted in \(H\) had incoming degree one, and the degree of all other nodes is preserved.
\end{proof}

\begin{corollary}[Tree Preservation] \label{corollary:treepreserving}
If \(G\) is an input graph and \(G \Rightarrow_{\mathcal{R}}^* H\), then \(G\) is a tree iff \(H\) is a tree.
\end{corollary}

\begin{proof}
Induction.
\end{proof}

\begin{lemma} \label{lem:treeroots}
The system \((\mathcal{L}, \mathcal{R})\) preserves the number of root nodes in graphs. That is, if \(G \Rightarrow_\mathcal{R}^* H\), then \(\abs{p_G^{-1}(\{1\})} = \abs{p_H^{-1}(\{1\})}\).
\end{lemma}

\begin{proof}
Each of the rules \(r_0, r_1, r_2\) preserves the number of roots, so by Corollary \ref{cor:invariantroots}, the number of root nodes is invariant across rule applications. Thus, the result holds by induction on derivation length.
\end{proof}

\begin{corollary} \label{cor:oneroot}
If \(G\) is an input graph and \(G \Rightarrow_{\mathcal{R}}^{*} H\), then \(H\) has exactly one root node. Moreover, there is no derivation sequence that derives the empty graph.
\end{corollary}

\begin{proof}
The first part follows from Lemma \ref{lem:treeroots} since input graphs have exactly one root node. To see that the empty graph cannot be derived, notice that each derivation reduces \(\# G\) by at most one, and no rules are applicable when \(\# G = 1\).
\end{proof}

\begin{remark}
In Bak and Plump's model (and hence GP\,2), Corollary \ref{cor:oneroot} is still true, however a more direct proof is needed. Since the root node in the LHS of each rule must be matched against a root node in the host graph, so the other non-roots can only be matched against non-roots.
\end{remark}

\begin{lemma} \label{lem:trianglechains}
If \(G\) is an input graph and \(G \Rightarrow_{\mathcal{R}}^{*} H\). Then, every \(\triangle\)-node in \(H\) either has a child \(\triangle\)-node or a root-node child.
\end{lemma}

\begin{proof}
Clearly \(G\) satisfies this, as there are no \(\triangle\)-nodes. We now proceed by induction. Suppose \(G \Rightarrow_{\mathcal{R}}^* H \Rightarrow_{\mathcal{R}} H'\) where \(H\) satisfies the condition. If \(r_0\) or \(r_1\) is applicable, we introduce no new \(\triangle\)-nodes. Additionally, in the case of \(r_1\), any \(\triangle\)-node parents of the image 1 are preserved. So \(H'\) satisfies the condition. Finally, if \(r_2\) is applied, then the new \(\triangle\)-node has a root-node child, and the \(\triangle\)-nodes in \(H' \setminus h(R)\) have the same children, so \(H'\) satisfies the condition.
\end{proof}

\begin{corollary} \label{lem:trianglechild}
Let \(G\) be an input tree and \(G \Rightarrow_{\mathcal{R}}^{*} H\). Then the root-node in \(H\) has no \(\triangle\)-node children.
\end{corollary}

\begin{proof}
By Corollary \ref{cor:oneroot}, \(H\) has exactly one root node, and by Lemma \ref{lem:trianglechains}, all chains of \(\triangle\)-nodes terminate with a root-node. If said root-node were to have a \(\triangle\)-node child, then we would have a cycle, which contradicts that \(H\) is a tree (Corollary \ref{corollary:treepreserving}).
\end{proof}

\begin{lemma}[Tree Progress] \label{lemma:treereduce}
Let \(G\) be an input tree and \(G \Rightarrow_{\mathcal{R}}^{*} H\). Then, either \(\abs{V_H} = 1\) or \(H\) is not in normal form.
\end{lemma}

\begin{proof}
By Corollary \ref{cor:oneroot}, \(\abs{V_H} \geq 1\). If \(\abs{V_G} = 1\), then \(G\) is in normal form. Otherwise, either the root node has no children, or it has at least one \(\Square\)-child. In the first case, \(r_0\) must be applicable, and in the second, \(r_2\).

Suppose \(G \Rightarrow_{\mathcal{R}}^{*} H\). If \(\abs{V_H} = 1\), then \(H\) is in normal form by the proof to Corollary \ref{cor:oneroot}. Otherwise, by Corollary \ref{corollary:treepreserving} \(H\) is a tree and \(\abs{V_H} > 1\). Now, the root-node in \(H\) (Corollary \ref{cor:oneroot}) must have a non-empty neighbourhood. If it has no children, then \(r_0\) or \(r_1\) must be applicable. Otherwise, \(r_2\) must be applicable, since by Corollary \ref{lem:trianglechild}, there must be a \(\Square\)-node child. So \(H\) is not in normal form.
\end{proof}

We now present the main result of this chapter:

\begin{theorem}[Tree Recognition] \label{thm:treerecthm}
Given an input graph \(G\), one may use the system \((\mathcal{L}, \mathcal{R})\) from \(G\) to find a normal form for \(G\), say \(H\). \(H\) is the single root-node graph labelled by \(\Square\) iff \([G^{\ominus}] \in \pmb{L}(\pmb{TREE})\). Moreover, for input graphs of bounded degree, we terminate in linear time.
\end{theorem}

\begin{proof}
By Lemma \ref{lem:treederlen}, our system is terminating and derivations have maximum length \(2 \# G\). By Corollary \ref{corollary:treepreserving} and Lemma \ref{lemma:treereduce}, \(G\) is a tree iff we can derive the singleton tree without backtracking. Finally, by Corollary \ref{corollary:lineargt}, the algorithm terminates in linear time, since our ruleset satisfies the necessary conditions.
\end{proof}

We will return to this example in Section \ref{sec:treerecrev}, where we see that the system is actually not confluent! We will show, however, that it is confluent \enquote{up to garbage} on a class containing the \enquote{input trees}.


\section{Recognising Full Binary Trees}

In the previous section, we gave a grammar defining the language of all unlabelled trees, and showed that it was possible to adapt it to efficiently recognise the language by using a \enquote{root} node. In this section, we will show that this technique can be adapted to recognise the language of all unlabelled full binary trees (Definition \ref{dfn:tree}) in linear time.

\vspace{-0.0em}
\begin{figure}[H]
\noindent
\centering
\input{fig/3/btree}
\vspace{-1.0em}
\caption{Full Binary Tree Rules}
\label{fig:btree}
\vspace{0.0em}
\end{figure}
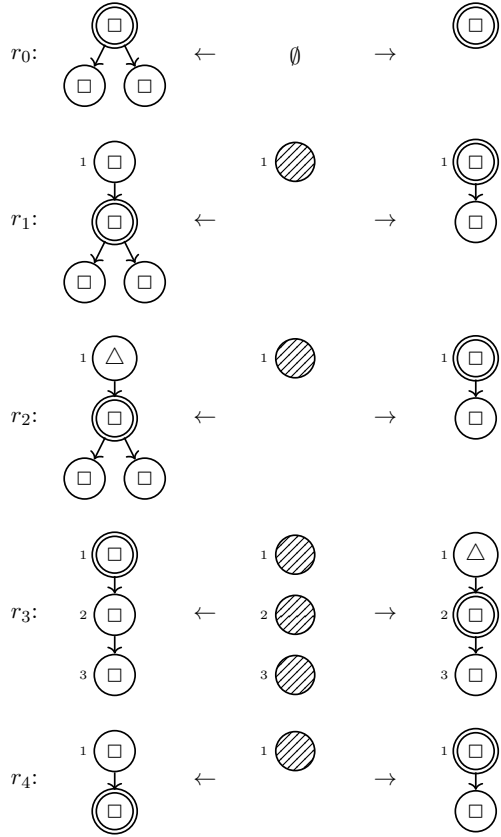

Let \(\mathcal{L} = (\{\Square, \triangle\}, \{\Square\})\), and \(\mathcal{R} = \{r_0, r_1, r_2, r_3, r_4\}\). We are going to show that \(\mathcal{R}\) induces a linear time algorithm for testing if a graph is a full binary tree (FBT). Intuitively, this works in a similar way to the reduction for trees, with the major difference that we only \enquote{prune} (\(r_0, r_1, r_2\)) whenever the root is in a position where it has exactly two children who are themselves leaf nodes. Rules \(r_3, r_4\) ensure that we can always prune, so that we only get stuck in the case that our input was not a full binary tree. Similarly to the previous system for trees, the triangle labels are necessary so that, in the case that the input graph is not a tree, we could \enquote{get stuck} in a directed cycle.

For the rest of this section, by \enquote{input graph}, we mean any TLRG containing exactly one \enquote{root} node, with edges and vertices all labelled \(\Square\). By \enquote{input FBT}, we mean an \enquote{input graph} that is also a FBT. The proof of complexity and correctness is similar to the previous system, so we won't give all the detail.

\begin{theorem}[FBT Recognition]
Given an input graph \(G\), one may use the system \((\mathcal{L}, \mathcal{R})\) from \(G\) to find a normal form for \(G\), say \(H\). \(H\) is the single root-node graph labelled by \(\Square\) iff \(G^{\ominus}\) is an unlabelled FBT. Moreover, for input graphs of bounded degree, we terminate in linear time.
\end{theorem}

\begin{proof}
The linear derivation length comes from the fact that if \(r_0\) appears in a derivation sequence from an input graph, then it must be the last derivation, and it must derive a normal form. Moreover, if \(r_5\) appears in a derivation sequence, then it must appear at the only at the very start of a derivation sequence from an input graph. The remaining \(r_1, r_2, r_3\) must appear in the middle of an derivation sequence from an input graph, at most a linear number of times. All of the rules in our GT system are of the form such that Corollary \ref{corollary:lineargt} can be applied, so we can conclude linear time complexity.

For correctness, one must show the analogous results to Corollary \ref{corollary:treepreserving} and Lemma \ref{lemma:treereduce}. That is, that FBTs derive only FBTs, and non-FBTs derive only non-FBTs, and that progress is always made, given an input FBT, until we reach the singleton FBT. The proof detail is not dissimilar from the existing proofs for trees, once one has taken care of the existing observations we have made about the location of rules \(r_0\) and \(r_5\) in derivation sequences starting from an input graph.
\end{proof}


\section{Recognising Trees in GP\,2}

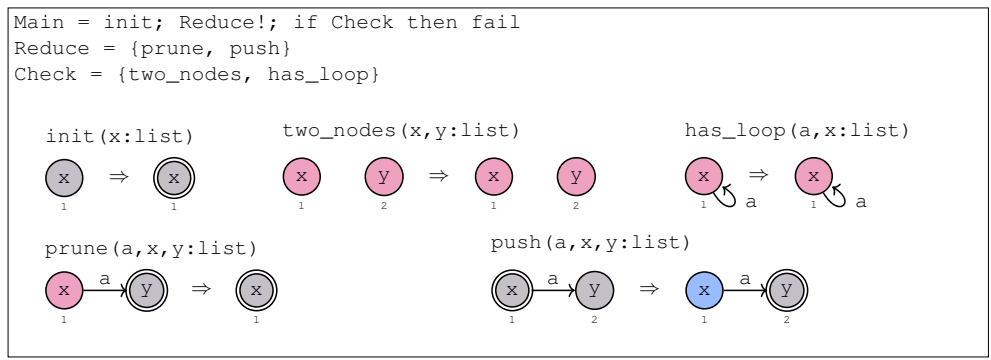
\begin{figure}[H]
\centering
\noindent
\scalebox{.73}{\input{fig/3/gp2}}
\caption{GP\,2 Implementation}
\label{fig:gp2-impl}
\end{figure}

Our algorithm from Section \ref{sec:linearrec} can be implemented in GP\,2. The program (Figure \ref{fig:gp2-impl}) expects an arbitrary labelled input graph with every node coloured grey, no \enquote{root} nodes, and no additional \enquote{marks}. It will fail iff the input is not a tree. Given an input graph of bounded degree, it will always terminate in linear time with respect to (w.r.t.) the number of nodes in the input graph.

\vspace{0.2em}

To see that the program is correct follows mostly from our existing proofs. Grey nodes encode the \(\Square\) label, and blue nodes, \(\triangle\). The \enquote{init} rule will fail if the input graph is empty, otherwise, it will make exactly one node rooted, in at most linear time. The \enquote{Reduce!} step is then exactly our previous GT system, which we have shown to be correct, and terminates in linear time. Finally, the \enquote{Check} step checks for garbage in linear time. There is no need to check the host graph is not equal to the empty graph (Corollary \ref{cor:oneroot}).

\begin{figure}[H]
\begin{subfigure}{.24\textwidth}
    \centering
    \input{fig/3/star}
    \vspace{0.2em}
    \caption{Star Graph}
\end{subfigure}
\begin{subfigure}{.25\textwidth}
    \centering
    \input{fig/3/grid}
    \vspace{0.2em}
    \caption{Grid Graph}
\end{subfigure}
\begin{subfigure}{.25\textwidth}
    \centering
    \input{fig/3/tree}
    \vspace{0.2em}
    \caption{Binary Tree}
\end{subfigure}
\begin{subfigure}{.24\textwidth}
    \centering
    \input{fig/3/list}
    \vspace{0.2em}
    \caption{Linked List}
\end{subfigure}
\vspace{0.8em}
\caption{Graph Classes}
\label{fig:graph-types}
\end{figure}
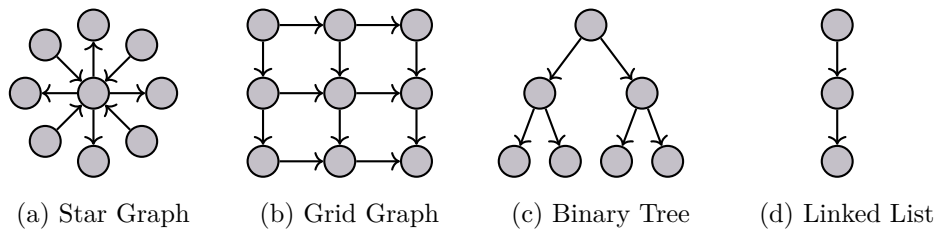

We have performed empirical benchmarking to verify the complexity of the program, testing it with linked lists, binary trees, grid graphs, and star graphs (Figure \ref{fig:graph-types}). Formal definitions of each of these graph classes can be found in Section \ref{sec:graphclasses}. We have exclusively used \enquote{perfect} binary trees, and \enquote{square} grid graphs in our testing.

\vspace{-0.5em}
\noindent

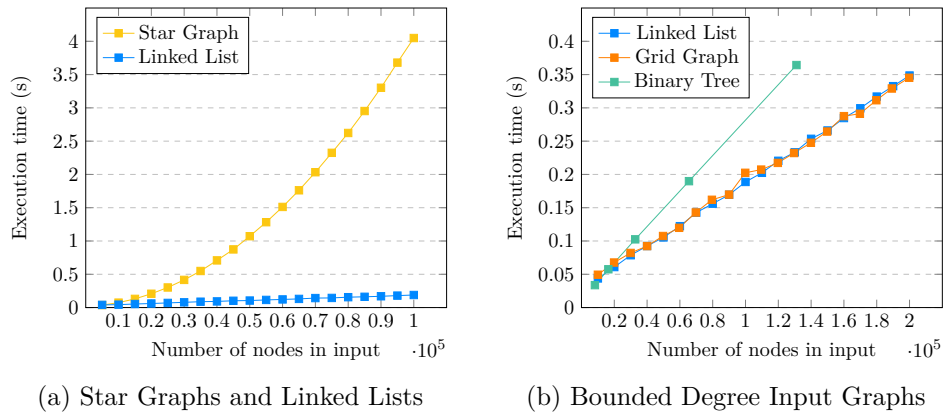
\begin{figure}[H]
\begin{subfigure}{.5\textwidth}
    \centering
    \input{fig/3/perf1}
    \caption{Star Graphs and Linked Lists}
\end{subfigure}
\begin{subfigure}{.5\textwidth}
    \centering
    \input{fig/3/perf2}
    \caption{Bounded Degree Input Graphs}
\end{subfigure}
\vspace{0.8em}
\caption{Measured Performance}
\label{fig:tree-bench}
\end{figure}

Star Graphs are not of bounded degree, so we saw quadratic time complexity as expected. The other graphs are of bounded degree, thus we observed linear time complexity (Figure \ref{fig:tree-bench}).

\vspace{0.4em}

We thus have empirical evidence, that in some sense, our complexity assumptions must be realistic, and that our proofs are not contradicted by the results. It remains future work to formalise complexity, in general, for GP\,2.

%% file: fig/3/complexity.tex
\begin{center}
\begin{tabular}{l|l|l}
Input       & Output                                                                                       & Time           \\ \hline
label \(l\) & The set \(Z\) of rooted nodes with label \(l\).                                              & \(O(\abs{Z})\) \\
label \(l\) & The set \(Z\) of unrooted nodes with label \(l\).                                            & \(O(\abs{V})\) \\
node \(v\)  & Values \(\operatorname{deg}(v)\), \(\operatorname{indeg}(v)\), \(\operatorname{outdeg}(v)\). & \(O(1)\)       \\
node \(v\)  & Set \(X\) of edges with source \(v\).                                                        & \(O(\abs{X})\) \\
node \(v\)  & Set \(X\) of edges with target \(v\).                                                        & \(O(\abs{X})\) \\
graph \(G\) & \(\abs{V_G}\) and \(\abs{E_G}\).                                                             & \(O(1)\)       \\
\end{tabular}
\end{center}

%% file: fig/3/tree1.tex
\begin{tikzpicture}[every node/.style={inner sep=0pt, text width=6.5mm, align=center}]
    \node (a) at (0.0,0) {$r$:};

    \node (b) at (1.0,0) [draw, circle, thick] {\(\Square\)};

    \node (c) at (2.0,0) {$\leftarrow$};

    \node (d) at (3.0,0) [draw, circle, thick] {\(\Square\)};

    \node (e) at (4.0,0) {$\rightarrow$};

    \node (f) at (5.0,0) [draw, circle, thick] {\(\Square\)};
    \node (g) at (6.5,0) [draw, circle, thick] {\(\Square\)};

    \node (B) at (1.0,-.52) {\tiny{1}};
    \node (D) at (3.0,-.52) {\tiny{1}};
    \node (F) at (5.0,-.52) {\tiny{1}};

    \draw (f) edge[->,thick] node[above, yshift=2.5pt] {\(\Square\)} (g);
\end{tikzpicture}

%% file: fig/3/tree2.tex
\scalebox{.8}{\begin{tikzpicture}[every node/.style={inner sep=0pt, text width=6.5mm, align=center}]
    \node (a) at (0.0,0) {$r_0$:};

    \node (b) at (1.0,0) [draw, circle, thick] {\(\Square\)};
    \node (c) at (2.5,0) [draw, circle, thick, double, double distance=0.4mm] {\(\Square\)};

    \node (d) at (3.5,0) {$\leftarrow$};

    \node (e) at (4.5,0) [draw, circle, thick, pattern=north east lines] {\,};

    \node (f) at (5.5,0) {$\rightarrow$};

    \node (g) at (6.5,0) [draw, circle, thick, double, double distance=0.4mm] {\(\Square\)};

    \node (B) at (1.0,-.52) {\tiny{1}};
    \node (E) at (4.5,-.52) {\tiny{1}};
    \node (G) at (6.5,-.52) {\tiny{1}};

    \draw (b) edge[->,thick] node[above, yshift=2.5pt] {\(\Square\)} (c);
\end{tikzpicture}}

\vspace{0.4em}

\scalebox{.8}{\begin{tikzpicture}[every node/.style={inner sep=0pt, text width=6.5mm, align=center}]
    \node (a) at (0.0,0) {$r_1$:};

    \node (b) at (1.0,0) [draw, circle, thick] {\(\triangle\)};
    \node (c) at (2.5,0) [draw, circle, thick, double, double distance=0.4mm] {\(\Square\)};

    \node (d) at (3.5,0) {$\leftarrow$};

    \node (e) at (4.5,0) [draw, circle, thick, pattern=north east lines] {\,};

    \node (f) at (5.5,0) {$\rightarrow$};

    \node (g) at (6.5,0) [draw, circle, thick, double, double distance=0.4mm] {\(\Square\)};

    \node (B) at (1.0,-.52) {\tiny{1}};
    \node (E) at (4.5,-.52) {\tiny{1}};
    \node (G) at (6.5,-.52) {\tiny{1}};

    \draw (b) edge[->,thick] node[above, yshift=2.5pt] {\(\Square\)} (c);
\end{tikzpicture}}

\vspace{0.4em}

\scalebox{.8}{\begin{tikzpicture}[every node/.style={inner sep=0pt, text width=6.5mm, align=center}]
    \node (a) at (0.0,0) {$r_2$:};

    \node (b) at (1.0,0) [draw, circle, thick, double, double distance=0.4mm] {\(\Square\)};
    \node (c) at (2.5,0) [draw, circle, thick] {\(\Square\)};

    \node (d) at (3.5,0) {$\leftarrow$};

    \node (e) at (4.5,0) [draw, circle, thick, pattern=north east lines] {\,};
    \node (f) at (6.0,0) [draw, circle, thick, pattern=north east lines] {\,};

    \node (g) at (7.0,0) {$\rightarrow$};

    \node (h) at (8.0,0) [draw, circle, thick] {\(\triangle\)};
    \node (i) at (9.5,0) [draw, circle, thick, double, double distance=0.4mm] {\(\Square\)};

    \node (B) at (1.0,-.52) {\tiny{1}};
    \node (C) at (2.5,-.52) {\tiny{2}};
    \node (E) at (4.5,-.52) {\tiny{1}};
    \node (F) at (6.0,-.52) {\tiny{2}};
    \node (H) at (8.0,-.52) {\tiny{1}};
    \node (I) at (9.5,-.52) {\tiny{2}};

    \draw (b) edge[->,thick] node[above, yshift=2.5pt] {\(\Square\)} (c)
          (h) edge[->,thick] node[above, yshift=2.5pt] {\(\Square\)} (i);
\end{tikzpicture}}

%% file: fig/3/reduction.tex
\scalebox{.75}{\begin{tikzpicture}[every node/.style={inner sep=0pt, text width=6.5mm, align=center}]

\node (a) at (0,0)     [draw,circle,thick] {\(\Square\)};
\node (b) at (-0.5,-1) [draw,circle,thick, double, double distance=0.4mm] {\(\Square\)};
\node (c) at (0.5,-1)  [draw,circle,thick] {\(\Square\)};
\node (d) at (-0.5,-2) [draw,circle,thick] {\(\Square\)};
\node (e) at (0.5,-2)  [draw,circle,thick] {\(\Square\)};

\draw (a) edge[->,thick] (b)
      (a) edge[->,thick] (c)
      (b) edge[->,thick] (d)
      (c) edge[->,thick] (e);

\node (t) at (1.35,-1) {$\Rightarrow_{r_2}$};

\node (a) at (2.7,0)  [draw,circle,thick] {\(\Square\)};
\node (b) at (2.2,-1) [draw,circle,thick] {\(\triangle\)};
\node (c) at (3.2,-1) [draw,circle,thick] {\(\Square\)};
\node (d) at (2.2,-2) [draw,circle,thick, double, double distance=0.4mm] {\(\Square\)};
\node (e) at (3.2,-2) [draw,circle,thick] {\(\Square\)};

\draw (a) edge[->,thick] (b)
      (a) edge[->,thick] (c)
      (b) edge[->,thick] (d)
      (c) edge[->,thick] (e);

\node (t) at (4.05,-1) {$\Rightarrow_{r_1}$};

\node (a) at (5.4,0)  [draw,circle,thick] {\(\Square\)};
\node (b) at (4.9,-1) [draw,circle,thick, double, double distance=0.4mm] {\(\Square\)};
\node (c) at (5.9,-1) [draw,circle,thick] {\(\Square\)};
\node (e) at (5.9,-2) [draw,circle,thick] {\(\Square\)};

\draw (a) edge[->,thick] (b)
      (a) edge[->,thick] (c)
      (c) edge[->,thick] (e);

\node (t) at (6.75,-1) {$\Rightarrow_{r_0}$};

\node (a) at (7.6,0)    [draw,circle,thick, double, double distance=0.4mm] {\(\Square\)};
\node (c) at (7.6,-1) [draw,circle,thick] {\(\Square\)};
\node (e) at (7.6,-2) [draw,circle,thick] {\(\Square\)};

\draw (a) edge[->,thick] (c)
      (c) edge[->,thick] (e);

\node (t) at (8.45,-1) {$\Rightarrow_{r_2}$};

\node (a) at (9.3,0)  [draw,circle,thick] {\(\triangle\)};
\node (c) at (9.3,-1) [draw,circle,thick, double, double distance=0.4mm] {\(\Square\)};
\node (e) at (9.3,-2) [draw,circle,thick] {\(\Square\)};

\draw (a) edge[->,thick] (c)
      (c) edge[->,thick] (e);

\node (t) at (10.15,-1) {$\Rightarrow_{r_2}$};

\node (a) at (11.0,0)  [draw,circle,thick] {\(\triangle\)};
\node (c) at (11.0,-1) [draw,circle,thick] {\(\triangle\)};
\node (e) at (11.0,-2) [draw,circle,thick, double, double distance=0.4mm] {\(\Square\)};

\draw (a) edge[->,thick] (c)
      (c) edge[->,thick] (e);

\node (t) at (11.85,-1) {$\Rightarrow_{r_1}$};

\node (a) at (12.7,0)  [draw,circle,thick] {\(\triangle\)};
\node (c) at (12.7,-1) [draw,circle,thick, double, double distance=0.4mm] {\(\Square\)};

\draw (a) edge[->,thick] (c);

\node (t) at (13.55,-1) {$\Rightarrow_{r_1}$};

\node (a) at (14.4,0)  [draw,circle,thick, double, double distance=0.4mm] {\(\Square\)};

\end{tikzpicture}}

%% file: fig/3/non-reduction1.tex
\scalebox{.75}{\begin{tikzpicture}[every node/.style={inner sep=0pt, text width=6.5mm, align=center}]

\node (a) at (0,0)     [draw,circle,thick] {\(\Square\)};
\node (b) at (-0.5,-1) [draw,circle,thick, double, double distance=0.4mm] {\(\Square\)};
\node (c) at (0.5,-1)  [draw,circle,thick] {\(\Square\)};

\draw (a) edge[->,thick] (b)
      (b) edge[->,thick] (c)
      (c) edge[->,thick] (a);

\node (t) at (1.35,-0.5) {$\Rightarrow_{r_2}$};

\node (a) at (2.7,0)  [draw,circle,thick] {\(\Square\)};
\node (b) at (2.2,-1) [draw,circle,thick] {\(\triangle\)};
\node (c) at (3.2,-1) [draw,circle,thick, double, double distance=0.4mm] {\(\Square\)};

\draw (a) edge[->,thick] (b)
      (b) edge[->,thick] (c)
      (c) edge[->,thick] (a);

\node (t) at (4.05,-0.5) {$\Rightarrow_{r_2}$};

\node (a) at (5.4,0)  [draw,circle,thick, double, double distance=0.4mm] {\(\Square\)};
\node (b) at (4.9,-1) [draw,circle,thick] {\(\triangle\)};
\node (c) at (5.9,-1) [draw,circle,thick] {\(\triangle\)};

\draw (a) edge[->,thick] (b)
      (b) edge[->,thick] (c)
      (c) edge[->,thick] (a);

\end{tikzpicture}}

%% file: fig/3/non-reduction2.tex
\scalebox{.75}{\begin{tikzpicture}[every node/.style={inner sep=0pt, text width=6.5mm, align=center}]

\node (a) at (-0.5,0)  [draw,circle,thick, double, double distance=0.4mm] {\(\Square\)};
\node (b) at (0.5,0)   [draw,circle,thick] {\(\Square\)};
\node (c) at (-0.5,-1) [draw,circle,thick] {\(\Square\)};
\node (d) at (0.5,-1)  [draw,circle,thick] {\(\Square\)};

\draw (a) edge[->,thick] (c)
      (b) edge[->,thick] (d);

\node (t) at (1.35,-0.5) {$\Rightarrow_{r_2}$};

\node (a) at (2.2,0)  [draw,circle,thick] {\(\triangle\)};
\node (b) at (3.2,0)  [draw,circle,thick] {\(\Square\)};
\node (c) at (2.2,-1) [draw,circle,thick, double, double distance=0.4mm] {\(\Square\)};
\node (d) at (3.2,-1) [draw,circle,thick] {\(\Square\)};

\draw (a) edge[->,thick] (c)
      (b) edge[->,thick] (d);

\node (t) at (4.05,-0.5) {$\Rightarrow_{r_1}$};

\node (a) at (4.9,0)  [draw,circle,thick, double, double distance=0.4mm] {\(\Square\)};
\node (b) at (5.9,0)  [draw,circle,thick] {\(\Square\)};
\node (d) at (5.9,-1) [draw,circle,thick] {\(\Square\)};

\draw (b) edge[->,thick] (d);

\end{tikzpicture}}

%% file: fig/3/btree.tex
\scalebox{.8}{\begin{tikzpicture}[every node/.style={inner sep=0pt, text width=6.5mm, align=center}]
    \node (a) at (0.0,0) {$r_0$:};

    \node (b) at (1.5,0.5)  [draw, circle, thick, double, double distance=0.4mm] {\(\Square\)};
    \node (c) at (1.0,-0.5) [draw, circle, thick] {\(\Square\)};
    \node (d) at (2.0,-0.5) [draw, circle, thick] {\(\Square\)};

    \node (e) at (3.0,0) {$\leftarrow$};

    \node (f) at (4.5,0) {$\emptyset$};

    \node (g) at (6.0,0) {$\rightarrow$};

    \node (h) at (7.5,0.5) [draw, circle, thick, double, double distance=0.4mm] {\(\Square\)};

    \draw (b) edge[->,thick] node[above, yshift=2.5pt] {\,} (c);
    \draw (b) edge[->,thick] node[above, yshift=2.5pt] {\,} (d);
\end{tikzpicture}}

\vspace{1.0em}

\scalebox{.8}{\begin{tikzpicture}[every node/.style={inner sep=0pt, text width=6.5mm, align=center}]
    \node (a) at (0.0,0) {$r_1$:};

    \node (b) at (1.5,1.0)  [draw, circle, thick] {\(\Square\)};
    \node (c) at (1.5,0.0)  [draw, circle, thick, double, double distance=0.4mm] {\(\Square\)};
    \node (d) at (1.0,-1.0) [draw, circle, thick] {\(\Square\)};
    \node (e) at (2.0,-1.0) [draw, circle, thick] {\(\Square\)};

    \node (f) at (3.0,0) {$\leftarrow$};

    \node (g) at (4.5,1.0) [draw, circle, thick, pattern=north east lines] {\,};

    \node (h) at (6.0,0) {$\rightarrow$};

    \node (i) at (7.5,1.0) [draw, circle, thick, double, double distance=0.4mm] {\(\Square\)};
    \node (j) at (7.5,0.0) [draw, circle, thick] {\(\Square\)};

    \node (B) at (0.98,1.0) {\tiny{1}};
    \node (G) at (3.98,1.0) {\tiny{1}};
    \node (I) at (6.98,1.0) {\tiny{1}};

    \draw (b) edge[->,thick] node[above, yshift=2.5pt] {\,} (c);
    \draw (c) edge[->,thick] node[above, yshift=2.5pt] {\,} (d);
    \draw (c) edge[->,thick] node[above, yshift=2.5pt] {\,} (e);
    \draw (i) edge[->,thick] node[above, yshift=2.5pt] {\,} (j);
\end{tikzpicture}}

\vspace{1.0em}

\scalebox{.8}{\begin{tikzpicture}[every node/.style={inner sep=0pt, text width=6.5mm, align=center}]
    \node (a) at (0.0,0) {$r_2$:};

    \node (b) at (1.5,1.0)  [draw, circle, thick] {\(\triangle\)};
    \node (c) at (1.5,0.0)  [draw, circle, thick, double, double distance=0.4mm] {\(\Square\)};
    \node (d) at (1.0,-1.0) [draw, circle, thick] {\(\Square\)};
    \node (e) at (2.0,-1.0) [draw, circle, thick] {\(\Square\)};

    \node (f) at (3.0,0) {$\leftarrow$};

    \node (g) at (4.5,1.0) [draw, circle, thick, pattern=north east lines] {\,};

    \node (h) at (6.0,0) {$\rightarrow$};

    \node (i) at (7.5,1.0) [draw, circle, thick, double, double distance=0.4mm] {\(\Square\)};
    \node (j) at (7.5,0.0) [draw, circle, thick] {\(\Square\)};

    \node (B) at (0.98,1.0) {\tiny{1}};
    \node (G) at (3.98,1.0) {\tiny{1}};
    \node (I) at (6.98,1.0) {\tiny{1}};

    \draw (b) edge[->,thick] node[above, yshift=2.5pt] {\,} (c);
    \draw (c) edge[->,thick] node[above, yshift=2.5pt] {\,} (d);
    \draw (c) edge[->,thick] node[above, yshift=2.5pt] {\,} (e);
    \draw (i) edge[->,thick] node[above, yshift=2.5pt] {\,} (j);
\end{tikzpicture}}

\vspace{1.0em}

\scalebox{.8}{\begin{tikzpicture}[every node/.style={inner sep=0pt, text width=6.5mm, align=center}]
    \node (a) at (0.0,0) {$r_3$:};

    \node (b) at (1.5,1.0)  [draw, circle, thick, double, double distance=0.4mm] {\(\Square\)};
    \node (c) at (1.5,0.0)  [draw, circle, thick] {\(\Square\)};
    \node (d) at (1.5,-1.0) [draw, circle, thick] {\(\Square\)};

    \node (e) at (3.0,0) {$\leftarrow$};

    \node (f) at (4.5,1.0) [draw, circle, thick, pattern=north east lines] {\,};
    \node (g) at (4.5,0.0) [draw, circle, thick, pattern=north east lines] {\,};
    \node (h) at (4.5,-1.0) [draw, circle, thick, pattern=north east lines] {\,};

    \node (i) at (6.0,0) {$\rightarrow$};

    \node (j) at (7.5,1.0)  [draw, circle, thick] {\(\triangle\)};
    \node (k) at (7.5,0.0)  [draw, circle, thick, double, double distance=0.4mm] {\(\Square\)};
    \node (l) at (7.5,-1.0) [draw, circle, thick] {\(\Square\)};

    \node (B) at (0.98,1.0)  {\tiny{1}};
    \node (C) at (0.98,0.0)  {\tiny{2}};
    \node (D) at (0.98,-1.0) {\tiny{3}};

    \node (F) at (3.98,1.0)  {\tiny{1}};
    \node (G) at (3.98,0.0)  {\tiny{2}};
    \node (H) at (3.98,-1.0) {\tiny{3}};

    \node (J) at (6.98,1.0)  {\tiny{1}};
    \node (K) at (6.98,0.0)  {\tiny{2}};
    \node (L) at (6.98,-1.0) {\tiny{3}};

    \draw (b) edge[->,thick] node[above, yshift=2.5pt] {\,} (c);
    \draw (c) edge[->,thick] node[above, yshift=2.5pt] {\,} (d);
    \draw (j) edge[->,thick] node[above, yshift=2.5pt] {\,} (k);
    \draw (k) edge[->,thick] node[above, yshift=2.5pt] {\,} (l);
\end{tikzpicture}}

\vspace{1.0em}

\scalebox{.8}{\begin{tikzpicture}[every node/.style={inner sep=0pt, text width=6.5mm, align=center}]
    \node (a) at (0.0,0) {$r_4$:};

    \node (b) at (1.5,0.5)  [draw, circle, thick] {\(\Square\)};
    \node (c) at (1.5,-0.5) [draw, circle, thick, double, double distance=0.4mm] {\(\Square\)};

    \node (d) at (3.0,0) {$\leftarrow$};

    \node (e) at (4.5,0.5) [draw, circle, thick, pattern=north east lines] {\,};

    \node (f) at (6.0,0) {$\rightarrow$};

    \node (g) at (7.5,0.5)  [draw, circle, thick, double, double distance=0.4mm] {\(\Square\)};
    \node (h) at (7.5,-0.5) [draw, circle, thick] {\(\Square\)};

    \node (B) at (0.98,0.5) {\tiny{1}};
    \node (G) at (3.98,0.5) {\tiny{1}};
    \node (I) at (6.98,0.5) {\tiny{1}};

    \draw (b) edge[->,thick] node[above, yshift=2.5pt] {\,} (c);
    \draw (g) edge[->,thick] node[above, yshift=2.5pt] {\,} (h);
\end{tikzpicture}}

\vspace{1.0em}

%% file: fig/3/gp2.tex
\fbox{\begin{minipage}{17.5cm}
\begin{allintypewriter}
Main = init; Reduce!; if Check then fail

Reduce = \{prune, push\}

Check = \{two\_nodes, has\_loop\}

\medskip
\setlength{\tabcolsep}{16pt}
\vspace{2.5mm}
\begin{tabular}{  p{3.2cm}  p{6.2cm}  p{4.6cm}  }
    
    \vspace{-1mm} init(x:list) & \vspace{-2mm} two\_nodes(x,y:list) & \vspace{-2mm} has\_loop(a,x:list) \\

    \vspace{-2mm}
    \adjustbox{valign=t}{\begin{tikzpicture}[every node/.style={inner sep=0pt, text width=6.5mm, align=center}]
        \node (a) at (0.0,0) [draw,circle,thick,fill=gp2grey] {x};

        \node (b) at (1.0,0) {$\Rightarrow$};
        
        \node (c) at (2.0,0) [draw,circle,thick,fill=gp2grey,double,double distance=0.4mm] {x};
        
        \node (A) at (0.0,-.52) {\tiny{1}};
        \node (C) at (2.0,-.52) {\tiny{1}};
        
    \end{tikzpicture}}
    &
    
    \vspace{-2mm}
    \adjustbox{valign=t}{\begin{tikzpicture}[every node/.style={inner sep=0pt, text width=6.5mm, align=center}]
        \node (a) at (0.0,0) [draw, circle, fill=gp2pink, thick] {x};
        \node (b) at (1.5,0) [draw, circle, fill=gp2pink, thick] {y};

        \node (c) at (2.5,0) {$\Rightarrow$};

        \node (d) at (3.5,0) [draw, circle, fill=gp2pink, thick] {x};
        \node (e) at (5.0,0) [draw, circle, fill=gp2pink, thick] {y};

        \node (A) at (0.0,-.52) {\tiny{1}};
        \node (B) at (1.5,-.52) {\tiny{2}};
        \node (D) at (3.5,-.52) {\tiny{1}};
        \node (E) at (5.0,-.52) {\tiny{2}};
    \end{tikzpicture}}
    &
    
    \vspace{-2mm}
    \adjustbox{valign=t}{\begin{tikzpicture}[every node/.style={inner sep=0pt, text width=6.5mm, align=center}]
        \node (a) at (0.0,0) [draw,circle,thick,fill=gp2pink] {x};
        
        \node (b) at (1.0,0) {$\Rightarrow$};
        
        \node (c) at (2.0,0) [draw,circle,thick,fill=gp2pink] {x};
        
        \node (A) at (0.0,-.52) {\tiny{1}};
        \node (C) at (2.0,-.52) {\tiny{1}};
        
        \draw (a) edge[->,in=-30,out=-60,loop,thick] node[right, yshift=1.5pt] {a} (a)
              (c) edge[->,in=-30,out=-60,loop,thick] node[right, yshift=1.5pt] {a} (c);
    \end{tikzpicture}}
    \\
\end{tabular}
\begin{tabular}{  p{7cm}  p{7cm}  }

    \vspace{-1mm} prune(a,x,y:list) & \vspace{-2mm} push(a,x,y:list) \\
    
    \vspace{-2mm}
    \adjustbox{valign=t}{\begin{tikzpicture}[every node/.style={inner sep=0pt, text width=6.5mm, align=center}]
        \node (a) at (0.0,0) [draw,circle,fill=gp2pink,thick] {x};
        \node (b) at (1.5,0) [draw,circle,fill=gp2grey,thick, double, double distance=0.4mm] {y};
        
        \node (c) at (2.5,0) {$\Rightarrow$};
        
        \node (d) at (3.5,0) [draw,circle,fill=gp2grey,thick, double, double distance=0.4mm] {x};
        
        \node (A) at (0.0,-.52) {\tiny{1}};
        \node (D) at (3.5,-.52) {\tiny{1}};
        
        \draw (a) edge[->,thick] node[above, yshift=2.5pt] {a} (b);
    \end{tikzpicture}}
    
    & 

    \vspace{-2mm}
    \adjustbox{valign=t}{\begin{tikzpicture}[every node/.style={inner sep=0pt, text width=6.5mm, align=center}]
    \node (a) at (0.0,0)     [draw, circle, fill=gp2grey, thick, double, double distance=0.4mm] {x};
    \node (b) at (1.5,0)     [draw, circle, fill=gp2grey, thick] {y};
    
    \node (c) at (2.5,0)     {$\Rightarrow$};
    
    \node (d) at (3.5,0)     [draw, circle, fill=gp2blue, thick] {x};
    \node (e) at (5,0)       [draw, circle, fill=gp2grey, thick, double, double distance=0.4mm] {y};
    
    \node (A) at (0.0,-.52) {\tiny{1}};
    \node (B) at (1.5,-.52) {\tiny{2}};
    \node (D) at (3.5,-.52) {\tiny{1}};
    \node (E) at (5.0,-.52) {\tiny{2}};
    
    \draw (a) edge[->,thick] node[above, yshift=2.5pt] {a} (b)
          (d) edge[->,thick] node[above, yshift=2.5pt] {a} (e);
    \end{tikzpicture}}
    \\
\end{tabular}
\end{allintypewriter}
\end{minipage}
}

%% file: fig/3/star.tex
\begin{tikzpicture}[scale=0.68]
    \node (a) at (0.000,0.000)   [draw,circle,thick,fill=gp2grey] {\,};
    \node (b) at (0.000,1.333)   [draw,circle,thick,fill=gp2grey] {\,};
    \node (c) at (0.943,0.943)   [draw,circle,thick,fill=gp2grey] {\,};
    \node (d) at (1.333,0.000)   [draw,circle,thick,fill=gp2grey] {\,};
    \node (e) at (0.943,-0.943)  [draw,circle,thick,fill=gp2grey] {\,};
    \node (f) at (0.000,-1.333)  [draw,circle,thick,fill=gp2grey] {\,};
    \node (g) at (-0.943,-0.943) [draw,circle,thick,fill=gp2grey] {\,};
    \node (h) at (-1.333,0.000)  [draw,circle,thick,fill=gp2grey] {\,};
    \node (i) at (-0.943,0.943)  [draw,circle,thick,fill=gp2grey] {\,};
    
    \draw (a) edge[->, thick] (b)
          (c) edge[->, thick] (a)
          (a) edge[->, thick] (d)
          (e) edge[->, thick] (a)
          (a) edge[->, thick] (f)
          (g) edge[->, thick] (a)
          (a) edge[->, thick] (h)
          (i) edge[->, thick] (a);
\end{tikzpicture}

%% file: fig/3/grid.tex
\begin{tikzpicture}[scale=0.68]
    \node (a) at (-1.500,1.333)  [draw,circle,thick,fill=gp2grey] {\,};
    \node (b) at (0.000,1.333)   [draw,circle,thick,fill=gp2grey] {\,};
    \node (c) at (1.500,1.333)   [draw,circle,thick,fill=gp2grey] {\,};
    \node (d) at (-1.500,0.000)  [draw,circle,thick,fill=gp2grey] {\,};
    \node (e) at (0.000,0.000)   [draw,circle,thick,fill=gp2grey] {\,};
    \node (f) at (1.500,0.000)   [draw,circle,thick,fill=gp2grey] {\,};
    \node (g) at (-1.500,-1.333) [draw,circle,thick,fill=gp2grey] {\,};
    \node (h) at (0.000,-1.333)  [draw,circle,thick,fill=gp2grey] {\,};
    \node (i) at (1.500,-1.333)  [draw,circle,thick,fill=gp2grey] {\,};
    
    \draw (a) edge[->, thick] (b)
          (a) edge[->, thick] (d)
          (b) edge[->, thick] (c)
          (b) edge[->, thick] (e)
          (c) edge[->, thick] (f)
          (d) edge[->, thick] (e)
          (d) edge[->, thick] (g)
          (e) edge[->, thick] (f)
          (e) edge[->, thick] (h)
          (f) edge[->, thick] (i)
          (g) edge[->, thick] (h)
          (h) edge[->, thick] (i);
\end{tikzpicture}

%% file: fig/3/tree.tex
\begin{tikzpicture}[scale=0.68]
    \node (a) at (0.000,1.333)   [draw,circle,thick,fill=gp2grey] {\,};
    \node (b) at (1.000,0.000)   [draw,circle,thick,fill=gp2grey] {\,};
    \node (c) at (-1.000,0.000)  [draw,circle,thick,fill=gp2grey] {\,};
    \node (d) at (1.500,-1.333)  [draw,circle,thick,fill=gp2grey] {\,};
    \node (e) at (0.500,-1.333)  [draw,circle,thick,fill=gp2grey] {\,};
    \node (f) at (-0.500,-1.333) [draw,circle,thick,fill=gp2grey] {\,};
    \node (g) at (-1.500,-1.333) [draw,circle,thick,fill=gp2grey] {\,};
    
    \draw (a) edge[->, thick] (b)
          (a) edge[->, thick] (c)
          (b) edge[->, thick] (d)
          (b) edge[->, thick] (e)
          (c) edge[->, thick] (f)
          (c) edge[->, thick] (g);
\end{tikzpicture}

%% file: fig/3/list.tex
\begin{tikzpicture}[scale=0.68]
    \node (a) at (0.000,1.333)  [draw,circle,thick,fill=gp2grey] {\,};
    \node (b) at (0.000,0.000)  [draw,circle,thick,fill=gp2grey] {\,};
    \node (c) at (0.000,-1.333) [draw,circle,thick,fill=gp2grey] {\,};
    
    \draw (a) edge[->, thick] (b)
          (b) edge[->, thick] (c);
\end{tikzpicture}

%% file: fig/3/perf1.tex
\begin{tikzpicture}[scale=0.7]
\begin{axis}[
xlabel={Number of nodes in input},
ylabel={Execution time (s)},
xmin=0, xmax=110000,
ymin=0, ymax=4.5,
xtick={10000,20000,30000,40000,50000,60000,70000,80000,90000,100000},
ytick={0.0,0.5,1.0,1.5,2.0,2.5,3.0,3.5,4.0},
legend pos=north west,
ymajorgrids=true,
grid style=dashed,
yticklabel style={/pgf/number format/fixed},
]
\addplot[color=performanceYellow, mark=square*] 
coordinates {
    (5000,0.03734)
    (10000,0.07405)
    (15000,0.13061)
    (20000,0.20765)
    (25000,0.30175)
    (30000,0.41571)
    (35000,0.54940)
    (40000,0.70865)
    (45000,0.87360)
    (50000,1.07186)
    (55000,1.28239)
    (60000,1.51184)
    (65000,1.76158)
    (70000,2.03321)
    (75000,2.32459)
    (80000,2.62288)
    (85000,2.95250)
    (90000,3.30216)
    (95000,3.67980)
    (100000,4.04914)
};
\addplot[color=performanceBlue, mark=square*] 
coordinates {
    (5000,0.04032)
    (10000,0.04375)
    (15000,0.05350)
    (20000,0.06085)
    (25000,0.06980)
    (30000,0.07863)
    (35000,0.08824)
    (40000,0.09247)
    (45000,0.10207)
    (50000,0.10537)
    (55000,0.11482)
    (60000,0.12228)
    (65000,0.13004)
    (70000,0.14266)
    (75000,0.14482)
    (80000,0.15604)
    (85000,0.16180)
    (90000,0.16944)
    (95000,0.17961)
    (100000,0.18871)
};
\addlegendentry{Star Graph}
\addlegendentry{Linked List}
\end{axis}
\end{tikzpicture}

%% file: fig/3/perf2.tex
\begin{tikzpicture}[scale=0.7]
\begin{axis}[
xlabel={Number of nodes in input},
ylabel={Execution time (s)},
xmin=0, xmax=220000,
ymin=0, ymax=0.45,
xtick={20000,40000,60000,80000,100000,120000,140000,160000,180000,200000},
ytick={0.00,0.05,0.10,0.15,0.20,0.25,0.30,0.35,0.40},
legend pos=north west,
ymajorgrids=true,
grid style=dashed,
yticklabel style={/pgf/number format/fixed},
]
\addplot[color=performanceBlue, mark=square*] 
coordinates {
    (10000,0.04375)
    (20000,0.06085)
    (30000,0.07863)
    (40000,0.09247)
    (50000,0.10537)
    (60000,0.12228)
    (70000,0.14266)
    (80000,0.15604)
    (90000,0.16944)
    (100000,0.18871)
    (110000,0.20241)
    (120000,0.22049)
    (130000,0.23322)
    (140000,0.25353)
    (150000,0.26601)
    (160000,0.28470)
    (170000,0.29935)
    (180000,0.31694)
    (190000,0.33262)
    (200000,0.34859)
};
\addplot[color=orange, mark=square*] 
coordinates {
    (10000,0.04936)
    (19881,0.06766)
    (29929,0.08206)
    (40000,0.09235)
    (49729,0.10750)
    (59536,0.11979)
    (69696,0.14284)
    (79524,0.16192)
    (90000,0.16971)
    (99856,0.20246)
    (109561,0.20725)
    (119716,0.21733)
    (129600,0.23215)
    (139876,0.24756)
    (149769,0.26442)
    (160000,0.28757)
    (169744,0.29114)
    (179776,0.31145)
    (189225,0.32908)
    (199809,0.34526)
};
\addplot[color=gp2green, mark=square*] 
coordinates {
    (8191,0.03370)
    (16383,0.05749)
    (32767,0.10249)
    (65535,0.18989)
    (131071,0.36431)
};
\addlegendentry{Linked List}
\addlegendentry{Grid Graph}
\addlegendentry{Binary Tree}
\end{axis}
\end{tikzpicture}

%% file: content/4.tex
\chapter{Confluence Analysis I} \label{chapter:confluenceanalysis1}

\begin{chapquote}{Donald Knuth, \textit{Notes on the van Emde Boas construction of priority deques: An instructive use of recursion (1977)}}
``Beware of bugs in the above code; I have only proved it correct, not tried it.''
\end{chapquote}

Efficient testing of language membership is an important problem in graph transformation \cite{Arnborg-Courcelle-Proskurowski-Seese93a,Bodlaender-Fluiter01a,Plump10a}. Our GT system for testing if a graph is a tree is actually not confluent, but if the input is a tree, then it has exactly one normal form, so it was in some sense confluent. We can formalise this with the new notion of \enquote{confluence up to garbage}.


\section{Definitions and Closedness} \label{sec:confmodgarb}

In this section, we shall be working with standard GT systems, as defined in Appendix \ref{appendix:transformation}, but without relabelling. That is, all graphs are totally labelled, including interface graphs. All the results in this section will actually generalise to systems with relabelling, or the systems defined in Chapter \ref{chapter:newtheory}.

\begin{definition}
Let \(T = (\mathcal{L}, \mathcal{R})\) be a GT system, and \(\pmb{D} \subseteq \mathcal{G}(\mathcal{L})\) be a set of abstract graphs. Then, a graph \(G\) is called \textbf{garbage} iff \([G] \not\in \pmb{D}\).
\end{definition}

\begin{definition}
Let \(T = (\mathcal{L}, \mathcal{R})\), and \(\pmb{D} \subseteq \mathcal{G}(\mathcal{L})\). \(\pmb{D}\) is \textbf{closed} under \(T\) iff for all \(G\), \(H\) such that \(G \Rightarrow_\mathcal{R} H\), if \([G] \in \pmb{D}\) then \([H] \in \pmb{D}\). \(\pmb{D}\) is \textbf{strongly closed} under \(T\) iff we have \([G] \in \pmb{D}\) iff \([H] \in \pmb{D}\).
\end{definition}

This set of abstract graphs \(\pmb{D}\) represents the \enquote{good input}, and the \enquote{garbage} is the graphs that are not in this set. \(\pmb{D}\) need not be explicitly generated by a graph grammar. For example, it could be defined by some (monadic second-order \cite{Courcelle89a}) logical formula.

There are a couple of immediately obvious results:

\begin{proposition} \label{lem:gsimpliesweak}
\textbf{Strong closedness} \(\Rightarrow\) \textbf{closedness}.
\end{proposition}

\begin{proposition} \label{lem:wgsclosure}
Given \(\pmb{D} \subseteq \mathcal{G}(\mathcal{L})\) \textbf{closed} under \(T = (\mathcal{L}, \mathcal{R})\), then for all graphs \(G\), \(H\) such that \(G \Rightarrow_\mathcal{R}^* H\), if \([G] \in \pmb{D}\), then \([H] \in \pmb{D}\).
\end{proposition}

\begin{figure}[H]
\centering
\noindent
\scalebox{.8}{\input{fig/4/eg1}}
\vspace{-1.3em}
\caption{Example Reduction Rules}
\label{fig:eg-reduct-rules}
\end{figure}
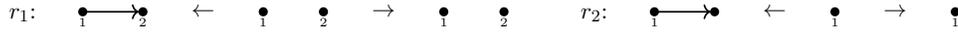

\begin{example}
Consider the reduction rules in Figure \ref{fig:eg-reduct-rules}. The language of acyclic graphs is \textbf{closed} under the GT system \(((\{\Square\}, \{\Square\}), \{r_1\})\), and the language of trees (forests) is \textbf{strongly closed} under \(((\{\Square\}, \{\Square\}), \{r_2\})\).
\end{example}

\vspace{-0.7em}

We can now define (\textbf{local}) \textbf{confluence up to garbage}, allowing us to say that, ignoring the garbage graphs, a system is (\textbf{locally}) \textbf{confluent}. Note that we previously called this property (\textbf{local}) \textbf{confluence modulo garbage}, however we have decided this name was confusing because we are not interested in confluence \enquote{modulo} an equivalence relation. Rather, we are just choosing to ignore certain start graphs.

\begin{definition}
Let \(T = (\mathcal{L}, \mathcal{R})\), \(\pmb{D} \subseteq \mathcal{G}(\mathcal{L})\). If for all graphs \(G\), \(H_1\), \(H_2\), such that \([G] \in \pmb{D}\), if \(H_1 \Leftarrow_{\mathcal{R}} G \Rightarrow_{\mathcal{R}} H_2\) implies that \(H_1\), \(H_2\) are \textbf{joinable}, then \(T\) is \textbf{locally confluent} on \(\pmb{D}\) (or \textbf{locally confluent up to garbage} w.r.t. \(\pmb{D}\)).
\end{definition}

\begin{definition}
Let \(T = (\mathcal{L}, \mathcal{R})\), \(\pmb{D} \subseteq \mathcal{G}(\mathcal{L})\). If for all graphs \(G\), \(H_1\), \(H_2\), such that \([G] \in \pmb{D}\), if \(H_1 \Leftarrow_{\mathcal{R}}^* G \Rightarrow_{\mathcal{R}}^* H_2\) implies that \(H_1\), \(H_2\) are \textbf{joinable}, then \(T\) is \textbf{confluent} on \(\pmb{D}\) (or \textbf{confluent up to garbage} w.r.t. \(\pmb{D}\)).
\end{definition}

\begin{definition}
Let \(T = (\mathcal{L}, \mathcal{R})\), \(\pmb{D} \subseteq \mathcal{G}(\mathcal{L})\). If there is no infinite derivation sequence \(G_0 \Rightarrow_{\mathcal{R}} G_1 \Rightarrow_{\mathcal{R}} G_2 \Rightarrow_{\mathcal{R}} \cdots\) such that \([G_0] \in \pmb{D}\), then \(T\) is \textbf{terminating} on \(\pmb{D}\) (or \textbf{terminating up to garbage} w.r.t. \(\pmb{D}\)).
\end{definition}

\begin{lemma} \label{lem:transitiveconfluence}
Let \(T = (\mathcal{L}, \mathcal{R})\), \(\pmb{D} \subseteq \mathcal{G}(\mathcal{L})\), \(\pmb{E} \subseteq D\). Then (\textbf{local}) \textbf{confluence} (\textbf{termination}) on \(\pmb{D}\) implies (\textbf{local}) \textbf{confluence} (\textbf{termination}) on \(\pmb{E}\).
\end{lemma}

\begin{proof}
Immediate consequence of set inclusion!
\end{proof}

\begin{corollary} \label{cor:garbageimplications}
Let \(T = (\mathcal{L}, \mathcal{R})\), \(\pmb{D} \subseteq \mathcal{G}(\mathcal{L})\). Then (\textbf{local}) \textbf{confluence} (\textbf{termination}) implies (\textbf{local}) \textbf{confluence} (\textbf{termination}) on \(\pmb{D}\).
\end{corollary}

\begin{proof}
Local confluence (confluence, termination) is exactly local confluence (confluence, termination) on \(\mathcal{G}(\mathcal{L})\).
\end{proof}

\begin{example}
Looking again at \(r_1\) and \(r_2\) from our first example, it is easy to see that \(r_1\) is in fact \textbf{terminating} and \textbf{confluent up to garbage} on the language of acyclic graphs. Similarly, \(r_2\) is \textbf{terminating} and \textbf{confluent up to garbage} on the language of trees.
\end{example}

\begin{lemma} \label{lem:garbagears}
Let \(T = (\mathcal{L}, \mathcal{R})\), \(\pmb{D} \subseteq \mathcal{G}(\mathcal{L})\). Then, if \(\pmb{D}\) is \textbf{closed} under \(T\), the \textbf{induced ARS} \((\pmb{D}, \rightarrow)\) where \([G] \rightarrow [H]\) iff \(G \Rightarrow_{\mathcal{R}} H\), is closed and well-defined. Moreover, it is (\textbf{locally}) \textbf{confluent} (\textbf{terminating}) whenever \(T\) is, \textbf{up to garbage} with respect to \(\pmb{D}\).
\end{lemma}

\begin{proof}
Since \(\pmb{D}\) is closed under \(T\), by Proposition \ref{lem:wgsclosure}, the induced ARS \((\pmb{D}, \rightarrow)\) where \([G] \rightarrow [H]\) iff \(G \Rightarrow_{\mathcal{R}} H\) is closed, and clearly it is well-defined due to the uniqueness of derivations up to isomorphism. Clearly this induced ARS is (locally) confluent (terminating) if \(T\) is (locally) confluent (terminating) up to garbage with respect to \(\pmb{D}\).
\end{proof}

\begin{definition}[Closedness Problem]~\\
\vspace{-1.2em}
\begin{itemize}[itemsep=-0.7ex,topsep=-0.7ex]
\setlength{\itemindent}{-1em}
\item[] \emph{Input}: A GT system \(T = (\mathcal{L}, \mathcal{R})\) and a graph grammar \(\pmb{G}\) over \(\mathcal{L}\).
\item[] \emph{Question}: Is \(\pmb{L}(\pmb{G})\) is \textbf{closed} under \(T\)?
\end{itemize}
\end{definition}

\begin{theorem}[Undecidable Closedness]
The \textbf{closedness problem} is \textbf{undecidable} in general, even for terminating GT systems \(T\) with only one rule, and \(\pmb{G}\) an edge replacement grammar (Section \ref{section:edgereplgram}).
\end{theorem}

\begin{proof}
This was shown in 1998 by Fradet and Le M{\'e}tayer \cite{Fradet-Metayer98a}.
\end{proof}

\begin{remark}
Closedness and language recognition has actually been considered before by Bakewell, Plump, and Runciman, in the context of languages specified by reduction systems without non-terminals \cite{Bakewell-Plump-Runciman03a,Bakewell-Plump-Runciman04a}.
\end{remark}

\begin{theorem}[Newman-Garbage Lemma] \label{thm:newmangarbage}
Let \(T = (\mathcal{L}, \mathcal{R})\), \(\pmb{D} \subseteq \mathcal{G}(\mathcal{L})\). If \(T\) is \textbf{terminating} on \(\pmb{D}\) and \(\pmb{D}\) is \textbf{closed} under \(T\), then it is \textbf{confluent} on \(\pmb{D}\) iff it is \textbf{locally confluent} on \(\pmb{D}\).
\end{theorem}

\begin{proof}
By Lemma \ref{lem:garbagears}, the induced ARS \((\pmb{D}, \rightarrow)\) is well-defined, closed, and terminating. Thus, by the original Newman's Lemma (Theorem \ref{thm:newmanlem}), the induced ARS is confluent iff it is locally confluent, as required.
\end{proof}

The requirement of closure is reasonable, even in the presence of termination on all graphs. Of course, there exists systems that are both locally confluent confluent up to garbage and terminating confluent up to garbage, which are not closed, but are confluent up to garbage, as in Example \ref{eg:closed-not-needed}.

\begin{example} \label{eg:closed-not-needed}
Choose some \(\pmb{D}\) of size at least 2, and a graph \(G\) which has a non-isomorphic successor under some GT system \(T\). Suppose \(\pmb{D}\) is closed under \(T\), \(T\) locally confluent and \(T\) terminating  on \(\pmb{D}\). Then construct \(\pmb{E} = \pmb{D} \setminus \{[H]\}\) where \(H \not\cong G\) is a successor of \(G\) under \(T\). Then, \(T\) is locally confluent and terminating on \(\pmb{E}\), but not closed.
\end{example}

However, in general, closure is needed. In Example \ref{eg:non-closed}, we show a system that is locally confluent up to garbage and terminating, yet is not confluent up to garbage.

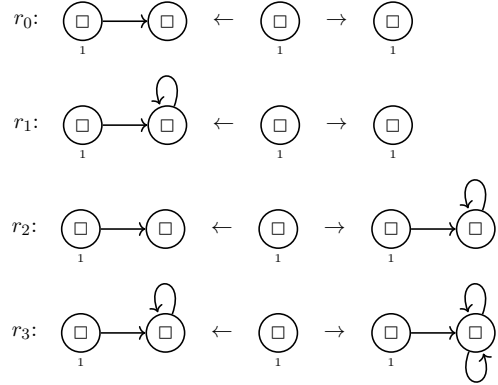
\begin{figure}[H]
\centering
\noindent
\input{fig/4/eg-closed}
\vspace{-0.9em}
\caption{Example Non-Closed Rules}
\label{fig:non-closed-rules}
\end{figure}

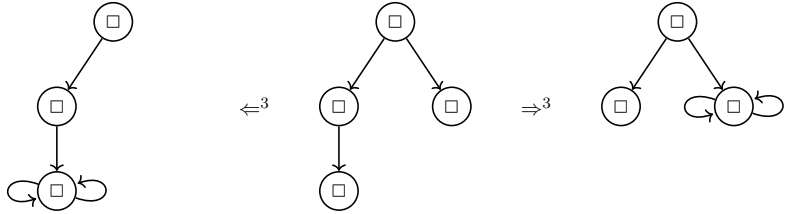
\begin{figure}[H]
\centering
\noindent
\input{fig/4/non-joinable-pair}
\vspace{-0.3em}
\caption{Example Non-Joinable Pair}
\label{fig:non-joinable-pair}
\end{figure}

\begin{example} \label{eg:non-closed}
Let \(\mathcal{L} = (\{\Square\}, \{\Square\})\) and \(\mathcal{R} = \{r_0, r_1, r_2, r_3\}\), where the rules are given in Figure \ref{fig:non-closed-rules}. Let \(\pmb{D} = \pmb{L}(\pmb{TREE})\) from Section \ref{sec:linearrec}.

Then, \((\mathcal{L}, \mathcal{R})\) is locally confluent on \(\pmb{D}\) and terminating on \(\mathcal{G}(\mathcal{L})\), however is not confluent on \(\pmb{D}\)! This was possible because we are not closed on \(\pmb{D}\). An example non-joinable pair is given in Figure \ref{fig:non-joinable-pair}, generated by moving three steps away from the start graph. In our diagrams, unlabelled edges are understood to be \(\Square\)-labelled.
\end{example}

\vspace{-0.5em}

\vspace{0.2em}

In this section, we have defined garbage, (local) confluence (termination) up to garbage, and closedness. We have shown that if we have closedness and termination up to garbage, then local confluence up to garbage implies confluence up to garbage, and an example where not having closedness fails, even in the presence of termination. Moreover, we have shown that closedness is undecidable in general.


\section{Non-Garbage Critical Pairs}

In 1970, Knuth and Bendix showed that confluence checking of terminating term rewriting systems is decidable \cite{Knuth-Bendix70a}. Moreover, it suffices to compute all \enquote{critical pairs} and check their joinability \cite{Huet80a,Baader-Nipkow98a,Terese03a}. Unfortunately, for (terminating) graph transformation systems, confluence is not decidable (Theorem \ref{thm:undecidablegtsrealdeal}), and joinability of critical pairs does not imply local confluence. In 1993, Plump showed that \enquote{strong joinability} of all critical pairs is sufficient but not necessary to show local confluence \cite{Plump93b,Plump05a}. We have summarised these results in Section \ref{section:critpairs}.

We would like to generalise Theorem \ref{thm:critpairlem} to allow us to determine when we have local confluence up to garbage. For this, we need to define a notion of subgraph closure and non-garbage critical pairs. In this section, we shall be working with standard GT systems, as defined in Appendix \ref{appendix:transformation}, but without relabelling. That is, all graphs are totally labelled, including interface graphs.

\begin{definition} \label{dfn:subgraphclosure}
Let \(\pmb{D} \subseteq \mathcal{G}(\mathcal{L})\) be a set of abstract graphs. Then \(\pmb{D}\) is \textbf{subgraph closed} iff for all graphs \(G\), \(H\), such that \(H \subseteq G\), if \([G] \in \pmb{D}\), then \([H] \in \pmb{D}\). The \textbf{subgraph closure} of \(\pmb{D}\), denoted \(\overbar{\pmb{D}}\), is the smallest set containing \(\pmb{D}\) that is \textbf{subgraph closed}.
\end{definition}

\begin{proposition}
Given \(\pmb{D} \subseteq \mathcal{G}(\mathcal{L})\), \(\overbar{\pmb{D}}\) always \textbf{exists}, and is \textbf{unique}. Moreover, \(\pmb{D} = \overbar{\pmb{D}}\) iff \(\pmb{D}\) is \textbf{subgraph closed}.
\end{proposition}

\begin{proof}
The key observations are that the subgraph relation is transitive, and each graph has only finitely many subgraphs. Clearly, the smallest possible set containing \(\pmb{D}\) is just the union of all subgraphs of the elements of \(\pmb{D}\), up to isomorphism. This is the unique subgraph closure of \(\pmb{D}\).
\end{proof}

\begin{remark}
\(\overbar{\pmb{D}}\) always exists, however it need not be decidable, even when \(\pmb{D}\) is! It is not obvious what conditions on \(\pmb{D}\) ensure that \(\overbar{\pmb{D}}\) is decidable. Interestingly, the classes of regular and context-free string languages are actually closed under substring closure \cite{Berstel79a}.
\end{remark}

\begin{example}
\(\emptyset\) and \(\mathcal{G}(\mathcal{L})\) are subgraph closed.
\end{example}

\begin{example}
The language of discrete graphs is subgraph closed.
\end{example}

\begin{example}
The subgraph closure of the language of trees is the language of forests. The subgraph closure of the language connected graphs is the language of all graphs.
\end{example}

\begin{definition}
Let \(T = (\mathcal{L}, \mathcal{R})\), \(\pmb{D} \subseteq \mathcal{G}(\mathcal{L})\). A \textbf{critical pair} (Definition \ref{dfn:critpair}) \(H_1 \Leftarrow G \Rightarrow H_2\) is \textbf{non-garbage} iff \([G] \in \overbar{\pmb{D}}\).
\end{definition}

\begin{lemma} \label{lem:fincritpairs}
Given a GT system \(T = (\mathcal{L}, \mathcal{R})\) and \(\pmb{D} \subseteq \mathcal{G}(\mathcal{L})\), then there are only finitely many \textbf{non-garbage critical pairs} up to isomorphism.
\end{lemma}

\begin{proof}
By Theorem \ref{thm:critpairlem} and Lemma \ref{lem:finitecritpairs}, there are only finitely many critical pairs for \(T\), up to isomorphism, and there exists a terminating procedure for generating them. Thus, there are only finitely many non-garbage critical pairs up to isomorphism.
\end{proof}

\begin{corollary} \label{cor:fincritpairs}
Given a GT system \(T = (\mathcal{L}, \mathcal{R})\) and \(\pmb{D} \subseteq \mathcal{G}(\mathcal{L})\) such that \(\overbar{\pmb{D}}\) is \textbf{decidable}, then one can find the finitely many \textbf{non-garbage critical pairs} in \textbf{finite time}.
\end{corollary}

\begin{proof}
By Lemma \ref{lem:fincritpairs}, we have only finitely many pairs to consider, and we can generate them in finite them. Since \(\overbar{\pmb{D}}\) has a computable membership function, we can test if the start graph in each pair is garbage in finite time.
\end{proof}

\begin{corollary} \label{cor:gencritpairs}
Let \(T = (\mathcal{L}, \mathcal{R})\), \(\pmb{D} \subseteq \mathcal{G}(\mathcal{L})\) be such that \(T\) is \textbf{terminating} on \(\pmb{D}\) and \(\overbar{\pmb{D}}\) is \textbf{decidable}. Then, one can \textbf{decide} if all the \textbf{non-garbage critical pairs} are \textbf{strongly joinable} (Definition \ref{dfn:strongjoin}).
\end{corollary}

\begin{proof}
By Corollary \ref{cor:fincritpairs}, we can find the finitely many pairs in finite time, and since \(T\) is terminating on \(\pmb{D}\) and finitely branching (Lemma \ref{lem:gtprops}), both sides of each pair have only finitely many successors (Lemma \ref{lem:arsbranching}), thus we can test for strong joinability in finite time.
\end{proof}

\begin{lemma}
Let \(T = (\mathcal{L}, \mathcal{R})\), \(\pmb{D} \subseteq \mathcal{G}(\mathcal{L})\). Then, the \textbf{non-garbage critical pairs} are \textbf{complete}. That is, for each pair of \textbf{parallelly independent} (Definition \ref{dfn:parindep}) direct derivations, \(H_1 \Leftarrow_{r_1,g_1} G \Rightarrow_{r_2,g_2} H_2\) such that \([G] \in \pmb{D}\), there is a \textbf{critical pair} \(P_1 \Leftarrow_{r_1,o_1} K \Rightarrow_{r_2,o_2} P_2\) with extension diagrams (1), (2), and an inclusion morphism \(m: K \to G\).
\end{lemma}

\vspace{-0.9em}
\begin{figure}[H]
\centering
\noindent
\input{fig/4/pairs}
\vspace{-0.3em}
\caption{Pair Factorisation Diagram}
\end{figure}
\vspace{0.2em}

\begin{proof}
By Lemma 6.22 in \cite{Ehrig-Ehrig-Prange-Taentzer06a}, critical pairs are complete when \(\pmb{D} = \mathcal{G}(\mathcal{L})\). If we only consider derivations from start graphs \(G\) such that \([G] \in \pmb{D} \subseteq \mathcal{G}(\mathcal{L})\), clearly all factorings with critical pairs are such that \(K\) can be embedded into \(G\), so \([K] \in \overbar{\pmb{D}}\). Thus, the non-garbage critical pairs are complete.
\end{proof}

\begin{theorem}[Non-Garbage Critical Pair Lemma] \label{thm:ngcritpairlem}
Let \(T = (\mathcal{L}, \mathcal{R})\), \(\pmb{D} \subseteq \mathcal{G}(\mathcal{L})\). If all its \textbf{non-garbage critical pairs} are \textbf{strongly joinable}, then \(T\) is \textbf{locally confluent} on \(\pmb{D}\).
\end{theorem}

\begin{proof}
By the proof of Theorem 6.28 in \cite{Ehrig-Ehrig-Prange-Taentzer06a}, strong joinability of critical pairs implies local confluence due to completeness. But, the non-garbage critical pairs are complete with respect to \(\pmb{D}\), so we have the result.
\end{proof}

\begin{corollary}
Let \(T = (\mathcal{L}, \mathcal{R})\), \(\pmb{D} \subseteq \mathcal{G}(\mathcal{L})\). If \(T\) is \textbf{terminating} on \(\pmb{D}\), \(\pmb{D}\) is \textbf{closed} under \(T\), and all \(T\)'s \textbf{non-garbage critical pairs} are \textbf{strongly joinable} then \(T\) is \textbf{confluent} on \(\pmb{D}\).
\end{corollary}

\begin{proof}
By the above theorem, \(T\) is \textbf{locally confluent up to garbage}, so by the Newman-Garbage Lemma (Theorem \ref{thm:newmangarbage}), \(T\) is \textbf{confluent up to garbage} as required.
\end{proof}

\begin{remark}
Obviously, testing for local confluence up to garbage is undecidable in general, even when \(\overbar{\pmb{D}}\) is decidable and the system is terminating and strongly closed. What is remarkable though, is that confluence up to garbage is actually undecidable general for a terminating non-length-increasing string rewriting systems and \(\pmb{D}\) a regular string language \cite{Caron91a}!
\end{remark}


\section{Extended Flow Diagrams}

In 1976, Farrow, Kennedy and Zucconi presented \textbf{semi-structured flow graphs}, defining a grammar with confluent reduction rules \cite{Farrow-Kennedy-Zucconi76a}. Plump has considered a restricted version of this language: \textbf{extended flow diagrams} (EFDs) \cite{Plump05a}. The reduction rules for \textbf{extended flow diagrams} efficiently recognise the EFDs, despite not being confluent. That is, we have an efficient mechanism for testing for language membership, since we need not \enquote{backtrack}, just like in Theorem \ref{thm:treerecthm}.

\begin{definition}[Efficient Recognition]
\normalfont
\(T = (\mathcal{L}, \mathcal{R})\) \textbf{efficiently recognises} \(\pmb{L}\) over \(\mathcal{P} \subseteq \mathcal{L}\) iff \(T\) recognises \(\pmb{L}\), \(T\) is terminating, and  \(T\) is confluent on \(\pmb{L}\).
\end{definition}

\begin{theorem}[Efficient Recognition Correctness]
Given an GT system \(T = (\mathcal{L}, \mathcal{R})\) \textbf{efficiently recognising} a language \(\pmb{L}\) over \(\mathcal{P} \subseteq \mathcal{L}\) and an input graph \(G\) over \(\mathcal{P}\), the following algorithm is correct: Compute a normal form of \(G\) by deriving successor graphs using \(T\) as long as possible. If the result graph is isomorphic to \(S\), the input graph is in the language. Otherwise, the graph is not in the language.
\end{theorem}

\begin{proof}
Suppose \(G\) is not \(\pmb{L}\). Then, since \(T\) is terminating our algorithm must be able to find a normal form of \(G\), say \(H\), and because \(T\) recognises \(\pmb{L}\), it must be the case that \(H\) is not isomorphic to \(S\), and so the algorithm correctly decides that \(G\) is not in \(\pmb{L}\).

Now, suppose that \(G\) is in \(\pmb{L}\). Then, because \(T\) is terminating, as before, we must be able to derive some normal form, \(H\). But then, since \(T\) is both confluent on \(\pmb{L}\) and recognises \(\pmb{L}\), it must be the case that \(H\) is isomorphic to \(S\), and so the algorithm correctly decides that \(G\) is in \(\pmb{L}\).
\end{proof}

\begin{figure}[H]
\centering
\vspace{-0.1em}
\includegraphics[totalheight=6.75cm]{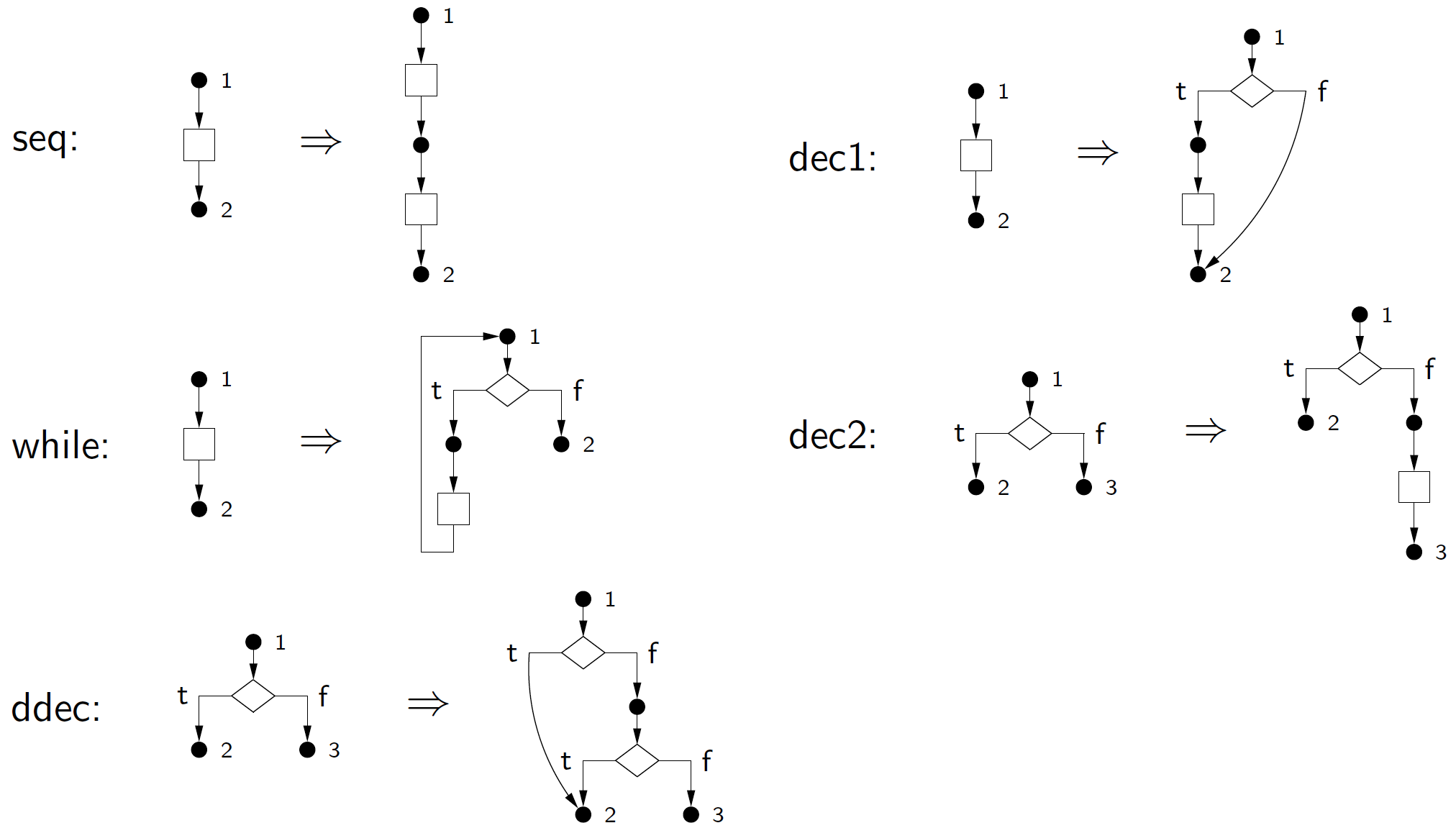}
\caption{EFD Grammar Rules}
\vspace{0.1em}
\end{figure}

\begin{definition}
The language of \textbf{extended flow diagrams} is generated by \(\pmb{EFD} = (\mathcal{L}, \mathcal{N}, \mathcal{R}, S)\) where \(\mathcal{L}_V = \{\bullet, \square, \Diamond\}\), \(\mathcal{L}_E = \{t, f, \square\}\), \(\mathcal{N}_V = \mathcal{N}_E = \emptyset\), \(\mathcal{R} = \{seq, while, ddec, dec1, dec2\}\), and \(S = \) \tikz[baseline]{ \tikzstyle{ann} = [draw=none,fill=none,right] \node (a) at (0.0,0.1) [draw, circle, thick, fill=black, scale=0.3] {}; \node[rectangle] (b) at (0.5,0.1) [minimum height=0.2cm,minimum width=0.2cm,draw] {}; \node (c) at (1.0,0.1) [draw, circle, thick, fill=black, scale=0.3] {}; \draw (a) edge[->,thick] (b) (b) edge[->,thick] (c);}.
\end{definition}

In the figure, the shorthand notation with the numbers under the nodes places such nodes in the interface graph of the rules. We assume that the interface graphs are discrete (have no edges).

\begin{lemma} \label{lem:efdcycles}
Every \textbf{directed cycle} in an extended flow diagram contains a \(t\)-labelled edge
\end{lemma}

\begin{proof}
Induction.
\end{proof}

\begin{theorem}[Efficient EFD Recognition]
\(EFD^{-1} = (\mathcal{L}, \mathcal{R}^{-1})\) \textbf{efficiently recognises} \(\pmb{L}(\pmb{EFD})\).
\end{theorem}

\begin{figure}[H]
\centering
\vspace{-0.2em}
\includegraphics[totalheight=2.55cm]{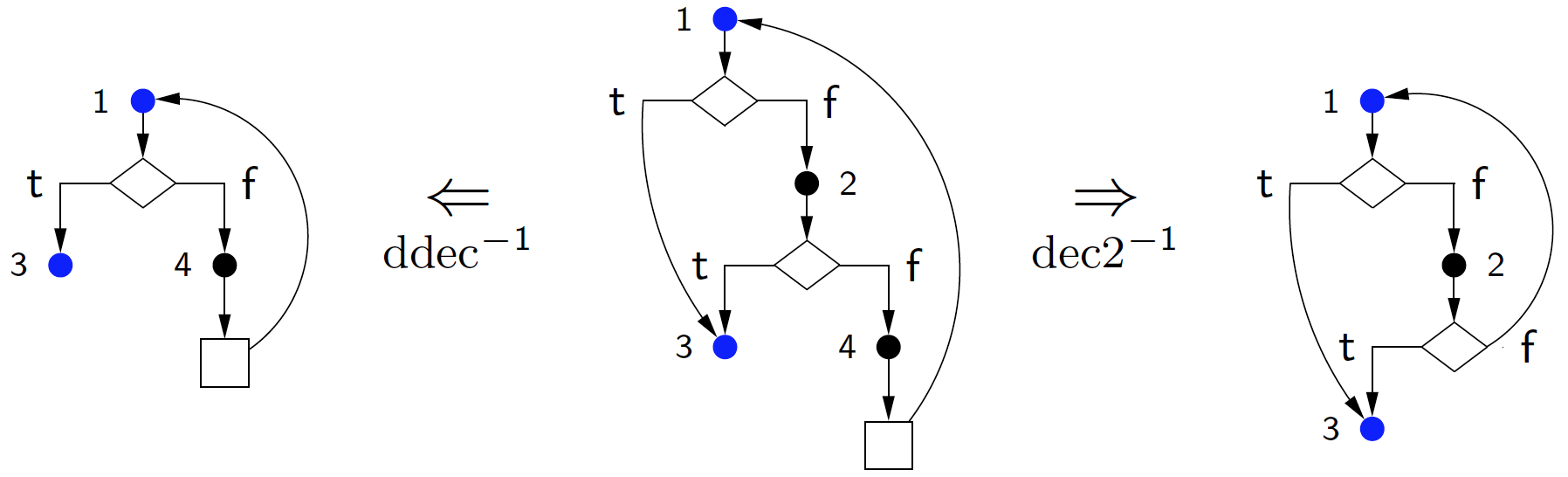}
\vspace{-0.1em}
\caption{Non-Joinable Critical Pair}
\label{fig:efd-pair}
\end{figure}

\begin{proof}
By Theorem \ref{thm:membershiptest}, \(EFD^{-1}\) recognises \(\pmb{L}(\pmb{EFD})\), and one can see that it is terminating since each rule is size reducing.

We now proceed by performing critical pair analysis on \(EFD^{-1}\). There are ten critical pairs, all but one of which are strongly joinable apart from one (Figure \ref{fig:efd-pair}). Now observe that Lemma \ref{lem:efdcycles} tells us that EFDs cannot contain such cycles. With this knowledge, we define \(D\) to be all graphs such that directed cycles contain at least one \(t\)-labelled edge. Clearly, \(D\) is subgraph closed, and then by our Non-Garbage Critical Pair Lemma (Theorem \ref{thm:ngcritpairlem}), we have that \(EFD^{-1}\) is locally confluent on \(D\).

Next, it is easy to see that \(D\) is closed on \(EFD^{-1}\), so we can use Newman-Garbage Lemma (Theorem \ref{thm:newmangarbage}) to conclude confluence on \(D\) and thus, by Lemma \ref{lem:transitiveconfluence}, \(EFD^{-1}\) is confluent on \(\pmb{L}(\pmb{EFD})\).

Thus, \(EFD^{-1}\) efficiently recognises \(\pmb{L}(\pmb{EFD})\), as required.
\end{proof}

\vspace{-0.1em}

%% file: fig/4/eg1.tex
\begin{tikzpicture}[every node/.style={align=center}]
    \node (a) at (0.0,-0.05) {$r_1$:};
    \node (b) at (1.0,0.0)   [draw, circle, thick, fill=black, scale=0.3] {\,};
    \node (c) at (2.0,0.0)   [draw, circle, thick, fill=black, scale=0.3] {\,};
    \node (d) at (3.0,0.0)   {$\leftarrow$};
    \node (e) at (4.0,0.0)   [draw, circle, thick, fill=black, scale=0.3] {\,};
    \node (f) at (5.0,0.0)   [draw, circle, thick, fill=black, scale=0.3] {\,};
    \node (g) at (6.0,0.0)   {$\rightarrow$};
    \node (h) at (7.0,0.0)   [draw, circle, thick, fill=black, scale=0.3] {\,};
    \node (i) at (8.0,0.0)   [draw, circle, thick, fill=black, scale=0.3] {\,};

    \node (j) at (9.5,-0.05) {$r_2$:};
    \node (k) at (10.5,0.0)  [draw, circle, thick, fill=black, scale=0.3] {\,};
    \node (l) at (11.5,0.0)  [draw, circle, thick, fill=black, scale=0.3] {\,};
    \node (m) at (12.5,0.0)  {$\leftarrow$};
    \node (n) at (13.5,0.0)  [draw, circle, thick, fill=black, scale=0.3] {\,};
    \node (o) at (14.5,0.0)  {$\rightarrow$};
    \node (p) at (15.5,0.0)  [draw, circle, thick, fill=black, scale=0.3] {\,};

    \node (B) at (1.0,-.18)  {\tiny{1}};
    \node (C) at (2.0,-.18)  {\tiny{2}};
    \node (E) at (4.0,-.18)  {\tiny{1}};
    \node (F) at (5.0,-.18)  {\tiny{2}};
    \node (H) at (7.0,-.18)  {\tiny{1}};
    \node (I) at (8.0,-.18)  {\tiny{2}};
    \node (K) at (10.5,-.18) {\tiny{1}};
    \node (N) at (13.5,-.18) {\tiny{1}};
    \node (P) at (15.5,-.18) {\tiny{1}};

    \draw (b) edge[->,thick] (c)
          (k) edge[->,thick] (l);
\end{tikzpicture}

%% file: fig/4/eg-closed.tex
\scalebox{.75}{\begin{tikzpicture}[every node/.style={inner sep=0pt, text width=6.5mm, align=center}]
    \node (a) at (0.0,0) {$r_0$:};

    \node (b) at (1.0,0) [draw,circle,thick] {\(\Square\)};
    \node (c) at (2.5,0) [draw,circle,thick] {\(\Square\)};

    \node (d) at (3.5,0) {$\leftarrow$};

    \node (e) at (4.5,0) [draw,circle,thick] {\(\Square\)};

    \node (f) at (5.5,0) {$\rightarrow$};

    \node (g) at (6.5,0) [draw,circle,thick] {\(\Square\)};
    \node (h) at (8.0,0) {\,};

    \node (B) at (1.0,-.52) {\tiny{1}};
    \node (E) at (4.5,-.52) {\tiny{1}};
    \node (G) at (6.5,-.52) {\tiny{1}};

    \draw (b) edge[->,thick] (c);
\end{tikzpicture}}

\vspace{0.3em}

\scalebox{.75}{\begin{tikzpicture}[every node/.style={inner sep=0pt, text width=6.5mm, align=center}]
    \node (a) at (0.0,0) {$r_1$:};

    \node (b) at (1.0,0) [draw,circle,thick] {\(\Square\)};
    \node (c) at (2.5,0) [draw,circle,thick] {\(\Square\)};

    \node (d) at (3.5,0) {$\leftarrow$};

    \node (e) at (4.5,0) [draw,circle,thick] {\(\Square\)};

    \node (f) at (5.5,0) {$\rightarrow$};

    \node (g) at (6.5,0) [draw,circle,thick] {\(\Square\)};
    \node (h) at (8.0,0) {\,};

    \node (B) at (1.0,-.52) {\tiny{1}};
    \node (E) at (4.5,-.52) {\tiny{1}};
    \node (G) at (6.5,-.52) {\tiny{1}};

    \draw (b) edge[->,thick] (c)
          (c) edge[->,in=110,out=70,loop,thick] (c);
\end{tikzpicture}}

\vspace{0.3em}

\scalebox{.75}{\begin{tikzpicture}[every node/.style={inner sep=0pt, text width=6.5mm, align=center}]
    \node (a) at (0.0,0) {$r_2$:};

    \node (b) at (1.0,0) [draw,circle,thick] {\(\Square\)};
    \node (c) at (2.5,0) [draw,circle,thick] {\(\Square\)};

    \node (d) at (3.5,0) {$\leftarrow$};

    \node (e) at (4.5,0) [draw,circle,thick] {\(\Square\)};

    \node (f) at (5.5,0) {$\rightarrow$};

    \node (g) at (6.5,0) [draw,circle,thick] {\(\Square\)};
    \node (h) at (8.0,0) [draw,circle,thick] {\(\Square\)};

    \node (B) at (1.0,-.52) {\tiny{1}};
    \node (E) at (4.5,-.52) {\tiny{1}};
    \node (G) at (6.5,-.52) {\tiny{1}};

    \draw (b) edge[->,thick] (c)
          (g) edge[->,thick] (h)
          (h) edge[->,in=110,out=70,loop,thick] (h);
\end{tikzpicture}}

\vspace{0.3em}

\scalebox{.75}{\begin{tikzpicture}[every node/.style={inner sep=0pt, text width=6.5mm, align=center}]
    \node (a) at (0.0,0) {$r_3$:};

    \node (b) at (1.0,0) [draw,circle,thick] {\(\Square\)};
    \node (c) at (2.5,0) [draw,circle,thick] {\(\Square\)};

    \node (d) at (3.5,0) {$\leftarrow$};

    \node (e) at (4.5,0) [draw,circle,thick] {\(\Square\)};

    \node (f) at (5.5,0) {$\rightarrow$};

    \node (g) at (6.5,0) [draw,circle,thick] {\(\Square\)};
    \node (h) at (8.0,0) [draw,circle,thick] {\(\Square\)};

    \node (B) at (1.0,-.52) {\tiny{1}};
    \node (E) at (4.5,-.52) {\tiny{1}};
    \node (G) at (6.5,-.52) {\tiny{1}};

    \draw (b) edge[->,thick] (c)
          (g) edge[->,thick] (h)
          (c) edge[->,in=110,out=70,loop,thick] (c)
          (h) edge[->,in=110,out=70,loop,thick] (h)
          (h) edge[->,in=-70,out=-110,loop,thick] (h);
\end{tikzpicture}}

\vspace{0.3em}

%% file: fig/4/non-joinable-pair.tex
\scalebox{.75}{\begin{tikzpicture}[every node/.style={inner sep=0pt, text width=6.5mm, align=center}]
    \node (a) at (2.0,0.0)   [draw, circle, thick] {\(\Square\)};
    \node (b) at (1.0,-1.5)  [draw, circle, thick] {\(\Square\)};
    \node (d) at (1.0,-3.0)  [draw, circle, thick] {\(\Square\)};

    \node (e) at (4.5,-1.5)  {$\Leftarrow^3$};

    \node (f) at (7.0,0.0)   [draw, circle, thick] {\(\Square\)};
    \node (g) at (6.0,-1.5)  [draw, circle, thick] {\(\Square\)};
    \node (h) at (8.0,-1.5)  [draw, circle, thick] {\(\Square\)};
    \node (i) at (6.0,-3.0)  [draw, circle, thick] {\(\Square\)};

    \node (j) at (9.5,-1.5)  {$\Rightarrow^3$};

    \node (k) at (12.0,0.0)  [draw, circle, thick] {\(\Square\)};
    \node (l) at (11.0,-1.5) [draw, circle, thick] {\(\Square\)};
    \node (m) at (13.0,-1.5) [draw, circle, thick] {\(\Square\)};

    \draw (a) edge[->,thick] (b)
          (b) edge[->,thick] (d)
          (d) edge[->,in=20,out=-20,loop,thick] (d)
          (d) edge[->,in=200,out=160,loop,thick] (d);

    \draw (f) edge[->,thick] (g)
          (f) edge[->,thick] (h)
          (g) edge[->,thick] (i);

    \draw (k) edge[->,thick] (l)
          (k) edge[->,thick] (m)
          (m) edge[->,in=20,out=-20,loop,thick] (m)
          (m) edge[->,in=200,out=160,loop,thick] (m);
\end{tikzpicture}}

%% file: fig/4/pairs.tex
\begin{tikzpicture}[every node/.style={inner sep=0pt, text width=6.5mm, align=center}]
    \node (a) at (0.0,0.0) {$P_1$};
    \node (b) at (1.0,0.0) {$\Longleftarrow$};
    \node (c) at (2.0,0.0) {$K$};
    \node (d) at (3.0,0.0) {$\Longrightarrow$};
    \node (e) at (4.0,0.0) {$P_2$};

    \node (f) at (0.0,-0.8) {$\big\downarrow$};
    \node (g) at (1.0,-0.8) {(1)};
    \node (h) at (2.0,-0.8) {$\big\downarrow$};
    \node (i) at (3.0,-0.8) {(2)};
    \node (j) at (4.0,-0.8) {$\big\downarrow$};

    \node (k) at (0.0,-1.6) {$H_1$};
    \node (l) at (1.0,-1.6) {$\Longleftarrow$};
    \node (m) at (2.0,-1.6) {$G$};
    \node (n) at (3.0,-1.6) {$\Longrightarrow$};
    \node (o) at (4.0,-1.6) {$H_2$};
\end{tikzpicture}

%% file: content/5.tex
\chapter{Confluence Analysis II} \label{chapter:confluenceanalysis2}

\begin{chapquote}{Kurt G{\"o}del, \textit{Opening of the paper introducing the undecidability theorem (1931)}}
``The development of mathematics towards greater precision has led, as is well known, to the formalisation of large tracts of it, so that one can prove any theorem using nothing but a few mechanical rules.''
\end{chapquote}

In the last chapter, we have seen that the notion of confluence up to garbage was sufficient to give efficient language recognition, however it is not yet known if these results transfer to the setting with root nodes and partial labelling. In this chapter, we will introduce definitions of sequential and parallel independence for such rules, and show that the (Non-Garbage) Critical Pair Lemma holds. We do this by showing that \enquote{conventional} totally labelled systems can simulate relabelling and root nodes.


\section{Encoding PLRGs} \label{encodinglabelling}

We can encode the graphs and morphisms given in Chapter \ref{chapter:newtheory} as totally labelled graphs and morphisms (in Appendix \ref{appendix:transformation}). Both of these are locally small categories, and we denote them \(\pmb{PLRG(\mathcal{L})}\) and \(\pmb{TLG(\mathcal{L})}\) respectively. We will refer to the graph transformation systems from Chapter \ref{chapter:newtheory} as \textbf{R-GT systems}, and those from Appendix \ref{appendix:transformation} without relabelling as \textbf{T-GT systems}.

Without loss of generality, we will assume that \(\mathcal{L} = (\mathcal{L}_V, \mathcal{L}_E)\) is an arbitrary label alphabet such that \(\mathcal{L}_V \cap \mathcal{L}_E = \emptyset\) and \(\{\Square, 0, 1\} \cap (\mathcal{L}_V \cup \mathcal{L}_E) = \emptyset\).

In this section we will see that this encoding is a actually fully faithful functor and that we have \textbf{derivation compatibility}, in a similar sense to Remark \ref{rem:compatderivations}.

\begin{definition}[Object Encoding]
Given \(G = (V, E, s, t, l, m, p) \in \pmb{PLRG(\mathcal{L})}\), define \(e(G) = (V', E', s', t', l', m')\) where:
\begin{enumerate}[itemsep=-0.4ex,topsep=-0.4ex]
\item \(V' = V\);
\item \(E' = E \times \{0\} \cup l^{-1}(\mathcal{L}_V) \times \{1\} \cup p^{-1}(\{0, 1\}) \times \{2\}\);
\item \(s'((e, 0)) = s(e)\) and \(s'((v, i)) = v\) for \(i = 1, 2\);
\item \(t'((e, 0)) = t(e)\) and \(t'((v, i)) = v\) for \(i = 1, 2\);
\item \(l'(v) = \Square\);
\item \(m'((e, 0)) = m(e)\), \(m'((v, 1)) = l(v)\), and \(m'((v, 2)) = p(v)\).
\end{enumerate}

We denote by \(e(X)\) where \(X\) is some collection of objects (or their isomorphism classes), the image of the objects under \(e\). In an abuse of notation, we will also denote the the label alphabet of the encoded objects by \(e(\mathcal{L}) = (\{\Square\}, \mathcal{L}_E \cup \mathcal{L}_V \cup \{0, 1\})\).
\end{definition}

\begin{proposition} \label{prop:objencode}
\(e(G) \in \pmb{TLG(e(\mathcal{L}))}\).
\end{proposition}

\begin{proof}
Clearly each graph has exactly one well-defined encoding, and it is totally labelled by construction.
\end{proof}

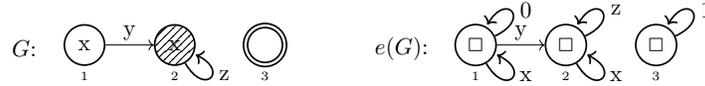
\begin{figure}[H]
\centering
\noindent
\scalebox{.8}{\input{fig/5/encoding}}
\vspace{-0.6em}
\caption{Example Encoded PLRG}
\label{fig:encoding}
\end{figure}

\begin{example}
Let \(\mathcal{L} = (\{x\}, \{y, z\})\). Then Figure \ref{fig:encoding} shows an example PLRG and its encoding as a totally labelled graph.
\end{example}

\vspace{-1.4em}

\begin{definition}[Arrow Encoding]
Given \(g \in \textrm{Hom}_{\pmb{PLRG(\mathcal{L})}}(G, H)\), define \(e(g) = (g_V', g_E')\) where:
\begin{enumerate}[itemsep=-0.4ex,topsep=-0.4ex]
\item \(g_V'(v) = g_V(v)\);
\item \(g_E'((e, 0)) = (g_E(e), 0)\) and \(g_E'((v, i)) = (g_V(v), i)\) for \(i = 1, 2\).
\end{enumerate}
\end{definition}

\begin{proposition} \label{prop:morencode}
\(e(g) \in \textrm{Hom}_{\pmb{TLG(e(\mathcal{L}))}}(e(G), e(H))\).
\end{proposition}

\begin{proof}
By construction.
\end{proof}

Now that we have setup our encoding function, we are ready to give our first important technical lemma, allowing us to then show that derivations are compatible (Theorem \ref{thm:compatderenc}).

\begin{lemma} \label{lem:fullyfaithful}
\(e\) is a \textbf{fully faithful functor} \(e: \pmb{PLRG(\mathcal{L})} \to \pmb{TLG(e(\mathcal{L}))}\).
\end{lemma}

\begin{proof}
By Propositions \ref{prop:objencode} and \ref{prop:morencode}, each graph and morphism has a well-defined encoding. The first axiom as given in Definition \ref{dfn:functor} is given by Proposition \ref{prop:morencode}, and the remaining 2 axioms are almost immediate. Thus we have a functor.

To see fullness, suppose that there is a morphism between two encoded graphs. Since morphisms are required to preserve labels, it is not possible to have confusion between the edges (that would introduce extra morphisms). That is, root-loops have to be mapped to root-loops, non-root-loops to non-root-loops, \(l\)-label-loops to \(l\)-label-loops for all \(l \in \mathcal{L}_V\), and \(m\)-label-loops to \(m\)-label-loops for all \(m \in \mathcal{L}_E\). We can then, decode such a morphism to a morphism between the original graphs in the original setting.

Faithfulness is explicit from the construction of encoded morphisms. Each morphism is mapped to a distinct encoded morphism.
\end{proof}

\begin{lemma}[Dangling Condition Compatibility] \label{lem:danglecompatrenc}
Given a TLRG \(G\) and an inclusion \(i: K \rightarrow L\) where \(L\) is a TLRG and \(K\) a PLRG, then the set of injective morphisms \(L \rightarrow G\) satisfying the dangling condition w.r.t. \(i\) (in the sense of R-GT systems) is in bijective correspondence via \(e\) with the set of in injective morphisms \(e(L) \rightarrow e(G)\) satisfying the dangling condition w.r.t. \(e(i)\) (in the sense of T-GT systems).
\end{lemma}

\begin{proof}
First, we show that the injective morphisms are in bijective correspondence. Suppose that \(g: L \to G\) is an injective morphism, then by construction, \(e(g): e(L) \to e(G)\) is injective. The reverse direction is a consequence of the elementary result that fully faithful functors reflect monomorphisms.

Suppose again that \(g: L \to G\) is an injective morphism. Then \(g\) satisfies the dangling condition w.r.t. \(i\) iff no edge in \(G \setminus g(L)\) is incident to a node in \(g(L \setminus K)\) iff no edge in \(e(G \setminus g(L))\) is incident to a node in \(e(g(L \setminus K))\) iff no edge in \(e(G) \setminus e(g)(e(L))\) is incident to a node in \(e(g)(e(L) \setminus e(K))\) iff \(e(g)\) satisfies the dangling condition w.r.t. \(e(i)\). The correctness of the movement of \(e\) comes from the explicit constructions.
\end{proof}

\begin{theorem}[Compatible Derivations] \label{thm:compatderenc}
Given a \textbf{rule} \(r = \langle L \leftarrow K \rightarrow R \rangle\) from a R-GT system, then for all TLRG \(G, H\), \(G \Rightarrow_r H\) iff \(e(G) \Rightarrow_{e(r)} e(H)\), where \(e(r)\) encodes the rule in the obvious way.
\end{theorem}

\begin{proof}
By Lemma \ref{lem:danglecompatrenc}, \(e\) reflects and preserves injective morphisms satisfying the danging condition. But, it is an elementary result that a fully faithful functor reflects all limits and colimits, so by Theorems \ref{theorem:uniquederivations} and \ref{thm:derivationtheorem} we have the result!
\end{proof}

We can define the track morphism and closedness in the same way as for T-GT systems. We have compatibility of tracks and also of closedness. Note, in particular, the special case when \(\pmb{D} = \mathcal{G}(\mathcal{L})\).

\begin{corollary}[Track Compatibility] \label{cor:trackcompat}
\(e(\mathit{tr}_{G \Rightarrow^* H}) = \mathit{tr}_{e(G) \Rightarrow^* e(H)}\).
\end{corollary}

\begin{proof}
Induction on derivation length using Theorem \ref{thm:compatderenc}.
\end{proof}

\begin{corollary}[Closedness Compatibility]
Let \(T = (\mathcal{L}, \mathcal{R})\) be some R-GT system, and \(\pmb{D} \subseteq \mathcal{G}^{\varoplus}(\mathcal{L})\). Then \(\pmb{D}\) is \textbf{closed} under \(T\) iff \(e(D)\) is \textbf{closed} under \(e(T) = (e(\mathcal{L}), e(\mathcal{R}))\).
\end{corollary}


\section{Local Church-Rosser and Independence}

The purpose of this chapter is to define sequential and parallel independence for R-GT systems, and show they are compatible with the T-GT encoding. Moreover, Theorems \ref{thm:churchros1} and \ref{thm:churchros2}, together, constitute the Local Church-Rosser Theorem, and so we have lifted this result into our new setting.

\begin{definition} \label{dfn:rseqindep}
The derivations \(G_1 \Rightarrow_{r_1,g_1} H \Rightarrow_{r_2,g_2} G_2\) are \textbf{sequentially independent} iff \((h_1(R_1) \cap g_2(L_2)) \subseteq (h_1(K_1) \cap g_2(K_2))\).
\end{definition}

\begin{proposition} \label{prop:seqindsim}
We have sequential independence of derivations iff their encoding is sequentially independent in the sense of Definition \ref{dfn:seqindep}.
\end{proposition}

\begin{proof}
Definition bashing.
\end{proof}

\begin{theorem}[Sequential Independence] \label{thm:churchros1}
If \(G_1 \Rightarrow_{r_1,g_1} H \Rightarrow_{r_2,g_2} G_2\) are \textbf{sequentially independent}, then there exists a graph \(H'\) and \textbf{sequentially independent} steps \(G \Rightarrow_{r_2} H' \Rightarrow_{r_1} G_2\).
\end{theorem}

\begin{proof}
By combining Theorem \ref{thm:compatderenc} and Proposition \ref{prop:seqindsim}, the result is immediate from Lemma \ref{lem:seqindep}.
\end{proof}

\begin{definition} \label{dfn:rparindep}
The derivations \(H_1 \Leftarrow_{r_1,g_1} G \Rightarrow_{r_2,g_2} H_2\) are \textbf{parallelly independent} iff \((g_1(L_1) \cap g_2(L_2)) \subseteq (g_1(K_1) \cap g_2(K_2))\).
\end{definition}

\begin{remark}
The derivations \(H_1 \Leftarrow_{r_1,g_1} G \Rightarrow_{r_2,g_2} H_2\) are \textbf{parallelly independent} iff \(H_1 \Rightarrow_{r_1^{-1},h_1} G \Rightarrow_{r_2,g_2} H_2\) are \textbf{sequentially independent}.
\end{remark}

\begin{proposition} \label{prop:parindsim}
We have parallel independence of derivations iff their encoding is parallelly independent in the sense of Definition \ref{dfn:parindep}.
\end{proposition}

\begin{proof}
Definition bashing.
\end{proof}

\begin{theorem}[Parallel Independence] \label{thm:churchros2}
If \(H_1 \Leftarrow_{r_1,g_1} G \Rightarrow_{r_2,g_2} H_2\) are \textbf{parallelly independent}, then there exists a graph \(G'\) and \textbf{direct derivations} \(H_1 \Rightarrow_{r_2} G' \Leftarrow_{r_1} H_2\) with \(G \Rightarrow_{r_1} H_1 \Rightarrow_{r_2} G'\) and \(G \Rightarrow_{r_2} H_2 \Rightarrow_{r_1} G'\) \textbf{sequentially independent}.
\end{theorem}

\begin{proof}
By combining Theorem \ref{thm:compatderenc} and Proposition \ref{prop:parindsim}, the result is immediate from Lemma \ref{lem:parindep}.
\end{proof}


\section{Confluence up to Garbage}

The ultimate goal of this chapter is to recover the (Non-Garbage) Critical Pair Lemma for R-GT systems. We will begin this section by noting that all the definitions from Section \ref{sec:confmodgarb} can be trivially re-formulated for R-GT systems. We will not give all the definitions and results again here, but we do need to state the relationship between (local) confluence (termination) of a R-GT system and that of its encoding.

We start by noting compatibility of confluence and termination.

\begin{lemma}[Compatible Properties]
Given a R-GT system \(T = (\mathcal{L}, \mathcal{R})\) and \(\pmb{D} \subseteq \mathcal{G}^{\varoplus}(\mathcal{L})\), then \(T\) is (\textbf{locally}) \textbf{confluent} (\textbf{terminating}) on \(\pmb{D}\) iff \(e(T) = (e(\mathcal{L}), e(\mathcal{R}))\) is (\textbf{locally}) \textbf{confluent} (\textbf{terminating}) on \(e(D)\).
\end{lemma}

Recall from Remark \ref{remark:subgraphdfns} that there are two obvious definitions of subgraph. The alternative definition that insists on the inclusion being undefinedness preserving will actually be most useful to us in this context, since we want to insist that all subgraphs of a TLRG are themselves TLRGs. We will thus use that definition of subgraph throughout the remainder of this chapter.

\begin{definition} \label{dfn:rsubgraphclosure}
Let \(\pmb{D} \subseteq \mathcal{G}^{\varoplus}(\mathcal{L})\) be a set of abstract TLRGs. Then \(\pmb{D}\) is \textbf{subgraph closed} iff for all TLRG \(G\), \(H\), such that \(H \subseteq G\), if \([G] \in \pmb{D}\), then \([H] \in \pmb{D}\). The \textbf{subgraph closure} of \(\pmb{D}\), denoted \(\overbar{\pmb{D}}\), is the smallest set containing \(\pmb{D}\) that is \textbf{subgraph closed}.
\end{definition}

\begin{proposition}
Given \(\pmb{D} \subseteq \mathcal{G}^{\varoplus}(\mathcal{L})\), \(\overbar{\pmb{D}}\) always \textbf{exists}, and is \textbf{unique}. Moreover, \(\pmb{D} = \overbar{\pmb{D}}\) iff \(\pmb{D}\) is \textbf{subgraph closed}.
\end{proposition}

We now must setup the definition of a (non-garbage) critical pair. We will show that this definition is compatible across encoding, and also that (strong) joinability is.

\begin{definition} \label{dfn:rcritpair}
A pair of \textbf{direct derivations} \(G_1 \Leftarrow_{r_1,g_1} H \Rightarrow_{r_2,g_2} G_2\) is a \textbf{critical pair} iff:
\begin{enumerate}[itemsep=-0.4ex,topsep=-0.4ex]
\item \(H = g_1(L_1) \cup g_2(L_2)\);
\item The steps are not \textbf{parallelly independent};
\item If \(r_1 = r_2\) then \(g_1 \neq g_2\).
\end{enumerate}
\end{definition}

\begin{definition}
A \textbf{critical pair} \(H_1 \Leftarrow G \Rightarrow H_2\) is \textbf{non-garbage} w.r.t. \(\pmb{D}\) iff \([G] \in \overbar{\pmb{D}}\).
\end{definition}

\begin{lemma}[Compatible Non-Garbage Critical Pairs] \label{lem:compatcirtpairs}
Given a R-GT system \(T = (\mathcal{L}, \mathcal{R})\) and \(\pmb{D} \subseteq \mathcal{G}^{\varoplus}(\mathcal{L})\), a pair of \textbf{direct derivations} \(G_1 \Leftarrow_{r_1,g_1} H \Rightarrow_{r_2,g_2} G_2\) is a \textbf{non-garbage critical pair} w.r.t. \(\pmb{D}\) iff its encoding via \(e\) is a non-garbage critical pair w.r.t. \(e(D)\) in the sense of a T-GT system.
\end{lemma}

\begin{proof}
Due to Proposition \ref{prop:parindsim} and Lemma \ref{lem:fullyfaithful}.
\end{proof}

\begin{corollary}[Compatible Critical Pairs]
Given a R-GT system \(T = (\mathcal{L}, \mathcal{R})\), a pair of \textbf{direct derivations} \(G_1 \Leftarrow_{r_1,g_1} H \Rightarrow_{r_2,g_2} G_2\) is a \textbf{critical pair} iff its encoding via \(e\) is a non-garbage critical pair w.r.t. \(e(\mathcal{G}^{\varoplus}(\mathcal{L}))\) in the sense of a T-GT system.
\end{corollary}

\begin{theorem}[Compatible Strong Joinability] \label{thm:compatstrongjoin}
Given a R-GT system \(T = (\mathcal{L}, \mathcal{R})\), a \textbf{critical pair} is (\textbf{strongly}) \textbf{joinable} iff its encoding is.
\end{theorem}

\begin{proof}
Due to Theorem \ref{thm:compatderenc}, Corollary \ref{cor:trackcompat} and Lemma \ref{lem:compatcirtpairs}.
\end{proof}

Just like in the previous case, there are only finitely many (non-garbage) critical pairs, and so long as the sub-graph closure of \(\pmb{D}\) is decidable, we can generate them all.

\begin{lemma} \label{lem:rfincritpairs}
Given a R-GT system \(T = (\mathcal{L}, \mathcal{R})\) and \(\pmb{D} \subseteq \mathcal{G}^{\varoplus}(\mathcal{L})\), then there are only finitely many \textbf{non-garbage critical pairs} up to isomorphism.
\end{lemma}

\begin{proof}
Lemma \ref{lem:fincritpairs} combined with Lemma \ref{lem:compatcirtpairs}.
\end{proof}

\begin{corollary} \label{cor:rfincritpairs}
Given a R-GT system \(T = (\mathcal{L}, \mathcal{R})\) and \(\pmb{D} \subseteq \mathcal{G}^{\varoplus}(\mathcal{L})\) such that \(\overbar{\pmb{D}}\) is \textbf{decidable}, then one can find the finitely many \textbf{non-garbage critical pairs} in \textbf{finite time}.
\end{corollary}

\begin{proof}
By Lemma \ref{lem:rfincritpairs}, we have have only finitely many pairs to consider. Continue as in the proof of Corollary \ref{cor:fincritpairs}.
\end{proof}

\begin{corollary}
Let \(T = (\mathcal{L}, \mathcal{R})\), \(\pmb{D} \subseteq \mathcal{G}^{\varoplus}(\mathcal{L})\) be such that \(T\) is \textbf{terminating} on \(\pmb{D}\) and \(\overbar{\pmb{D}}\) is \textbf{decidable}. Then, one can \textbf{decide} if all the \textbf{non-garbage critical pairs} are \textbf{strongly joinable} (Definition \ref{dfn:strongjoin}).
\end{corollary}

\begin{proof}
By Corollary \ref{cor:rfincritpairs}, we can find the finitely many pairs in finite time. Continue as in the proof of Corollary  \ref{cor:gencritpairs}.
\end{proof}

\begin{remark}
Clearly \(e(\mathcal{G}^{\varoplus}(\mathcal{L}))\) is always decidable, so it is always possible to generate all the genuine critical pairs of a R-GT system.
\end{remark}

We are now, finally, ready to reproduce the (Non-Garbage) Critical Pair Lemma.

\begin{theorem}[Non-Garbage Critical Pair Lemma] \label{thm:rngcritpairlem}
Let \(T = (\mathcal{L}, \mathcal{R})\), \(\pmb{D} \subseteq \mathcal{G}^{\varoplus}(\mathcal{L})\). If all its \textbf{non-garbage critical pairs} are \textbf{strongly joinable}, then \(T\) is \textbf{locally confluent} on \(\pmb{D}\).
\end{theorem}

\begin{proof}
By compatibility Lemma \ref{lem:compatcirtpairs} and Theorem \ref{thm:compatstrongjoin}, we can simply apply the original Non-Garbage Critical Pair Lemma (Theorem \ref{thm:ngcritpairlem}).
\end{proof}

\begin{corollary}
Let \(T = (\mathcal{L}, \mathcal{R})\), \(\pmb{D} \subseteq \mathcal{G}^{\varoplus}(\mathcal{L})\). If \(T\) is \textbf{terminating} on \(\pmb{D}\), \(\pmb{D}\) is \textbf{closed} under \(T\), and all \(T\)'s \textbf{non-garbage critical pairs} are \textbf{strongly joinable} then \(T\) is \textbf{confluent} on \(\pmb{D}\).
\end{corollary}

\begin{proof}
By the above theorem, \(T\) is \textbf{locally confluent up to garbage}, so by the Newman-Garbage Lemma, \(T\) is \textbf{confluent up to garbage}.
\end{proof}


\section{Tree Recognition Revisited} \label{sec:treerecrev}

It is possible to rephrase the results from Section \ref{sec:linearrec} in terms of our new notion of garbage, where the rules \(r_0, r_1, r_2\) are from Figure \ref{fig:tree2}:

\begin{proposition} \label{prop:treegarb}
Let \(\mathcal{L} = (\{\Square, \triangle\}, \{\Square\})\) and \(\mathcal{R} = \{r_0, r_1, r_2\}\). Then, \(\pmb{D} = \{[G] \in \mathcal{G}^{\varoplus}(\mathcal{L}) \mid [G^{\varominus}] \in \pmb{L}(\pmb{TREE}), \abs{p_G^{-1}(\{1\})} = 1\}\) is \textbf{strongly closed} under \(T = (\mathcal{L}, \mathcal{R})\) and \(T\) is \textbf{confluent} on \(\pmb{E} = \{[G] \in \pmb{D} \mid l_G(V_G) = \{\Square\}\}\).
\end{proposition}

\begin{proof}
Strong closedness is due to Lemma \ref{lem:treegarbagesep} and confluence up to garbage due to Theorem \ref{thm:treerecthm}.
\end{proof}

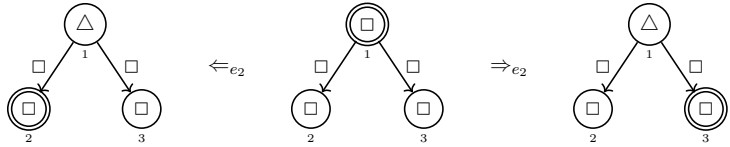
\begin{figure}[H]
\centering
\noindent
\input{fig/5/tree-pair}
\vspace{-0.3em}
\caption{Non-Strongly Joinable Critical Pair}
\label{fig:tree-pair}
\end{figure}

One would hope that we could then perform non-garbage critical pair analysis w.r.t. \(\pmb{D}\) (where \(\pmb{D}\) is as in Proposition \ref{prop:treegarb}) in order to demonstrate local confluence on \(\pmb{D}\). It turns out that every non-garbage critical pair is joinable, but unfortunately, one of them is not strongly joinable (Figure \ref{fig:tree-pair}), so we are unable to make any conclusion about local confluence up to garbage using the Non-Garbage Critical Pair Lemma (Theorem \ref{thm:rngcritpairlem}).

\vspace{0em}

Thus, just like Plump's original Critical Pair Lemma, strong joinability of (non-garbage) critical pairs is sufficient, but not necessary to imply local confluence (up to garbage). As discussed in the next chapter, it remains future work to develop stronger (non-garbage) critical pair analysis theorems.

%% file: fig/5/encoding.tex
\begin{tikzpicture}[every node/.style={inner sep=0pt, text width=6.5mm, align=center}]
    \node (a) at (0.0,-0.05) {$G$:};
    \node (b) at (1.0,0.0)   [draw, circle, thick] {x};
    \node (c) at (2.5,0.0)   [draw, circle, thick, pattern=north east lines] {x};
    \node (d) at (4.0,0.0)   [draw, circle, thick, double, double distance=0.4mm] {\,};

    \node (e) at (6.15,-0.05) {$e(G)$:};
    \node (f) at (7.5,0.0)   [draw, circle, thick] {$\Square$};
    \node (g) at (9.0,0.0)   [draw, circle, thick] {$\Square$};
    \node (h) at (10.5,0.0)  [draw, circle, thick] {$\Square$};

    \draw (b) edge[->] node[above, yshift=2pt] {y} (c)
          (c) edge[->,in=-25,out=-60,loop,thick] node [right, yshift=1pt, xshift=-3pt] {z} (c);

    \node (B) at (1.0,-0.52)  {\tiny{1}};
    \node (C) at (2.5,-0.52)  {\tiny{2}};
    \node (D) at (4.0,-0.52)  {\tiny{3}};
    \node (F) at (7.5,-0.52)  {\tiny{1}};
    \node (G) at (9.0,-0.52)  {\tiny{2}};
    \node (H) at (10.5,-0.52) {\tiny{3}};

    \draw (f) edge[->] node[above, yshift=2pt] {y} (g)
          (f) edge[->,in=-25,out=-60,loop,thick] node [right, yshift=1pt, xshift=-3pt] {x} (f)
          (f) edge[->,in=60,out=25,loop,thick] node [right, yshift=1pt, xshift=-3pt] {0} (f)
          (g) edge[->,in=-25,out=-60,loop,thick] node [right, yshift=1pt, xshift=-3pt] {x} (g)
          (g) edge[->,in=60,out=25,loop,thick] node [right, yshift=1pt, xshift=-3pt] {z} (g)
          (h) edge[->,in=60,out=25,loop,thick] node [right, yshift=1pt, xshift=-3pt] {1} (h);
\end{tikzpicture}

%% file: fig/5/tree-pair.tex
\scalebox{.75}{\begin{tikzpicture}[every node/.style={inner sep=0pt, text width=6.5mm, align=center}]
    \node (a) at (2.0,0.0)   [draw, circle, thick] {\(\triangle\)};
    \node (b) at (1.0,-1.5)  [draw, circle, thick, double, double distance=0.4mm] {\(\Square\)};
    \node (c) at (3.0,-1.5)  [draw, circle, thick] {\(\Square\)};

    \node (d) at (4.5,-0.8)  {$\Leftarrow_{e_2}$};

    \node (e) at (7.0,0.0)   [draw, circle, thick, double, double distance=0.4mm] {\(\Square\)};
    \node (f) at (6.0,-1.5)  [draw, circle, thick] {\(\Square\)};
    \node (g) at (8.0,-1.5)  [draw, circle, thick] {\(\Square\)};

    \node (h) at (9.5,-0.8)  {$\Rightarrow_{e_2}$};

    \node (i) at (12.0,0.0)  [draw, circle, thick] {\(\triangle\)};
    \node (j) at (11.0,-1.5) [draw, circle, thick] {\(\Square\)};
    \node (k) at (13.0,-1.5) [draw, circle, thick, double, double distance=0.4mm] {\(\Square\)};

    \node (A) at (2.0,-0.52)  {\tiny{1}};
    \node (B) at (1.0,-2.02)  {\tiny{2}};
    \node (C) at (3.0,-2.02)  {\tiny{3}};

    \node (E) at (7.0,-0.52)  {\tiny{1}};
    \node (F) at (6.0,-2.02)  {\tiny{2}};
    \node (G) at (8.0,-2.02)  {\tiny{3}};

    \node (I) at (12.0,-0.52) {\tiny{1}};
    \node (J) at (11.0,-2.02) {\tiny{2}};
    \node (K) at (13.0,-2.02) {\tiny{3}};

    \draw (a) edge[->,thick] node[above, yshift=-3pt, xshift=-9pt] {\(\Square\)} (b)
          (a) edge[->,thick] node[above, yshift=-3pt, xshift=9pt] {\(\Square\)} (c)
          (e) edge[->,thick] node[above, yshift=-3pt, xshift=-9pt] {\(\Square\)} (f)
          (e) edge[->,thick] node[above, yshift=-3pt, xshift=9pt] {\(\Square\)} (g)
          (i) edge[->,thick] node[above, yshift=-3pt, xshift=-9pt] {\(\Square\)} (j)
          (i) edge[->,thick] node[above, yshift=-3pt, xshift=9pt] {\(\Square\)} (k);
\end{tikzpicture}}

%% file: content/6.tex
\chapter{Conclusion}

\begin{chapquote}{Felix Klein, \textit{Unknown source}}
``In a sense, mathematics has been most advanced by those who distinguished themselves by intuition rather than by rigorous proofs.''
\end{chapquote}

We have reviewed the current state of graph transformation, with a particular focus on the \enquote{injective DPO} approach with relabelling and graph programming languages, establishing issues with the current approach to rooted graph transformation. We developed a new type of graph transformation system that supports relabelling and root nodes, but where derivations are invertible, and looked at a case study, showing that rooted graph transformation systems can recognise trees in linear time. This work on tree recognition was presented at CALCO 2019 \cite{Campbell-Courtehoute-Plump19b}. We have also defined some notions of equivalence for our new type of graph transformation system, and briefly discussed a possible theory of refinement.

In Chapter \ref{chapter:confluenceanalysis1}, we have introduced the new notion of confluence up to garbage for graph transformation systems, that allows us to have confluence, except in the cases we do not care about. Moreover, we have shown that it is sufficient to only analyse the non-garbage critical pairs to establish confluence up to garbage, and if said system is closed and terminating up to garbage, then we may conclude confluence up to garbage. We have shown closedness is undecidable in general, and applied our results to see that EFDs can be recognised by a system that is confluent up to garbage. Finally, in Chapter \ref{chapter:confluenceanalysis2}, we recovered notions of independence of derivations in our new notion of rooted GT systems with relabelling, along with the Local Church-Rosser Theorem and (Non-Garbage) Critical Pair Lemma, by encoding our rooted systems as conventional systems via a fully faithful functor.


\section{Evaluation}

We regard this project as a success, having achieved our four original goals as detailed in the Executive Summary. Our first goal was to review rooted DPO graph transformation with relabelling. We have done this in Chapter \ref{chapter:theoryintro}, looking at labelled GT systems with the DPO approach with injective matching, and how relabelling and root nodes have been implemented, providing further detail in Appendices \ref{appendix:ars} and \ref{appendix:transformation}. We also briefly reviewed graph languages and DPO-based graph programming languages.

Our second goal was to address the problem that the current theory of rooted graph transformation does not have invertible derivations. We have fixed this problem in Chapter \ref{chapter:newtheory} by defining rootedness using a partial function onto a two-point set rather than pointing graphs with root nodes. We have shown rule application corresponds to NDPOs, how Dodds' complexity theory applies in our system, and briefly discussed the equivalence of and refinement of GT systems.

Our third goal was to show a new example of how rooted graph transformation can be applied. We showed a new result that the graph class of trees can be recognised by a rooted GT system in linear time, given an input graph of bounded degree. Moreover, we have given empirical evidence by implementing the algorithm in GP\,2 and collecting timing results. The results are in the CALCO 2019 post-proceedings \cite{Campbell-Courtehoute-Plump19b}.

Our final goal was to develop new confluence analysis theory. We have defined a new notion of confluence up to garbage and non-garbage critical pairs, and shown that it is sufficient to require strong joinability of only the non-garbage critical pairs to establish confluence up to garbage, applying this theory to EFDs. Moreover, we have recovered the Local Church-Rosser Theorem and (Non-Garbage) Critical Pair Lemma for our new notion of rooted graph transformation with relabelling. We look to publish this work.


\section{Future Work}

Developing a fully-fledged theory of correctness and refinement for (rooted) GT systems remains future work, extending the work from Section \ref{sec:gtequiv}. Additionally, extending this notion to GP\,2, or other graph transformation based languages, and looking at the automated introduction of root nodes in order to improve time complexity remains open. Overcoming the restriction of host graphs to be of bounded degree in Theorem \ref{thm:fastderivations} remains open too.

Further exploring the relationship between (local) confluence up to garbage and closedness remains open work. In fact, confluence analysis of GT systems remains an underexplored area in general. Developing a stronger version of the Non-Garbage Critical Pair Lemma that allows for the detection of persistent nodes that need not be identified in the joined graph would allow conclusions of confluence up to garbage where it was previously not determined, remains future work. It is also unclear under what conditions one can decide if a graph is in the subgraph closure of a language, specified by a grammar, or otherwise.

Additional future work in the foundations of our new theory of rooted graph transformation would be to continue to establish the foundational theorems. We have shown the Local Church-Rosser Theorem and Critical Pair Lemma, however, the Parallelism and Concurrency Theorems \cite{Ehrig-Golas-Habel-Lambers-Orejas14a}, which have applications in database systems \cite{Ehrig-Kreowski80a}, algebraic specifications \cite{ParisiPresicce89a}, and logic programming \cite{Corradini-Rossi-ParisiPresicce91a}, remain future work. One may also want to recover a formalism of the Extension Theorem, which we didn't need when proving the Critical Pair Lemma \cite{Ehrig-Golas-Habel-Lambers-Orejas12a}.

Finally, it remains open research, to explore the overlap between graph transformation systems and the study of \enquote{reversible computation} \cite{Frank05a}. Our new foundations of rooted graph transformation allows for the specification of both efficient and reversible GT systems. Since graph transformation is a uniform way of expressing many problems in computer science, it is only natural that its applications in reversible computation is explored.

%% file: content/a.tex
\chapter{Basic Mathematical Notions} \label{appendix:notation}

There is not time to give an account of an axiomatization of set theory. Instead, we will simply state that we are using Von Neumann-Bernays-G\"odel Set Theory (NBG) \cite{Godel40a,Mendelson15a}. For the most part, a naive approach will suffice. We will use the word \enquote{collection} to refer, informally, to something that may or may not be a set. We will avoid the use of the word \enquote{class} to mitigate confusion with equivalence relations. When it comes to the definition of decidable (for sets of graphs) we will tacitly assuming a G{\"o}del numbering \cite{Godel31a}.


\section{Sets I}

We split the \enquote{Sets} section into two halves. This section is derived from Chapters 2 and 3 of \cite{Sutherland09a} and Chapter 3 of \cite{Awodey10a}.

\begin{definition}
We let \(\emptyset\) denote the \textbf{empty set}. If \(A\) is a \textbf{set}, then we write \(a \in A\) to say that \(a\) \enquote{belongs to} \(A\). We say that \(B\) is a \textbf{subset} of \(A\), \(B \subseteq A\) iff \(\forall x \in B, x \in A\). We say \(A = B\) iff \(A \subseteq B\) and \(B \subseteq A\).
\end{definition}

\begin{definition}
If \(A, B\) are sets, then we define:
\begin{enumerate}[itemsep=-0.4ex,topsep=-0.4ex]
\item \textbf{Set union}: \(A \cup B = \{x \mid x \in A \text{ or } x \in B\}\);
\item \textbf{Set intersection}: \(A \cap B = \{x \mid x \in A \text{ and } x \in B\}\);
\item \textbf{Set difference}: \(A \setminus B = \{x \mid x \in A \text{ and } x \not\in B\}\);
\item \textbf{Cartesian product}: \(A \times B = \{(a, b) \mid a \in A \text{ and } b \in B\}\);
\item \textbf{Power set}: \(\mathcal{P}(A) = \{X \mid X \subseteq A\}\), \(\mathcal{P}_1(A) = \mathcal{P}(A) \setminus \emptyset\).
\end{enumerate}
\end{definition}

\begin{definition}
We define the \textbf{disjoint union} of sets \(A, B\) to be their \textbf{coproduct} in the category of sets (Example \ref{eg:catsets}). It can be convenient to define this as \(A + B = (A \times \{1\}) \cup (B \times \{2\})\), or simply to assume that \(A \cap B = \emptyset\) and define \(A + B = A \cup B\).
\end{definition}

\begin{definition}
Let \(\mathbb{N} = \{0, 1, 2, \ldots\}\), and \(\mathbb{Z} = \{\ldots, -1, 0, 1, \ldots\}\).
\end{definition}


\section{Functions}

This section is derived from Chapters 2 and 3 of \cite{Sutherland09a} and Chapter 1 of \cite{Howie95a}. We use the conventional order of composition.

\begin{definition}
Let \(A, B\) be sets. A \textbf{function} \(f\) from \textbf{domain} \(A\) to \textbf{codomain} \(B\) is a rule which assigns to each \(a \in A\) a \textbf{unique} \(b \in B\). We write \(b = f(a)\), \(f: A \to B\), and call \(a\) the \textbf{argument} of \(f\). Formally, a \textbf{function} from \(A\) to \(B\) is a subset of \(A \times B\) such that for each \(a \in A\) there is \textbf{exactly one} element \((a, b)\) in \(f\).
\end{definition}

\begin{definition}
Let \(A, B, C, D\) be sets. If \(f: A \to B\), \(g: C \to D\) are \textbf{functions}, then \(f\) and \(g\) are \textbf{equal} (\(f = g\)) iff they are equal as sets.
\end{definition}

\begin{definition}
Let \(A, B, C\) be sets. If \(f: A \to B\), \(g: B \to C\) are \textbf{functions}, then we form a new \textbf{function} \((g \circ f): A \to C\) the \textbf{composite} of \(f\) and \(g\) by the rule \((g \circ f)(a) = g(f(a))\).
\end{definition}

\begin{proposition}
\textbf{Composition} of functions is \textbf{associative}. That is, given \(f: A \to B\), \(g: B \to C\), \(h: C \to D\), then \(h \circ (g \circ f) = (h \circ g) \circ f\).
\end{proposition}

\begin{definition}
For any set \(A\), the \textbf{identity function} on \(A\), \(id_A: A \to A\) is defined by \(\forall a \in A, id_A(a) = a\).
\end{definition}

\begin{proposition}
If \(f: B \to A\), then \(id_A \circ f = f\). If \(g: A \to C\), \(g \circ id_A= g\).
\end{proposition}

\begin{definition}
Let \(f: A \to B\) be a \textbf{function}. Then a \textbf{function} \(g: B \to A\) is the \textbf{inverse} of \(f\) iff \(g \circ f = id_A\) and \(f \circ g = id_B\)
\end{definition}

\begin{proposition}
Let \(f: A \to B\) be a \textbf{function}. Then, if an \textbf{inverse} exists, it is unique, and is denoted \(f^{-1}: B \to A\).
\end{proposition}

\begin{definition}
Let \(f: A \to B\) be a \textbf{function}. Then \(f\) is \textbf{injective} iff \(\forall a, b \in A, f(a) = f(b)\) implies \(a = b\). \(f\) is \textbf{surjective} iff \(\forall a \in A, \exists b \in B, f(a) = b\). If \(f\) satisfies both properties, then it is \textbf{bijective}.
\end{definition}

\begin{lemma}
A \textbf{function} has an \textbf{inverse} iff it is a \textbf{bijection}.
\end{lemma}

\begin{definition}
Let \(f: A \to B\) be a \textbf{function}, \(X \subseteq A\), and \(Y \subseteq B\). Then the \textbf{image} of \(A\) under \(f\) is \(f(A) = \{f(a) \mid a \in A\} \subseteq B\), and the \textbf{preimage} of \(B\) is \(f^{-1}(B) = \{a \in A \mid f(a) \in B\} \subseteq A\).
\end{definition}

\begin{remark}
This does not imply the existence of an inverse, but if it does exist, then preimage of \(f\) coincides with the image of \(f^{-1}\).
\end{remark}

\begin{definition}
Let \(f: A \to B\) be a \textbf{function}, and \(X \subseteq A\). Then the \textbf{restriction} of \(f\) to \(X\) is \(\restr{f}{X}: X \to B\) is defined by \(\forall x \in X, \restr{f}{X}(x) = f(x)\).
\end{definition}

\begin{definition}
A \textbf{partial function} \(f: A \to B\) is a subset \(f\) of \(A \times B\) such that there is \textbf{at most one} element \((a, b)\) in \(f\).
\end{definition}


\section{Binary Relations}

This section is derived from Chapter 2 of \cite{Sutherland09a}, Chapter 1 of \cite{Howie95a} and Appendix A of \cite{Baader-Nipkow98a}.

\begin{definition} \label{def:binrel}
Let \(A\) be a set. Then a \textbf{binary relation} on \(A\) is a subset \(R\) of \(A \times A\). For any \(a, b \in A\), we write \(a R b\) iff \((a, b) \in R\).
\end{definition}

\begin{definition}
Let \(A\) be a set. Then, the \textbf{identity relation} on \(A\) is \(\iota_A = \{(a, a) \mid a \in A\}\), and the \textbf{universal relation} on \(A\) is \(\omega_A = A \times A\).
\end{definition}

\begin{definition}
Let \(A\) be a set. Then we call a \textbf{binary relation} \(R\) on \(A\) \textbf{functional} iff for any \(a, b, c \in A\), \(a R b\) and \(a R c\) implies \(b = c\).
\end{definition}

\begin{definition} \label{dfn:relcomposition}
Let \(A\) be a set, and \(R, S\) be \textbf{binary relations} on \(A\). Then the \textbf{composition} of \(R\) and \(S\) is \(S \circ R = \{(x, y) \in A \times A \mid \exists z \in A \text{ with } x R z \text{ and } z S y\}\). Define \(R^0 = \iota_A\), and \(\forall n \in \mathbb{N}^+, R^n = R \circ R^{n-1}\).
\end{definition}

\begin{definition}
Let \(A\) be a set. Then the \textbf{inverse} of a \textbf{binary relation} \(R\) on \(A\) is \(R^{-1} = \{(b, a) \in A \times A \mid a R b\}\).
\end{definition}

\begin{proposition}
When considered as \textbf{binary relations}, \textbf{functions} and \textbf{partial functions} are \textbf{functional}. Moreover, the definitions of \textbf{composition} and \textbf{inverses} coincide.
\end{proposition}

\begin{definition}
A \textbf{binary relation} \(R\) on \(A\) is:
\begin{enumerate}[itemsep=-0.4ex,topsep=-0.4ex]
\item \textbf{Reflexive} iff \(\iota_A \subseteq R\);
\item \textbf{Irreflexive} iff \(\iota_A \cap R = \emptyset\);
\item \textbf{Symmetric} iff \(R = R^{-1}\);
\item \textbf{Antisymmetric} iff \(R \cap R^{-1} \subseteq \iota_A\);
\item \textbf{Transitive} iff \(R \circ R \subseteq R\);
\item \textbf{Connex} iff \(\omega_A \setminus \, \iota_A \subseteq R \cup R^{-1}\).
\end{enumerate}
\end{definition}

\begin{definition} \label{dfn:preorder}
A \textbf{binary relation} is a \textbf{preorder} iff it is \textbf{reflexive} and \textbf{transitive}. A \textbf{symmetric preorder} is called an \textbf{equivalence}.
\end{definition}

\begin{proposition}
The \textbf{classes} \([a] = \{b \in A \mid a \sim b\}\) of an \textbf{equivalence} \(\sim\) on \(A\) \textbf{partition} \(A\) into a union of pairwise disjoint non-empty subsets.
\end{proposition}

\begin{definition} \label{dfn:quotient}
Given an \textbf{equivalence} \(\sim\), define \(A /\!\!\sim \,= \{[a] \mid a \in A\}\).
\end{definition}


\section{Orders}

This section is derived from Chapter 1 of \cite{Howie95a} and Appendix A of \cite{Baader-Nipkow98a}.

\begin{definition}
An \textbf{antisymmetric preorder} \(\leq\) on \(X\) is called a \textbf{partial order}, and we call \((X, \leq)\) a \textbf{partially ordered set} (\textbf{poset}).
\end{definition}

\begin{definition}
A \textbf{partial order} satisfying the \textbf{connex} property is called a \textbf{total order}, giving a \textbf{totally ordered set}.
\end{definition}

\begin{definition}
A \textbf{strict order} is an \textbf{irreflexive}, \textbf{transitive} relation.
\end{definition}

\begin{proposition}
Every \textbf{partial order} \(\leq\) induces a \textbf{strict order} \(\leq \setminus \,\, \iota\), and every \textbf{strict order} \(<\) induces a \textbf{partial order} \(< \cup \,\, \iota\).
\end{proposition}

\begin{definition}
Let \((X, \leq)\) be a poset, and \(\emptyset \neq Y \subseteq X\). Then:
\begin{enumerate}[itemsep=-0.4ex,topsep=-0.4ex]
\item \(a \in Y\) is \textbf{minimal} iff \(\forall y \in Y, y \leq a\) implies \(y = a\);
\item \(b \in Y\) is the \textbf{minimum} iff \(\forall y \in b \leq y\);
\item \(c \in X\) is a \textbf{lower bound} for \(Y\) iff \(\forall y \in Y, c \leq y\).
\end{enumerate}
\end{definition}

\begin{proposition}
Let \((X, \leq)\) be a poset, and \(\emptyset \neq Y \subseteq X\). Then every \textbf{minimum} element of \(Y\) is \textbf{minimal}, \(Y\) and has \textbf{at most} one \textbf{minimum}.
\end{proposition}

\begin{definition} \label{dfn:wellfounded}
We say that the poset \((X, \leq)\) satisfies the \textbf{minimal condition} (\textbf{well-founded}) iff every non-empty subset of \(X\) has a \textbf{minimal} element. If \(\leq\) is also a \textbf{total order}, then we say it is \textbf{well-ordered}.
\end{definition}

\begin{definition} \label{dfn:monotone}
Let \((X, \leq_X)\), \((Y, \leq_Y)\) be posets. Then a function \(\varphi: X \to Y\) is called \textbf{monotone} iff \(a \leq_X b\) implies \(\varphi(a) \leq_Y \varphi(b)\).
\end{definition}


\section{Sets II}

This section is derived from Chapter 7 of \cite{Weiss08a}, Part I of \cite{Munkres18a}, Chapter 8 of \cite{Martin11a}, and Chapter 1 of \cite{Weihrauch13a}.

\begin{theorem}[Well-Ordered Sets]
Every set can be \textbf{well-ordered}, and every \textbf{well-ordered} set is \textbf{isomorphic} to an \textbf{ordinal} (see \cite{Weiss08a} for details).
\end{theorem}

\begin{proposition}
We define the \textbf{cardinality} of \(A\) (\(\abs{A}\)), to be the \textbf{least ordinal} \(\kappa\) such that there is some \textbf{bijection} \(f: A \to \kappa\). Every set has \textbf{unique} cardinality. All sets with \textbf{cardinality} \(\leq\) to that of \(\mathbb{N}\) are \textbf{countable}.
\end{proposition}

\begin{theorem}[Countable Sets] \label{thm:setprod}
The \textbf{Cartesian product} of two \textbf{countable} sets is \textbf{countable}, and a \textbf{countable union} of \textbf{countable} sets is \textbf{countable}. If \(A\) is \textbf{finite}, then \(\mathcal{P}(A)\) is \textbf{finite}. The set \(\mathcal{P}(\mathbb{N})\) is \textbf{uncountable}.
\end{theorem}

\begin{definition}
A \textbf{partial function} \(f: \mathbb{N} \to \mathbb{N}\) is \textbf{computable} iff there exists a \textbf{Turing Machine} that computes \(f\) (see \cite{Martin11a} for details).
\end{definition}

\begin{definition} \label{dfn:decidable}
A \textbf{countable} set \(A \subseteq \mathbb{N}\) has \textbf{characteristic function} \(\chi_A: \mathbb{N} \to \{0, 1\}\) defined by \(\forall x \in \mathbb{N}, \chi_A(x) = 1\) iff \(x \in A\). \(A\) is \textbf{decidable} or \textbf{recursive} iff \(\chi_A\) is \textbf{computable}. Otherwise, \(A\) is \textbf{undecidable}.
\end{definition}

\begin{definition} \label{dfn:semidecidable}
\(A \subseteq \mathbb{N}\) is \textbf{semidecidable} or \textbf{recursively enumerable} iff it is the \textbf{domain} of a \textbf{computable} function.
\end{definition}

%% file: content/b.tex
\chapter{Category Theory} \label{appendix:cats}

The definitions and theorems in this appendix are derived Chapter 1 of \cite{Lane78a}, Appendix A of \cite{Ehrig-Ehrig-Prange-Taentzer06a}, Chapter 1 of \cite{Awodey10a}, and Chapter 1 of \cite{Leinster14a}.


\section{Foundations}

\begin{definition} \label{def:cat}
A \textbf{category} \(\pmb{C}\) consists of the following data:
\begin{enumerate}[itemsep=-0.4ex,topsep=-0.4ex]
\item \textbf{Objects}: \(A\), \(B\), \(C\), \dots
\item \textbf{Arrows}: \(f\), \(g\), \(h\), \dots
\item For each arrow \(f\), there are given objects \(\operatorname{dom}(f)\), \(\operatorname{cod}(f)\), and we write \(f: A \to B\) to indicate that \(A = \operatorname{dom}(f)\), \(B = \operatorname{cod}(f)\);
\item Given arrows \(f: A \to B\), \(g: B \to C\), \(g \circ f\) is an arrow such that \(A = \operatorname{dom}(g \circ f)\), \(C = \operatorname{cod}(g \circ f)\);
\item For each object \(A\) there is given an arrow \(1_A: A \to A\);
\end{enumerate}
such that for all arrows \(f: A \to B, g: B \to C, h: C \to D\), we have \(h \circ (g \circ f) = (h \circ g) \circ f\) and \(f \circ 1_A = f = 1_B \circ f\). We write \(A \in \pmb{C}\) when \(A\) is an object in \(\pmb{C}\).
\end{definition}

\begin{definition}
We call an arrow \(f: A \to B\) \textbf{monic} (\textbf{epic}) iff it is \textbf{left} (\textbf{right}) \textbf{cancellative}. We call \(f\) an \textbf{isomorphism} and write \(A \cong B\) iff \(f\) admits a two-sided inverse.
\end{definition}

\begin{definition} \label{def:catsmall}
A category \(\pmb{C}\) is called \textbf{small} iff the collection of all arrows is a \textbf{set}. Otherwise, it is called \textbf{large}. \(\pmb{C}\) is called \textbf{locally small} iff the collection of arrows between any two objects (\(\textrm{Hom}_{\pmb{C}}(X, Y)\)) is a \textbf{set}.
\end{definition}

\begin{proposition}
Every \textbf{small} category is \textbf{locally small}. Moreover, a \textbf{locally small} category is \textbf{small} iff the collection of all objects is a \textbf{set}.
\end{proposition}

\begin{definition} \label{def:catfinitary}
A category is \textbf{finitary} iff there are only finitely many \textbf{subobjects} up to isomorphism, where \(A\) is a \textbf{subobject} of \(B\) iff there is a monomorphism \(A \to B\).
\end{definition}

\begin{example} \label{eg:catsets}
The collection of all sets as objects and all functions between them as arrows forms a locally small category. The collection of all finite sets and functions between them is a locally small, finitary category.
\end{example}

\begin{definition} \label{dfn:functor}
A \textbf{functor} \(F: \pmb{C} \to \pmb{D}\) between categories \(\pmb{C}\), \(\pmb{D}\) is a mapping such that:
\begin{enumerate}[itemsep=-0.4ex,topsep=-0.4ex]
\item \(F(f\!: A \to B) = F(f)\!: F(A) \to F(B)\);
\item \(F(1_A) = 1_{F(A)}\);
\item \(F(g \circ f) = F(g) \circ F(f)\).
\end{enumerate}
\end{definition}

\begin{definition}
A (covariant) \textbf{functor} \(F: \pmb{C} \to \pmb{D}\) between two locally small categories \(\pmb{C}, \pmb{D}\) is \textbf{faithful} (\textbf{full}) if the induced function \(F_{X,Y}: \textrm{Hom}_{\pmb{C}}(X, Y) \to \textrm{Hom}_{\pmb{D}}(F(X), F(Y))\) is \textbf{injective} (\textbf{surjective}) for all objects \(X, Y\) in \(\pmb{C}\).
\end{definition}


\section{Pushouts and Pullbacks}

\vspace{-0.5em}
\begin{figure}[H]
\centering
\noindent
\input{fig/b/limits}
\vspace{-1.4em}
\caption{Pushout and Pullback}
\vspace{0.1em}
\end{figure}

\begin{definition} \label{dfn:pushout}
Given arrows \(A \to B\) and \(A \to C\), an object \(D\) together with arrows \(B \to D\) and \(C \to D\) is a \textbf{pushout} iff:
\begin{enumerate}[itemsep=-0.4ex,topsep=-0.4ex]
\item \textbf{Commutativity}: \(A \to B \to D = A \to C \to D\).
\item \textbf{Universal property}: For all arrows \(B \to D'\), \(C \to D'\) such that \(A \to B \to D' = A \to C \to D'\), there is a unique arrow \(D \to D'\) such that \(B \to D \to D' = B \to D'\) and \(C \to D \to D' = C \to D'\).
\end{enumerate}
\end{definition}

\begin{definition} \label{dfn:pullback}
Given arrows \(B \to D\) and \(C \to D\), an object \(A\) together with arrows \(A \to B\) and \(A \to C\) is a \textbf{pullback} iff:
\begin{enumerate}[itemsep=-0.4ex,topsep=-0.4ex]
\item \textbf{Commutativity}: \(A \to B \to D = A \to C \to D\).
\item \textbf{Universal property}: For all arrows \(A' \to B\), \(A' \to C\) such that \(A' \to B \to D = A' \to C \to D\), there is a unique arrow \(A' \to A\) such that \(A' \to A \to B = A' \to B\) and \(A' \to A \to C = A' \to C\).
\end{enumerate}
\end{definition}

\begin{definition} \label{dfn:npo}
A \textbf{pushout} that is a \textbf{pullback} is called a \textbf{natural pushout}.
\end{definition}

\begin{theorem}[Limit Uniqueness]
If they exist, a \textbf{pushout} (\textbf{pullback}), \(D\) (\(A\)) are unique up isomorphism.
\end{theorem}

\begin{definition}
Given arrows \(A \to B\) and \(B \to D\), a (natural) \textbf{pushout complement} is an object \(C\) together with arrows \(A \to C\) and \(C \to D\) such that the resulting square is a (natural) \textbf{pushout}.
\end{definition}

%% file: fig/b/limits.tex
\begin{equation*}
\begin{tikzcd}
  A \arrow[r] \arrow[d]
    & B \arrow[d] \arrow[ddr, bend left=20]
    & \text{}
    & A' \arrow[drr, bend left=20] \arrow[ddr, bend right=20] \arrow[dr, dashed]
    & \text{}
    & \text{} \\
  C \arrow[r] \arrow[drr, bend right=20]
    & D \arrow[dr, dashed]
    & \text{}
    & \text{}
    & A \arrow[r] \arrow[d]
    & B \arrow[d] \\
  \text{}
    & \text{}
    & D'
    & \text{}
    & C \arrow[r]
    & D
\end{tikzcd}
\end{equation*}

%% file: content/c.tex
\chapter{Abstract Reduction Systems} \label{appendix:ars}

The definitions and theorems in this appendix are derived from Chapter 2 of \cite{Baader-Nipkow98a}, Section 2.2 of \cite{Plump99a}, and Section 1.1 of \cite{Book-Otto93a}.


\section{Foundations}

\begin{definition} \label{dfn:ars0}
An \textbf{abstract reduction system} (ARS) is a pair \((A, \rightarrow)\) where \(A\) is a \textbf{set} and \(\rightarrow\) a \textbf{binary relation} on \(A\).
\end{definition}

\begin{definition} \label{dfn:ars1}
Let \((A, \rightarrow)\) be an ARS. We define the notation:
\begin{enumerate}[itemsep=-0.4ex,topsep=-0.4ex]
\item \textbf{Composition}: \(\xrightarrow{n} \defeq \rightarrow^n\) (\(n \geq 0\));
\item \textbf{Transitive closure}: \(\xrightarrow{+} \defeq \bigcup_{n \geq 1} \xrightarrow{n}\);
\item \textbf{Reflexive transitive closure}: \(\xrightarrow{*} \defeq \xrightarrow{+} \cup \xrightarrow{0}\);
\item \textbf{Reflexive closure}: \(\xrightarrow{=} \defeq \rightarrow \cup \xrightarrow{0}\);
\item \textbf{Inverse}: \(\leftarrow \defeq \rightarrow^{-1}\);
\item \textbf{Symmetric closure}: \(\leftrightarrow \defeq \rightarrow \cup \leftarrow\).
\end{enumerate}
\end{definition}

\begin{remark}
It is usual, in practice, that \(\rightarrow\) is \textbf{decidable}. This \textbf{does not} imply that \(\xrightarrow{+}\) is \textbf{decidable}, only that \(\xrightarrow{n}\) is \textbf{decidable}.
\end{remark}

\begin{definition} \label{dfn:ars2}
Let \((A, \rightarrow)\) be an ARS. We say that:
\begin{enumerate}[itemsep=-0.4ex,topsep=-0.4ex]
\item \(x\) is \textbf{reducible} iff there is a \(y\) s.t. \(x \rightarrow y\);
\item \(x\) is \textbf{in normal form} iff \(x\) is not reducible;
\item \(y\) is \textbf{a normal form} of \(x\) iff \(x \xrightarrow{*} y\) and \(y\) is in normal form. If \(x\) has a \textbf{unique normal form}, it is denoted \(x\!\!\downarrow\);
\item \(y\) is a \textbf{successor} to \(x\) iff \(x \xrightarrow{+} y\), and a \textbf{direct successor} iff \(x \rightarrow y\);
\item \(x\) and \(y\) are \textbf{joinable} iff there is a \(z\) s.t. \(x \xrightarrow{*} z \xleftarrow{*} y\). We write \(x \downarrow y\).
\end{enumerate}
\end{definition}


\section{Termination}

\begin{definition} \label{dfn:arsterm}
Let \((A, \rightarrow)\) be an ARS. Then \(\rightarrow\) is called \textbf{terminating} iff there is no infinite descending chain \(x_0 \rightarrow x_1 \rightarrow \dots\).
\end{definition}

\begin{remark}
Other texts call a \textbf{terminating} reduction \textbf{uniformly terminating} or \textbf{Noetherian}, or say it satisfies the \textbf{descending chain condition}.
\end{remark}

The principle of \textbf{Noetherian induction} (\textbf{well-founded induction}) is a generalisation of induction from \((\mathbb{N}, >)\) to any \textbf{terminating} reduction system.

\begin{definition}
Let \((A, \rightarrow)\) be an ARS, and \(P\) is some property of the elements of \(A\). Then the inference rule for \textbf{Noetherian Induction} is:
\noindent
\vspace{-0.2em}
\begin{align*}
\infer{\forall x \in A, P(x)}{\forall x \in A, (\forall y \in A, x \xrightarrow{+} y \Rightarrow P(y)) \Rightarrow P(x)}
\end{align*}
\end{definition}

\begin{theorem}[Noetherian Induction] \label{thm:ninduct}
Let \((A, \rightarrow)\) The following are equivalent for an ARS:
\begin{enumerate}[itemsep=-0.4ex,topsep=-0.4ex]
\item The principle of \textbf{Noetherian induction} holds;
\item \(\rightarrow\) is \textbf{well-founded} (Definition \ref{dfn:wellfounded});
\item \(\rightarrow\) is \textbf{terminating} (Definition \ref{dfn:arsterm}).
\end{enumerate}
\end{theorem}

\begin{definition} \label{dfn:branching}
Let \((A, \rightarrow)\) be an ARS. Then \(\rightarrow\) is called
\begin{enumerate}[itemsep=-0.4ex,topsep=-0.4ex]
\item \textbf{Finitely branching} iff each \(a\) has only finitely many direct successors;
\item \textbf{Globally finite} iff each \(a\) has only finitely many successors;
\item \textbf{Acyclic} iff there is no \(a\) such that \(a \xrightarrow{+} a\).
\end{enumerate}
\end{definition}

\begin{lemma} \label{lem:arsbranching}
Let \((A, \rightarrow)\) be an ARS. Then:
\begin{enumerate}[itemsep=-0.4ex,topsep=-0.4ex]
\item If \(\rightarrow\) is \textbf{finitely branching} and \textbf{terminating}, then it is \textbf{globally finite};
\item If \(\rightarrow\) is \textbf{acyclic} and \textbf{globally finite}, then it is \textbf{terminating};
\item \(\rightarrow\) is \textbf{acyclic} iff \(\xrightarrow{+}\) is a \textbf{strict order}.
\end{enumerate}
\end{lemma}

\begin{definition}
Let \((A, \rightarrow)\) be an ARS. Then \(\rightarrow\) is called \textbf{normalising} iff every element has a \textbf{normal form} (Definition \ref{dfn:ars2}).
\end{definition}

\begin{lemma}
Let \((A, \rightarrow)\) be an ARS. If \(\rightarrow\) is \textbf{terminating}, then every element has at least one \textbf{normal form}.
\end{lemma}

\begin{lemma} \label{lem:monoembedding}
A \textbf{finitely branching} reduction \textbf{terminates} iff there is a \textbf{monotone} (Definition \ref{dfn:monotone}) embedding into \((\mathbb{N}, >)\).
\end{lemma}

\begin{remark}
The \textbf{union} of two \textbf{terminating} reductions need not be \textbf{terminating}.
\end{remark}


\section{Confluence}

\vspace{-0.3em}
\begin{figure}[H]
\begin{subfigure}{.5\textwidth}
    \centering
    \includegraphics[totalheight=3.0cm]{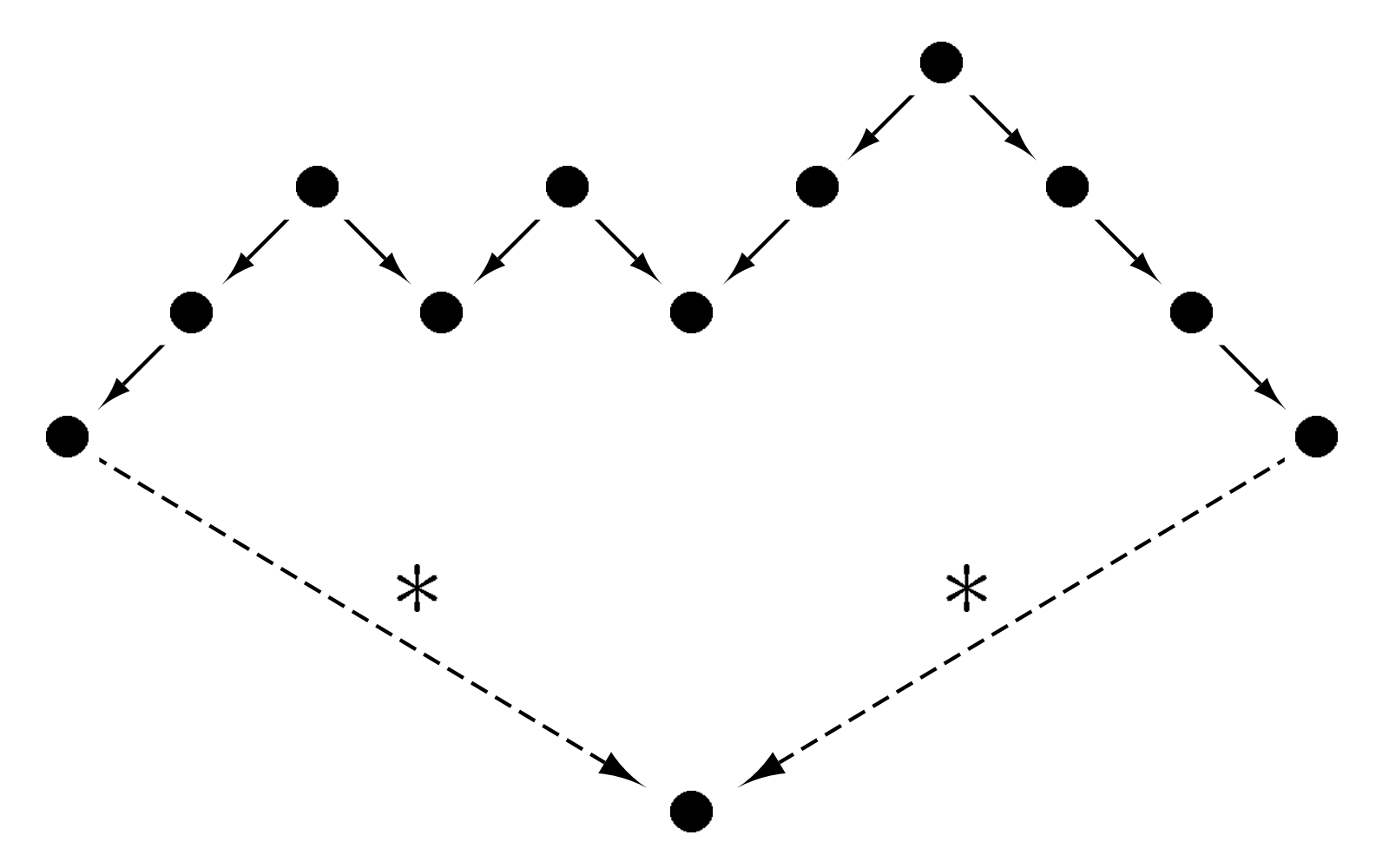}
    \vspace{-0.2em}
    \caption{Church-Rosser}
\end{subfigure}
\begin{subfigure}{.5\textwidth}
    \centering
    \includegraphics[totalheight=3.0cm]{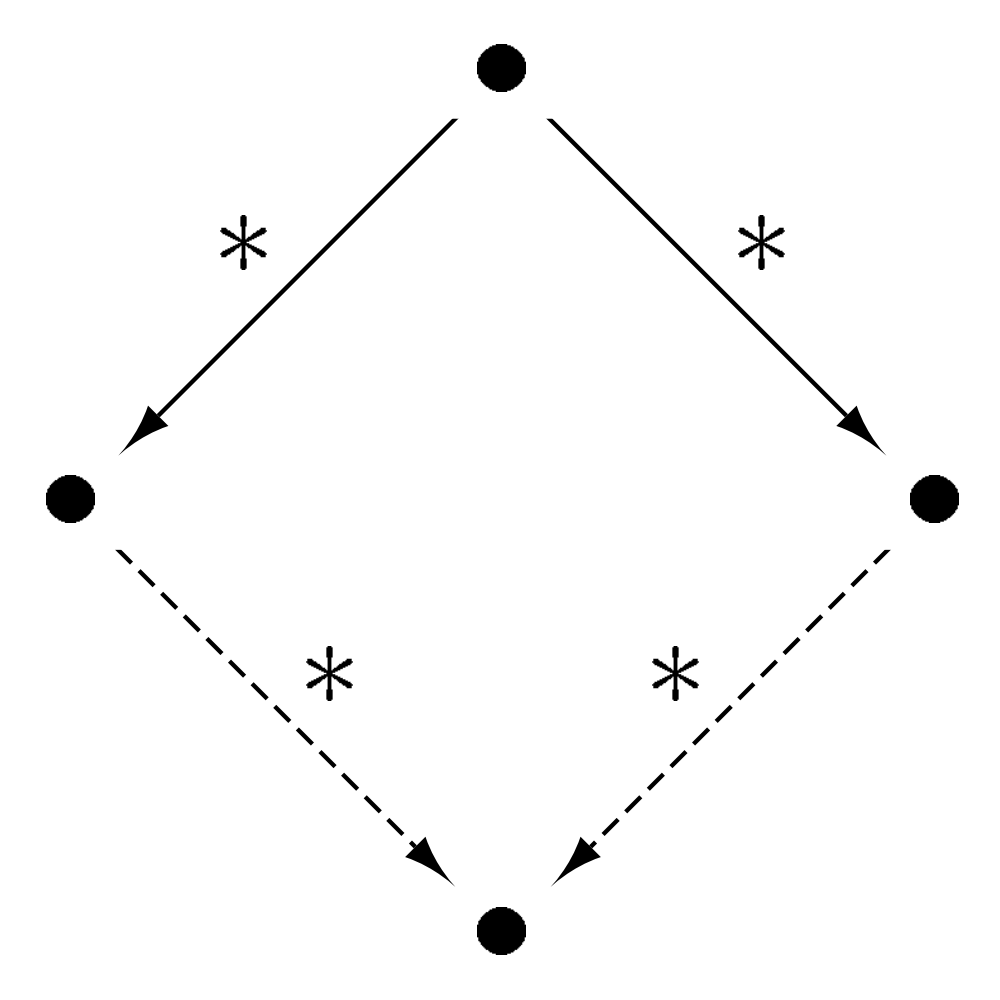}
    \vspace{-0.2em}
    \caption{Confluence}
\end{subfigure}
\begin{subfigure}{.5\textwidth}
    \centering
    \vspace{2.2em}
    \includegraphics[totalheight=3.0cm]{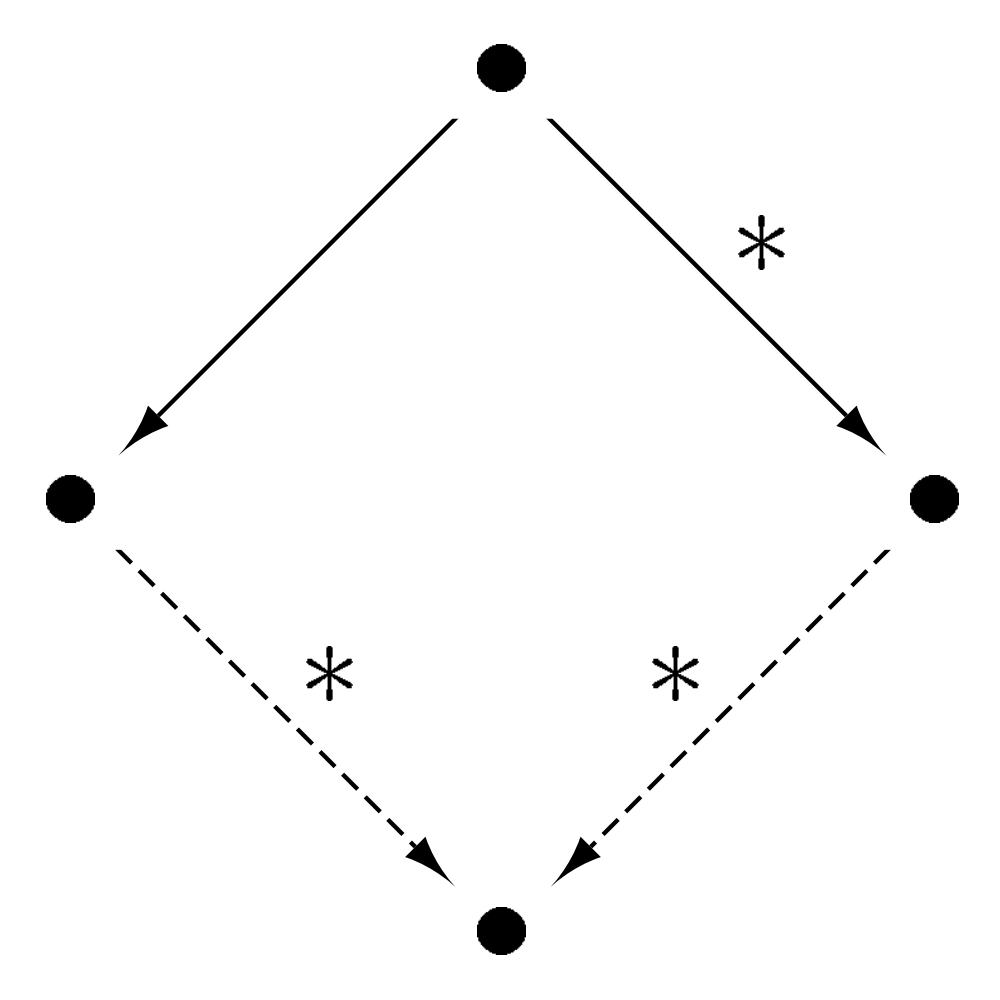}
    \vspace{-0.2em}
    \caption{Semi-Confluence}
\end{subfigure}
\begin{subfigure}{.5\textwidth}
    \centering
    \vspace{2.2em}
    \includegraphics[totalheight=3.0cm]{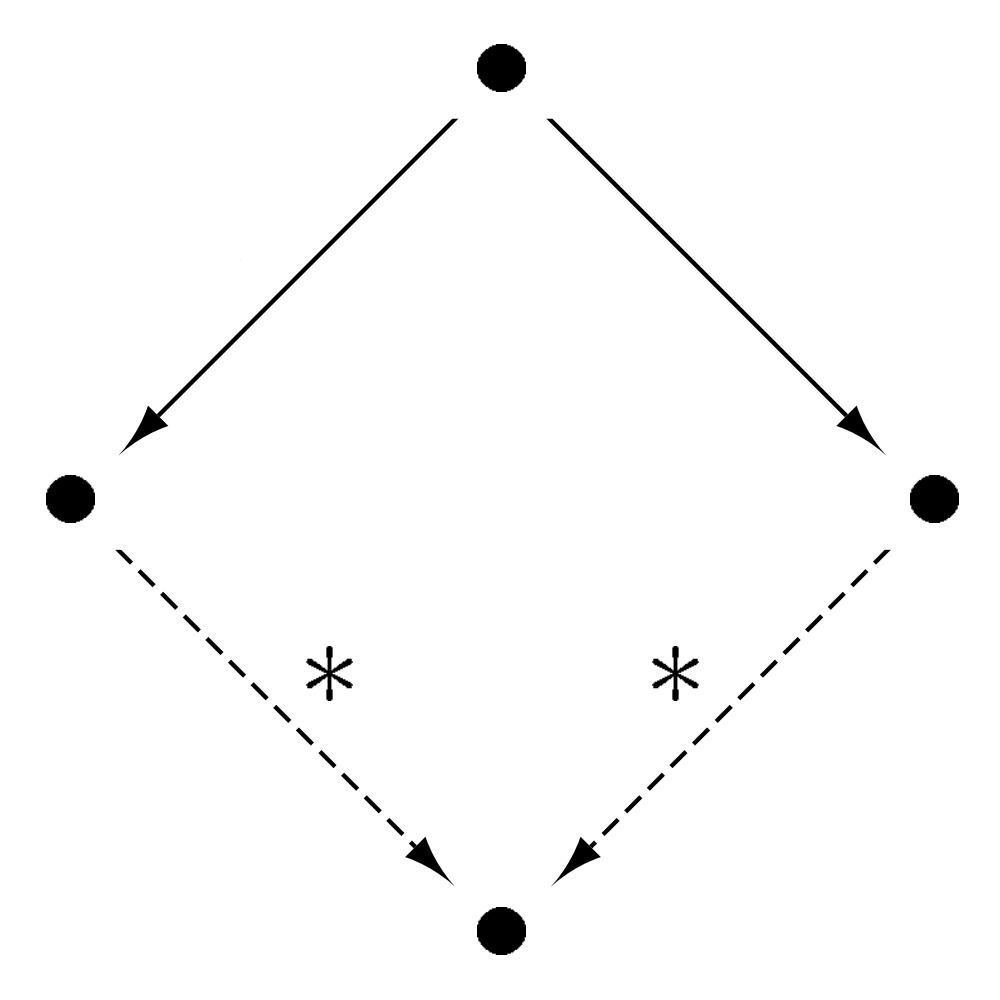}
    \vspace{-0.2em}
    \caption{Local Confluence}
\end{subfigure}
\vspace{1.0em}
\caption{Confluence Properties}
\vspace{0.2em}
\end{figure}

\begin{definition} \label{dfn:arsconf}
Let \((A, \rightarrow)\) be an ARS. Then \(\rightarrow\) is called:
\begin{enumerate}[itemsep=-0.4ex,topsep=-0.4ex]
\item \textbf{Church-Rosser} iff \(x \xleftrightarrow{*} y\) implies \(x \downarrow y\);
\item \textbf{Confluent} iff \(y_1 \xleftarrow{*} x \xrightarrow{*} y_2\) implies \(y_1 \downarrow y_2\);
\item \textbf{Semi-confluent} iff \(y_1 \leftarrow x \xrightarrow{*} y_2\) implies \(y_1 \downarrow y_2\);
\item \textbf{Locally confluent} iff \(y_1 \leftarrow x \rightarrow y_2\) implies \(y_1 \downarrow y_2\);
\item \textbf{Convergent} iff it is both confluent and terminating.
\end{enumerate}
\end{definition}

\begin{theorem}[Church-Rosser]
Let \((A, \rightarrow)\) be an ARS. Then, the following are equivalent:
\begin{enumerate}[itemsep=-0.4ex,topsep=-0.4ex]
\item \(\rightarrow\) is \textbf{Church-Rosser};
\item \(\rightarrow\) is \textbf{semi-confluent};
\item \(\rightarrow\) is \textbf{confluent}.
\end{enumerate}
\end{theorem}

\begin{lemma}
Let \((A, \rightarrow)\) be an ARS. If \(\rightarrow\) is \textbf{confluent}, then:
\begin{enumerate}[itemsep=-0.4ex,topsep=-0.4ex]
\item Every element has at most one \textbf{normal form};
\item If \(\rightarrow\) \textbf{normalising}, then \(x \xleftrightarrow{*} y\) iff \(x\!\!\downarrow = y\!\!\downarrow\).
\end{enumerate}
\end{lemma}

\begin{theorem}[Newman's Lemma] \label{thm:newmanlem}
A \textbf{terminating} reduction is \textbf{confluent} iff it is \textbf{locally confluent}.
\end{theorem}

%% file: content/d.tex
\chapter{Graph Transformation} \label{appendix:transformation}

We give a quick introduction to the theory of algebraic graph transformation derived from my earlier literature review \cite{Campbell18a}, which is in turn derived from \cite{Ehrig-Ehrig-Prange-Taentzer06a}. This Appendix should be read along side Chapter \ref{chapter:theoryintro}, since definitions are split between that Chapter and this Appendix, such as the definition of an unlabelled graph (Section \ref{sec:maingraphs}).

Compared to \cite{Campbell18a}, we generalise to \textbf{partially labelled graphs} using \cite{Habel-Plump02a} and \cite{Plump19a}, in that we allow \textbf{relabelling} of \textbf{totally labelled graphs}. The additional sections on pushouts and pullbacks and critical pair analysis is derived from \cite{Ehrig-Ehrig-Prange-Taentzer06a} and \cite{Plump19a}, the section on rooted graphs from \cite{Bak-Plump12a}, and the section on edge replacement grammars from \cite{Habel-Kreowski83a} and \cite{Drewes93a}. The proof of Theorem \ref{theorem:uniquederivations} is by Habel and Plump \cite{Habel-Plump02a}, and of Theorem \ref{thm:undecidablegts} is given by the proof of Theorem \ref{thm:undecidablegtsrealdeal} which is due to Plump \cite{Plump93b,Plump98a,Plump05a}.


\section{Partially Labelled Graphs} \label{sec:graphpl}

\begin{definition} \label{dfn:labelalphabet}
A label alphabet \(\mathcal{L} = (\mathcal{L}_V, \mathcal{L}_E)\) consists of \textbf{finite} sets of \textbf{node labels} \(\mathcal{L}_V\) and \textbf{edge labels} \(\mathcal{L}_E\).
\end{definition}

\begin{definition} \label{dfn:plgraph}
A \textbf{concrete partially labelled graph} over a label alphabet \(\mathcal{L}\) is a \textbf{concrete graph} equipped with two \textbf{partial} label maps \(l: V \to \mathcal{L}_V\), \(m: E \to \mathcal{L}_E\): \(G = (V, E, s, t, l, m)\).
\end{definition}

\vspace{-1.4em}
\begin{figure}[H]
\centering
\noindent
\input{fig/d/labels}
\vspace{-1.4em}
\caption{Partially Labelled Graph Diagram}
\end{figure}

\begin{remark}
By this definition, we \textbf{do not} work with the free monoid on the alphabet, as in string rewriting systems. Nodes and edges are labelled exactly with the elements from the respective alphabets.
\end{remark}

\begin{definition}
We say that a \textbf{partially labelled graph} \(G\) is \textbf{totally labelled} iff \(l_G\) is total.
\end{definition}

\begin{definition}
Given a common \(\mathcal{L}\), a \textbf{partially labelled graph morphism} \(g: G \to H\) is a graph morphism on the underlying concrete graphs, with the extra constraint that labels must be preserved, if defined. That is:
\begin{enumerate}[itemsep=-0.4ex,topsep=-0.4ex]
\item \(\forall e \in E_G, \, g_V(s_G(e))= s_H(g_E(e))\);                 \tabto*{20em} [Sources]
\item \(\forall e \in E_G, \, g_V(t_G(e)) = t_H(g_E(e))\);                \tabto*{20em} [Targets]
\item \(\forall e \in E_G, \, m_G(e) = m_H(g_E(e))\);                     \tabto*{20em} [Edge Labels]
\item \(\forall v \in l_G^{-1}(\mathcal{L}_V), \, l_G(v) = l_H(g_V(v))\). \tabto*{20em} [Node Labels]
\end{enumerate}
\end{definition}

\begin{definition}
Given a common \(\mathcal{L}\), a partially labelled graph morphism \(g: G \to H\) is \textbf{undefinedness preserving} iff \(l_G = l_H \circ g_V\).
\end{definition}

\begin{definition}
Given a common \(\mathcal{L}\), a partially labelled graph morphism \(g: G \to H\) is \textbf{injective}/\textbf{surjective} iff the underlying graph morphism is injective/surjective.
\end{definition}

\begin{definition}
Given a common \(\mathcal{L}\), we say \(H\) is a \textbf{subgraph} of \(G\) iff there exists an \textbf{inclusion morphism} \(H \hookrightarrow G\). This happens iff \(V_H \subseteq V_G\), \(E_H \subseteq E_G\), \(s_H = \restr{s_G}{E_H}\), \(t_H = \restr{t_G}{E_H}\), \(m_H = \restr{m_G}{E_G}\), \(l_H \subseteq l_G\).
\end{definition}

\begin{remark}
Given a \textbf{totally labelled graph} \(G\), and \(H\) \textbf{partially labelled}. If there exists a \textbf{surjective morphism} \(G \to H\), then \(H\) is \textbf{totally labelled}.
\end{remark}

\begin{definition}
We say that graphs \(G, H\) are \textbf{isomorphic} iff there exists an \textbf{injective}, \textbf{surjective} graph morphism \(g: G \to H\) such that \(g^{-1}: H \to G\) is a graph morphism. We write \(G \cong H\), and call \(g\) an \textbf{isomorphism}. This naturally gives rise to \textbf{equivalence classes} \([G]\): the \textbf{countably} many partially labelled abstract graphs over some fixed \(\mathcal{L}\).
\end{definition}


\section{Typed Graphs} \label{section:typed}

\begin{definition}
A \textbf{typed graph} is the tuple \(G_T = (G, type_G)\) where \(G\) is an \textbf{unlabelled graph}, and \(type_G\) is a graph morphism \(G \to TG\) where \(TG\) is an \textbf{unlabelled graph} called a \textbf{type graph}. The vertices and edges of \(TG\) are called the \textbf{node alphabet} and \textbf{edge alphabet}.
\end{definition}

\begin{definition}
Given two \textbf{typed graphs} \(G_T, H_T\), a \textbf{typed graph morphism} is an \textbf{unlabelled graph morphism} \(f: G \to H\) such that \(type_H \circ f = type_G\).
\end{definition}

\begin{theorem}[Typed-Labelled Graph Correspondence]
There is a \textbf{bijective correspondence} between the \textbf{totally labelled graphs} over some fixed \textbf{label alphabet} \(\mathcal{L}\) and the \textbf{typed graphs} over \(\mathcal{L}\).
\end{theorem}


\section{Pushouts and Pullbacks}

The following propositions hold in the category of unlabelled (totally labelled) graphs, but not necessarily for partially labelled graphs.

\begin{proposition}
Every pushout satisfies the following:
\begin{enumerate}[itemsep=-0.4ex,topsep=-0.4ex]
\item \textbf{No junk}: Each item in \(D\) has a preimage in \(B\) or \(C\).
\item \textbf{No confusion}: If \(A \to B\), \(A \to C\) monic, then \(B \to D\), \(C \to D\) monic and an item from \(B\) is merged in \(D\) with an item from \(C\) only if the items have a common preimage in \(A\).
\end{enumerate}
\end{proposition}

\begin{proposition}
A \textbf{pushout} is \textbf{natural} if \(A \to B\) is \textbf{monic}.
\end{proposition}

\begin{theorem}[Limit Existence]
\textbf{Pushouts}, \textbf{pushout complements}, and \textbf{pullbacks} always exist.
\end{theorem}


\section{Rules and Derivations} \label{section:gt1}

Let \(\mathcal{L} = (\mathcal{L}_V, \mathcal{L}_E)\) be the ambient label alphabet, and graphs be concrete.

\begin{definition}
A \textbf{rule} \(r = \langle L \leftarrow K \rightarrow R \rangle\) consists of \textbf{totally labelled graphs} \(L\), \(R\) over \(\mathcal{L}\), the \textbf{partially labelled graph} \(K\) over \(\mathcal{L}\), and \textbf{inclusions} \(K \hookrightarrow L\) and \(K \hookrightarrow R\).
\end{definition}

\begin{definition}
We define the \textbf{inverse rule} to be \(r^{-1} = \langle R \leftarrow K \rightarrow L \rangle\).
\end{definition}

\begin{definition}
If \(r = \langle L \leftarrow K \rightarrow R \rangle\) is a \textbf{rule}, then \(\abs{r} = max \{\abs{L},\abs{R}\}\).
\end{definition}

\begin{definition}
Given a \textbf{rule} \(r = \langle L \leftarrow K \rightarrow R \rangle\) and a \textbf{totally labelled graph} \(G\), we say that an \textbf{injective} morphism \(g: L \hookrightarrow G\) satisfies the \textbf{dangling condition} iff no edge in \(G \setminus g(L)\) is incident to a node in \(g(L \setminus K)\).
\end{definition}

\vspace{-1.0em}
\begin{figure}[H]
\centering
\noindent
\input{fig/d/derivation}
\vspace{-1.2em}
\caption{Direct Derivation}
\label{fig:directderivation}
\end{figure}
\vspace{-0.4em}

\begin{definition} \label{def:rukeapp}
To \textbf{apply} a rule \(r = \langle L \leftarrow K \rightarrow R \rangle\) to some \textbf{totally labelled graph} \(G\), find an \textbf{injective} graph morphism \(g: L \hookrightarrow G\) satisfying the \textbf{dangling condition}, then:

\begin{enumerate}[itemsep=-0.4ex,topsep=-0.4ex]
\item Delete \(g(L \setminus K)\) from \(G\), and for each unlabelled node \(v\) in \(K\), make \(g_V(v)\) unlabelled, giving the \textbf{intermediate graph} \(D\);
\item Add disjointly \(R \setminus K\) to D, keeping their labels, and for each unlabelled node \(v\) in \(K\), label \(g_V(v)\) with \(l_R(v)\), giving the \textbf{result graph} \(H\).
\end{enumerate}

\noindent
If the \textbf{dangling condition} fails, then the rule is not applicable using the \textbf{match} \(g\). We can exhaustively check all matches to determine applicability.
\end{definition}

\begin{definition}
We write \(G \Rightarrow_{r,g} M\) for a successful application of \(r\) to \(G\) using match \(g\), obtaining result \(M \cong H\). We call Figure \ref {fig:directderivation} a \textbf{direct derivation}, and the injective morphism \(h\) the \textbf{comatch}.
\end{definition}

\begin{theorem}[Derivation Theorem] \label{theorem:uniquederivations}
It turns out that \textbf{deletions} are \textbf{natural pushout complements} and \textbf{gluings} are \textbf{natural pushouts} in the category of partially labelled graphs. Moreover, direct derivations are \textbf{natural double pushouts}, \(D\) and \(H\) are \textbf{unique up to isomorphism}, and \(H\) is \textbf{totally labelled}. Moreover, derivations \(G \Rightarrow_{r,g} H\) are \textbf{invertible}.
\end{theorem}

\begin{definition}
Given a rule set \(\mathcal{R}\), we define \(\mathcal{R}^{-1} = \{r^{-1} \mid r \in \mathcal{R}\}\).
\end{definition}

\begin{definition} \label{def:directderives}
For a given set of rules \(\mathcal{R}\), we write \(G \Rightarrow_{\mathcal{R}} H\) iff \(H\) is \textbf{directly derived} from \(G\) using any of the rules from \(\mathcal{R}\).
\end{definition}

\begin{definition} \label{def:derives}
We write \(G \Rightarrow_{\mathcal{R}}^{+} H\) iff \(H\) is \textbf{derived} from \(G\) in one or more \textbf{direct derivations}, and \(G \Rightarrow_{\mathcal{R}}^{*} H\) iff \(G \cong H\) or \(G \Rightarrow_{\mathcal{R}}^{+} H\).
\end{definition}


\section{Transformation Systems} \label{section:gt2}

\begin{definition}
A \textbf{graph transformation system} \(T = (\mathcal{L}, \mathcal{R})\), consists of a label alphabet \(\mathcal{L} = (\mathcal{L}_V, \mathcal{L}_E)\), and a \textbf{finite} set \(\mathcal{R}\) of rules over \(\mathcal{L}\).
\end{definition}

\begin{proposition}
Given a \textbf{graph transformation system} \(T = (\mathcal{L}, \mathcal{R})\), then one can always decide if \(G \Rightarrow_{\mathcal{R}} H\).
\end{proposition}

\begin{definition}
Given a \textbf{graph transformation system} \(T = (\mathcal{L}, \mathcal{R})\), we define the inverse system \(T^{-1} = (\mathcal{L}, \mathcal{R}^{-1})\).
\end{definition}

\begin{definition}
Given a label alphabet \(\mathcal{L} = (\mathcal{L}_V, \mathcal{L}_E)\), \(\mathcal{P} = (\mathcal{P}_V, \mathcal{P}_E)\) is a \textbf{subalphabet} of \(\mathcal{L}\) iff \(\mathcal{P}_V \subseteq \mathcal{L}_V\) and \(\mathcal{P}_E \subseteq \mathcal{L}_E\). We define the other standard set operations pairwise too, such as union, intersection, and difference of alphabets.
\end{definition}

\begin{definition}
Given a \textbf{graph transformation system} \(T = (\mathcal{L}, \mathcal{R})\), a subalphabet of \textbf{non-terminals} \(\mathcal{N}\), and a \textbf{start graph} \(S\) over \(\mathcal{L}\), then a \textbf{graph grammar} is the system \(\pmb{G} = (\mathcal{L}, \mathcal{N}, \mathcal{R}, S)\).
\end{definition}

\begin{definition} \label{dfn:graphgrammar}
Given a \textbf{graph grammar} \(\pmb{G}\) as defined above, we say that a graph \(G\) is \textbf{terminally labelled} iff \(l(V) \cap \mathcal{N}_V = \emptyset\) and \(m(E) \cap \mathcal{N}_E = \emptyset\). Thus, we can define the \textbf{graph language} generated by \(\pmb{G}\):
\begin{align*}
\pmb{L}(\pmb{G}) = \{[G] \mid S \Rightarrow_{\mathcal{R}}^{*} G, G \text{ terminally labelled}\}
\end{align*}
\end{definition}

\begin{proposition} \label{prop:inversegram}
Given a \textbf{graph grammar} \(\pmb{G} = (\mathcal{L}, \mathcal{N}, \mathcal{R}, S)\), \(G \Rightarrow_{r} H\) iff \(H \Rightarrow_{r^{-1}} G\), for some \(r \in \mathcal{R}\) (simply use the comatch). Moreover, \([G] \in \pmb{L}(\pmb{G})\) iff \(G \Rightarrow_{\mathcal{R}^{-1}}^* S\) and \(G\) is terminally labelled.
\end{proposition}

\begin{remark}
Graph languages need not be finite. In fact, graph grammars are as powerful as unrestricted string grammars. As such, many questions like if the language is empty, are undecidable in general.
\end{remark}


\section{Confluence and Termination}

Let \(T = (\mathcal{L}, \mathcal{R})\) be a graph transformation system.

\begin{definition}
The graphs \(H_1\), \(H_2\) are \textbf{joinable} iff there is a graph \(M\) such that \(H_1 \Rightarrow_{\mathcal{R}}^{*} M \Leftarrow_{\mathcal{R}}^{*} H_2\).
\end{definition}

\begin{definition}
\(T\) is \textbf{locally confluent} iff for all graphs \(G\), \(H_1\), \(H_2\) such that \(H_1 \Leftarrow_{\mathcal{R}} G \Rightarrow_{\mathcal{R}} H_2\), \(H_1\) and \(H_2\) are \textbf{joinable}.
\end{definition}

\begin{definition}
\(T\) is \textbf{confluent} iff for all graphs \(G\), \(H_1\), \(H_2\) such that \(H_1 \Leftarrow_{\mathcal{R}}^{*} G \Rightarrow_{\mathcal{R}}^{*} H_2\), \(H_1\) and \(H_2\) are \textbf{joinable}.
\end{definition}

\begin{definition}
\(T\) is \textbf{terminating} iff there is no infinite derivation sequence \(G_0 \Rightarrow_{\mathcal{R}} G_1 \Rightarrow_{\mathcal{R}} G_2 \Rightarrow_{\mathcal{R}} G_3 \Rightarrow_{\mathcal{R}} \cdots\).
\end{definition}

\begin{theorem}[Property Undecidability] \label{thm:undecidablegts}
Testing if \(T\) has (\textbf{local}) \textbf{confluence} or is \textbf{terminating} is \textbf{undecidable} in general.
\end{theorem}


\section{Critical Pair Analysis} \label{section:critpairs}

Throughout this section, we fix some common label alphabet \(\mathcal{L} = (\mathcal{L}_V, \mathcal{L}_E)\), and also require that the \textbf{interface} in all rules to be \textbf{totally labelled}.

\begin{definition} \label{dfn:seqindep}
The derivations \(G_1 \Rightarrow_{r_1,g_1} H \Rightarrow_{r_2,g_2} G_2\) are \textbf{sequentially independent} iff \((h_1(R_1) \cap g_2(L_2)) \subseteq (h_1(K_1) \cap g_2(K_2))\).
\end{definition}

\vspace{-1.0em}
\begin{figure}[H]
\centering
\noindent
\input{fig/d/seq}
\vspace{-1.0em}
\caption{Sequential Independence Diagram}
\end{figure}
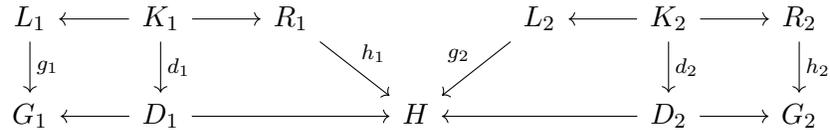
\vspace{-0.3em}

\begin{lemma} \label{lem:seqindep}
The derivations \(G_1 \Rightarrow_{r_1,g_1} H \Rightarrow_{r_2,g_2} G_2\) are \textbf{sequentially independent} iff there exist morphisms \(R_1 \to D_2\) and \(L_2 \to D_1\) with \(R_1 \to D_1 \to H = R_1 \to H\) and \(L_1 \to D_2 \to H = L_2 \to H\).
\end{lemma}

\begin{theorem}[Sequential Independence]
If \(G_1 \Rightarrow_{r_1,g_1} H \Rightarrow_{r_2,g_2} G_2\) are \textbf{sequentially independent}, then there exists a graph \(H'\) and \textbf{sequentially independent} steps \(G \Rightarrow_{r_2} H' \Rightarrow_{r_1} G_2\).
\end{theorem}

\begin{definition} \label{dfn:parindep}
The derivations \(H_1 \Leftarrow_{r_1,g_1} G \Rightarrow_{r_2,g_2} H_2\) are \textbf{parallelly independent} iff \((g_1(L_1) \cap g_2(L_2)) \subseteq (g_1(K_1) \cap g_2(K_2))\).
\end{definition}

\vspace{-0.4em}
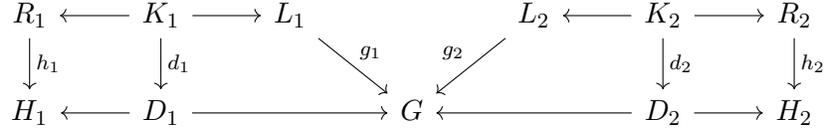
\begin{figure}[H]
\centering
\noindent
\input{fig/d/par}
\vspace{-1.0em}
\caption{Parallel Independence Diagram}
\end{figure}
\vspace{-0.3em}

\begin{lemma} \label{lem:parindep}
The derivations \(H_1 \Leftarrow_{r_1,g_1} G \Rightarrow_{r_2,g_2} H_2\) are \textbf{parallelly independent} iff there exist morphisms \(L_1 \to D_2\) and \(L_2 \to D_1\) with \(L_1 \to D_2 \to G = L_1 \to G\) and \(L_2 \to D_1 \to G = L_2 \to G\).
\end{lemma}

\begin{remark}
The derivations \(H_1 \Leftarrow_{r_1,g_1} G \Rightarrow_{r_2,g_2} H_2\) are \textbf{parallelly independent} iff \(H_1 \Rightarrow_{r_1^{-1},h_1} G \Rightarrow_{r_2,g_2} H_2\) are \textbf{sequentially independent}.
\end{remark}

\begin{theorem}[Parallel Independence]
If \(H_1 \Leftarrow_{r_1,g_1} G \Rightarrow_{r_2,g_2} H_2\) are \textbf{parallelly independent}, then there exists a graph \(G'\) and \textbf{direct derivations} \(H_1 \Rightarrow_{r_2} G' \Leftarrow_{r_1} H_2\) with \(G \Rightarrow_{r_1} H_1 \Rightarrow_{r_2} G'\) and \(G \Rightarrow_{r_2} H_2 \Rightarrow_{r_1} G'\) \textbf{sequentially independent}.
\end{theorem}

\begin{definition} \label{dfn:critpair}
A pair of \textbf{direct derivations} \(G_1 \Leftarrow_{r_1,g_1} H \Rightarrow_{r_2,g_2} G_2\) is a \textbf{critical pair} iff \(H = g_1(L_1) \cup g_2(L_2)\), the steps are not \textbf{parallelly independent}, and if \(r_1 = r_2\) then \(g_1 \neq g_2\).
\end{definition}

\begin{lemma} \label{lem:finitecritpairs}
Every graph transformation system has, up to isomorphism, only finitely many critical pairs.
\end{lemma}

\begin{definition}
Let \(G \Rightarrow H\) be a \textbf{direct derivation}. Then the \textbf{track morphism} is defined to be the partial morphism \(\mathit{tr}_{G \Rightarrow H} = \mathit{in}' \circ \mathit{in}^{-1}\), where \(\mathit{in}\) and \(\mathit{in}'\) are the bottom left and right morphisms in Figure \ref{fig:directderivation}, respectively. We define \(\mathit{tr}_{G \Rightarrow^* H}\) inductively as the composition of track morphisms.
\end{definition}

\begin{definition}
The set of \textbf{persistent nodes} of a critical pair \(\Phi : H_1 \Leftarrow G \Rightarrow H_2\) is \(\mathit{Persist}_{\Phi} = \{v \in G_V \mid \mathit{tr}_{G \Rightarrow H_1}(\{v\}), \mathit{tr}_{G \Rightarrow H_2}(\{v\}) \neq \emptyset\}\). That is, those nodes that are are not deleted by the application of either rule.
\end{definition}

\begin{definition} \label{dfn:strongjoin}
A critical pair \(\Phi : H_1 \Leftarrow G \Rightarrow H_2\) is \textbf{strongly joinable} iff it is \textbf{joinable} without deleting any of the persistent nodes, and the persistent nodes are identified. That is, there is a graph \(M\) and derivations \(H_1 \Rightarrow_{\mathcal{R}}^* M\) \(\Leftarrow_{\mathcal{R}}^* H_2\) such that \(\forall v \in \mathit{Persist}_{\Phi}, \mathit{tr}_{G \Rightarrow H_1 \Rightarrow^* M}(\{v\}) = \mathit{tr}_{G \Rightarrow H_2 \Rightarrow^* M}(\{v\}) \neq \emptyset\).
\end{definition}

\begin{theorem}[Critical Pair Lemma] \label{thm:critpairlem}
A graph transformation system \(T\) is \textbf{locally confluent} if all its \textbf{critical pairs} are \textbf{strongly joinable}.
\end{theorem}

\begin{remark} 
The reverse direction of this theorem is false, as this would contradict the undecidability of checking for confluence. If we find a non-joinable critical pair, we have non-confluence, and if we have all all critical pairs joinable, but not all strongly, we can draw no conclusions.
\end{remark}


\section{Rooted Graph Transformation} \label{section:rootedgt}

We fix some common label alphabet \(\mathcal{L} = (\mathcal{L}_V, \mathcal{L}_E)\), and allow rules to have a partially labelled interface again.

\begin{definition}
Let \(G\) be a \textbf{partially labelled graph}, and \(P_G \subseteq V_G\) be a set of \textbf{root nodes}. Then a \textbf{rooted partially labelled graph} is the tuple \(\widehat{G} = (G, P_G)\).
\end{definition}

\begin{definition}
Given two \textbf{rooted partially labelled graphs} \(\widehat{G}, \widehat{H}\), a \textbf{partially labelled graph morphism} \(g: G \to H\) is a \textbf{rooted labelled graph morphism} \(\widehat{G} \to \widehat{H}\) iff \(g_V(P_G) \subseteq P_H\). A morphism \(g: \widehat{G} \to \widehat{H}\) is \textbf{injective}/\textbf{surjective} iff the underlying graph morphism is injective/surjective. \textbf{Inclusion morphisms} and \textbf{subgraphs} are defined in the obvious way.
\end{definition}

\begin{definition}
We say that \textbf{rooted partially labelled graphs} \(\widehat{G}, \widehat{H}\) are \textbf{isomorphic} iff there exists an \textbf{injective}, \textbf{surjective} morphism \(g: \widehat{G} \to \widehat{H}\) such that \(g^{-1}: \widehat{H} \to \widehat{G}\) is also a morphism, and we write \(\widehat{G} \cong \widehat{H}\). This naturally gives rise to \textbf{equivalence classes} \([\widehat{G}]\): the \textbf{countably} many rooted partially labelled abstract graphs over some fixed \(\mathcal{L}\).
\end{definition}

\begin{definition}
\textbf{Direct derivations} on \textbf{rooted totally labelled graphs} are defined analogously as for \textbf{totally labelled graphs}, but with the following modifications to the rule application process (Definition \ref{def:rukeapp}):
\begin{enumerate}[itemsep=-0.4ex,topsep=-0.4ex]
  \item The root nodes of the \textbf{intermediate graph} are \(P_G \setminus g_V(P_L \setminus P_K)\).
  \item The root nodes of the \textbf{result graph} are \(P_D \cup h_V(P_R \setminus P_K)\).
\end{enumerate}
We write \(\widehat{G} \Rightarrow_{r,g} \widehat{M}\) for a successful application of \(r\) to \(\widehat{G}\) using match \(g\), obtaining result \(\widehat{H} \cong \widehat{M}\). We call this a \textbf{direct derivation}. Definitions \ref{def:directderives} and \ref{def:derives} are analogous.
\end{definition}

\begin{theorem}[Rooted Derivation Uniqueness] \label{thm:uniquederivations}
The \textbf{result graph} of a \textbf{direct derivation} is \textbf{unique up to isomorphism} and is \textbf{totally labelled}.
\end{theorem}

\begin{definition}
A \textbf{rooted graph transformation system} \(\widehat{T} = (\mathcal{L}, \widehat{\mathcal{R}})\), consists of a label alphabet \(\mathcal{L}\), and a \textbf{finite} set \(\mathcal{\widehat{R}}\) of rules over \(\mathcal{L}\).
\end{definition}

\begin{proposition}
Given a \textbf{rooted graph transformation system} \(\widehat{T} = (\mathcal{L}, \widehat{\mathcal{R}})\), then one can always decide if \(\widehat{G} \Rightarrow_{\widehat{\mathcal{R}}} \widehat{H}\).
\end{proposition}

\begin{definition}
Given a \textbf{rooted graph transformation system} \(\widehat{T} = (\mathcal{L}, \widehat{\mathcal{R}})\), a subalphabet of \textbf{non-terminals} \(\mathcal{N}\), and a \textbf{start graph} \(\widehat{S}\) over \(\mathcal{L}\), then a \textbf{rooted graph grammar} is the system \(\pmb{\widehat{G}} = (\mathcal{L}, \mathcal{N}, \widehat{\mathcal{R}}, \widehat{S})\).
\end{definition}

\begin{definition}
Given a \textbf{rooted graph grammar} \(\pmb{\widehat{G}}\) as defined above, we say that a graph \(\widehat{G}\) is \textbf{terminally labelled} iff \(l(V) \cap \mathcal{N}_V = \emptyset\) and \(m(E) \cap \mathcal{N}_E = \emptyset\). Thus, we can define the \textbf{graph language}:
\begin{align*}
\pmb{L}(\pmb{\widehat{G}}) = \{[\widehat{G}] \mid \widehat{S} \Rightarrow_{\mathcal{\widehat{R}}}^{*} \widehat{G}, \widehat{G} \text{ terminally labelled}\}
\end{align*}
\end{definition}


\section{Edge Replacement Grammars} \label{section:edgereplgram}

One can describe edge replacement grammars in terms of DPO-grammars.

\begin{definition}
An \textbf{edge replacement} (DPO-)grammar is a tuple \(\pmb{G} = (\mathcal{L}, \mathcal{N}, \mathcal{R}, S)\) where \(\mathcal{L}_V = \{\Square\}\), \(\mathcal{N}_V = \emptyset\), \(S = \) \tikz[baseline]{\node (a) at (0.0,0.1) [draw, circle, thick, fill=black, scale=0.3] {\,}; \node (b) at (0.5,0.1) [draw, circle, thick, fill=black, scale=0.3] {\,}; \node (A) at (0.2,0.2) {\tiny{a}}; \draw (a) edge[->,thick] (b);} for some \(a \in \mathcal{N}_E\), and the rules in \(\mathcal{R}\) are of the form given in Figure \ref{fig:edgereplrule} for some \(a \in \mathcal{N}_E\) and arbitrary graph \(R\) over \(\mathcal{L}\).
\end{definition}

\vspace{-0.4em}
\begin{figure}[H]
\centering
\noindent
\input{fig/d/grammar}
\vspace{-0.2em}
\caption{Edge Replacement Rule}
\label{fig:edgereplrule}
\end{figure}
\vspace{0.0em}

\begin{theorem}[Decidable Membership Problem]
The \textbf{membership problem} for \textbf{edge replacement} grammars is in \(NP\). Moreover, there exists a grammar such that the problem is \(NP\)-complete.
\end{theorem}

\begin{corollary}
There exists a \textbf{recursively enumerable} set of graphs that cannot be generated by an \textbf{edge replacement} grammar. Moreover, the \textbf{edge replacement} grammars generate a \textbf{recursive} class of languages.
\end{corollary}

%% file: fig/d/labels.tex
\begin{equation*}
\begin{tikzcd}
  \mathcal{L}_E
    & E \arrow[l, "m"] \arrow[r, shift left=1ex, "s"] \arrow[r, shift right=1ex, "t"]
    & V \arrow[r, "l"]
    & \mathcal{L}_V
\end{tikzcd}
\end{equation*}

%% file: fig/d/derivation.tex
\begin{equation*}
\begin{tikzcd}
  L \arrow[d, "g"]
  & K \arrow[l, ""] \arrow[d, "d"] \arrow[r, ""]
  & R \arrow[d, "h"] \\
  G 
    & D \arrow[l, ""] \arrow[r, ""]
    & H
\end{tikzcd}
\end{equation*}

%% file: fig/d/seq.tex
\begin{equation*}
\adjustbox{scale=1.0,center}{
\begin{tikzcd}
  L_1 \arrow[d, "g_1"]
  & K_1 \arrow[l, ""] \arrow[d, "d_1"] \arrow[r, ""]
  & R_1 \arrow[dr, "h_1"]
  & 
  & L_2 \arrow[dl, swap, "g_2"]
  & K_2 \arrow[l, ""] \arrow[d, "d_2"] \arrow[r, ""]
  & R_2 \arrow[d, "h_2"] \\
  G_1
  & D_1 \arrow[l, ""] \arrow[rr, ""]
  &
  & H
  &
  & D_2 \arrow[ll, ""] \arrow[r, ""]
  & G_2
\end{tikzcd}
}
\end{equation*}

%% file: fig/d/par.tex
\begin{equation*}
\adjustbox{scale=1.0,center}{
\begin{tikzcd}
  R_1 \arrow[d, "h_1"]
  & K_1 \arrow[l, ""] \arrow[d, "d_1"] \arrow[r, ""]
  & L_1 \arrow[dr, "g_1"]
  & 
  & L_2 \arrow[dl, swap, "g_2"]
  & K_2 \arrow[l, ""] \arrow[d, "d_2"] \arrow[r, ""]
  & R_2 \arrow[d, "h_2"] \\
  H_1
  & D_1 \arrow[l, ""] \arrow[rr, ""]
  &
  & G
  &
  & D_2 \arrow[ll, ""] \arrow[r, ""]
  & H_2
\end{tikzcd}
}
\end{equation*}

%% file: fig/d/grammar.tex
\begin{tikzpicture}[every node/.style={inner sep=0pt, text width=6.5mm, align=center}]
    \node (a) at (0.0,0) [draw, circle, thick] {\(\Square\)};
    \node (b) at (1.5,0) [draw, circle, thick] {\(\Square\)};

    \node (c) at (2.5,0) {$\leftarrow$};

    \node (d) at (3.5,0) [draw, circle, thick] {\(\Square\)};
    \node (e) at (5.0,0) [draw, circle, thick] {\(\Square\)};

    \node (f) at (6.0,0) {$\rightarrow$};

    \node (g) at (7.0,0) {R};

    \node (A) at (0.0,-.52) {\tiny{1}};
    \node (B) at (1.5,-.52) {\tiny{2}};
    \node (D) at (3.5,-.52) {\tiny{1}};
    \node (E) at (5.0,-.52) {\tiny{2}};

    \draw (a) edge[->,thick] node[above, yshift=2.5pt] {a} (b);
\end{tikzpicture}

%% file: content/e.tex
\chapter{Graph Theory} \label{appendix:graphtheory}

In standard literature, \enquote{graph theory} is the mathematical study of \enquote{graphs}, where in this context a graph is a finite set of vertices with (directed) edges between them, without parallel edges. We will present this theory in terms of the more general notion of a (labelled) graph from Appendix \ref{appendix:transformation}. The definitions and theorems in this appendix have been adapted from \cite{Johnson18a}, \cite{Lozin18a}, Chapter 1 of \cite{BangJensen-Gutin09a}, and Chapter 3 of \cite{Skiena08a}.


\section{Basic Definitions}

\begin{definition} \label{def:graphprops}
Given a concrete graph \(G\), \(v \in V_G\), we define the:
\begin{enumerate}[itemsep=-0.4ex,topsep=-0.4ex]
\item \textbf{Incoming degree}: \(\operatorname{indeg}_G(v) = \abs{{t_G}^{-1}(\{v\})}\);
\item \textbf{Outgoing degree}: \(\operatorname{outdeg}_G(v) = \abs{{s_G}^{-1}(\{v\} )}\);
\item \textbf{Degree}: \(\operatorname{deg}_G(v) = \operatorname{indeg}_G(v) + \operatorname{outdeg}_G(v)\);
\item \textbf{Neighbourhood}: \(\operatorname{N}_G(v) = s_G({t_G}^{-1}(\{v\})) \cup t_G({s_G}^{-1}(\{v\}))\);
\item \textbf{Closed neighbourhood}: \(\operatorname{N}_G[v] = \operatorname{N}_G(v) \cup \{v\}\).
\end{enumerate}
\end{definition}

\begin{definition} \label{dfn:leafnode}
Given a concrete graph \(G\), \(v \in V_G\), we:
\begin{enumerate}[itemsep=-0.4ex,topsep=-0.4ex]
\item Say \(v \in V_G\) is a \textbf{leaf node} iff \(\operatorname{outdeg}_G(v) = 0\);
\item Say \(u, v \in V_G\) are \textbf{adjacent} iff \(\{u, v\} \subseteq \operatorname{N}[u] \cap \operatorname{N}[v]\);
\item Say \(e \in E_G\) is \textbf{proper} iff \(s_G(e) \neq t_G(e)\).
\end{enumerate}
\end{definition}

\begin{definition}
We say two proper edges \(e, f \in E_G\) are \textbf{parallel} iff [\(s_G(e) = s_G(f)\) and \(t_G(e) = t_G(f)\)] or [\(s_G(e) = t_G(f)\) and \(s_G(e) = t_G(f)\)].
\end{definition}

\begin{definition} \label{dfn:walksconnected}
Let \(G\) be a concrete graph. Then:
\begin{enumerate}[itemsep=-0.4ex,topsep=-0.4ex]
\item An \textbf{undirected walk} of length \(k\) is a non-empty, finite sequence of alternating vertices and edges in \(G\): \(\langle v_0, e_0, v_1, e_1, \dots, e_{k-1}, v_k \rangle\), such that for each \(e_i\) (\(0 \neq i < k\)), [\(s_G(e_i) = v_i\) and \(t_G(e_i) = v_{i+1}\)] or [\(s_G(e_i) = v_{i+}\) and \(t_G(e_i) = v_{i}\)];
\item A \textbf{walk} is an undirected walk such that for each \(e_i\) (\(0 \leq i < k\)), \(s_G(e_i) = v_i\) and \(t_G(e_i) = v_{i+1}\);
\item We call a (undirected) walk \textbf{closed} iff \(v_0 = v_k\);
\item If the vertices \(v_i\) of a \textbf{walk} are all \textbf{distinct} (except possibly \(v_0 = v_k\)), we call the walk a \textbf{path};
\item A \textbf{closed walk} is called a \textbf{cycle}; a graph with no cycles is \textbf{acyclic}. Similarly, a \textbf{closed undirected walk} is called an \textbf{undirected cycle}.
\end{enumerate}
\end{definition}

\begin{definition} \label{dfn:connectedcomponent}
A graph is called \textbf{connected} iff there is an undirected walk between every pair of distinct vertices. A \textbf{connected component} of a concrete graph \(G\) is a \textbf{maximal} connected subgraph.
\end{definition}

\begin{theorem}[Graph Decomposition]
Every concrete graph \(G\) has a unique decomposition into \textbf{connected components}.
\end{theorem}

\begin{definition}
Given a concrete graph \(G\), \(v \in V_G\) we define the:
\begin{enumerate}[itemsep=-0.4ex,topsep=-0.4ex]
\item \textbf{Children}: \(\operatorname{children}_G(v) = t_G({s_G}^{-1}(\{v\}))\);
\item \textbf{Parents}: \(\operatorname{parents}_G(v) = s_G({t_G}^{-1}(\{v\}))\).
\end{enumerate}
\(u\) is a \textbf{child} of \(v\) iff \(u \in \operatorname{children}_G(v)\), and a \textbf{parent} iff \(u \in \operatorname{parents}_G(v)\).
\end{definition}

\begin{proposition}
Given a concrete graph \(G\), \(v \in V_G\). Then:
\begin{enumerate}[itemsep=-0.4ex,topsep=-0.4ex]
\item \(\operatorname{children}_G(v) \subseteq \operatorname{N}_G(v)\) and \(\operatorname{parents}_G(v) \subseteq \operatorname{N}_G(v)\);
\item \(\abs{\operatorname{children}_G(v)} \leq \operatorname{outdeg}_G(v)\) and \(\abs{\operatorname{parents}_G(v)} \leq \operatorname{indeg}_G(v)\).
\end{enumerate}
\end{proposition}


\section{Classes of Graphs} \label{sec:graphclasses}

\begin{definition} \label{dfn:boundeddegree}
We say a graph class is of \textbf{bounded degree} iff there exists some \(b \in \mathbb{N}\) such that for all graphs \(G\) in the class, \(\forall v \in V_G, \operatorname{deg}(v) \leq b\).
\end{definition}

\begin{definition} \label{dfn:discretegraph}
A graph is called \textbf{discrete} iff it has no edges.
\end{definition}

\begin{definition} \label{dfn:tree}
A \textbf{tree} is a non-empty connected graph without undirected cycles such that every node has at most one incoming edge. Moreover:
\begin{enumerate}[itemsep=-0.4ex,topsep=-0.4ex]
\item A \textbf{linked list} is a \textbf{tree} such that every node has outgoing degree at most \(1\);
\item A \textbf{binary tree} is a \textbf{tree} such that every node has outgoing degree at most \(2\);
\item A \textbf{full binary tree} is a \textbf{binary tree} such that every node has either \(0\) or \(2\) children;
\item A \textbf{perfect binary tree} is a \textbf{full binary tree} such that every \textbf{maximal path} is the same length;
\item A \textbf{forest} is a graph where each \textbf{connected component} is a \textbf{tree}.
\end{enumerate}
\end{definition}

\begin{proposition} \label{prop:treedfn}
A \textbf{tree} is a graph containing a node from which there is a unique \textbf{directed path} to each node in the graph.
\end{proposition}

\begin{definition}
A \(n \times m\)-\textbf{grid graph} is a graph with underlying unlabelled graph isomorphic to \((V, E, s, t)\) where \(V = \mathbb{Z}_n \times \mathbb{Z}_m\), \(E = (\mathbb{Z}_2 \times V) \setminus \{(0, i, m-1), (1, n-1, j) \mid i \in \mathbb{Z}_n, j \in \mathbb{Z}_m\}\), \(s(d, i, j) = (i, j)\), and \(t(d, i, j) = (i+d, j+1-d)\). We call such a graph \textbf{square} iff \(n = m\).
\end{definition}

\begin{definition}
An \(n\)-\textbf{star graph} is a graph with underlying unlabelled graph isomorphic to \((V, E, s, t)\) where \(V = \mathbb{Z}_{n+1}\), \(E = \mathbb{Z}_{n}\), and:

\vspace{-2em}
\begin{multicols}{2}
\[
s(i) =
\begin{cases}
  n & \text{ if } i \equiv 0 \text{ mod } 2 \\
  i & \text{ otherwise}
\end{cases}
\]

\[
t(i) =
\begin{cases}
  n & \text{ if } i \equiv 1 \text{ mod } 2 \\
  i & \text{ otherwise}
\end{cases}
\]
\end{multicols}
\end{definition}

An example linked list, perfect binary tree, square grid graph, and star graph can be found in Figure \ref{fig:graph-types}.